\documentclass[twocolumn]{aastex631}
\turnoffediting
\renewcommand\edit[2]{#2}

\usepackage[T1]{fontenc}

\newcommand\chisq{\ifmmode{\chi\sp{2}}\else\math{\chi\sp{2}}\fi}
\newcommand\redchisq{\ifmmode{ \chi\sp{2}\sub{\rm red}}
                    \else\math{\chi\sp{2}\sub{\rm red}}\fi}

\newcommand\mjup{$M\sub{\rm J}$}
\newcommand\rjup{$R\sub{\rm J}$}

\newcommand\rsun{$R\sub{\odot}$}

\DeclareSymbolFont{UPM}{U}{eur}{m}{n}
\DeclareMathSymbol{\umu}{0}{UPM}{"16}
\let\oldumu=\umu
\renewcommand\umu{\ifmmode\oldumu\else\math{\oldumu}\fi}
\newcommand\micro{\umu}
\renewcommand\micron{\micro m}
\newcommand\microns{\micron}

\let\oldsim=\sim
\renewcommand\sim{\ifmmode\oldsim\else\math{\oldsim}\fi}
\let\oldpm=\pm
\renewcommand\pm{\ifmmode\oldpm\else\math{\oldpm}\fi}
\newcommand\by{\ifmmode\times\else\math{\times}\fi}
\newcommand\ttt[1]{10\sp{#1}}
\newcommand\tttt[1]{\by\ttt{#1}}

\newbox{\wdbox}
\renewcommand\c{\setbox\wdbox=\hbox{,}\hspace{\wd\wdbox}}
\renewcommand\i{\setbox\wdbox=\hbox{i}\hspace{\wd\wdbox}}
\newcommand\n{\hspace{0.5em}}

\newcount\timect
\newcount\hourct
\newcount\minct
\newcommand\now{\timect=\time \divide\timect by 60
         \hourct=\timect \multiply\hourct by 60
         \minct=\time \advance\minct by -\hourct
         \number\timect:\ifnum \minct < 10 0\fi\number\minct}

\catcode`@=11

\newcommand\comment[1]{}

\newcommand\commenton{\catcode`\%=14}
\newcommand\commentoff{\catcode`\%=12}

\renewcommand\math[1]{$#1$}
\newcommand\mathshifton{\catcode`\$=3}
\newcommand\mathshiftoff{\catcode`\$=12}

\comment{alignment tab}
\let\atab=&
\newcommand\atabon{\catcode`\&=4}
\newcommand\ataboff{\catcode`\&=12}

\let\oldmsp=\sp
\let\oldmsb=\sb
\def\sp#1{\ifmmode
           \oldmsp{#1}%
         \else\strut\raise.85ex\hbox{\scriptsize #1}\fi}
\def\sub#1{\ifmmode
           \oldmsb{#1}%
         \else\strut\raise-.54ex\hbox{\scriptsize #1}\fi}
\newbox\@sp
\newbox\@sb
\def\sbp#1#2{\ifmmode%
           \oldmsb{#1}\oldmsp{#2}%
         \else
           \setbox\@sb=\hbox{\sub{#1}}%
           \setbox\@sp=\hbox{\sp{#2}}%
           \rlap{\copy\@sb}\copy\@sp
           \ifdim \wd\@sb >\wd\@sp
             \hskip -\wd\@sp \hskip \wd\@sb
           \fi
        \fi}
\def\msp#1{\ifmmode
           \oldmsp{#1}
         \else \math{\oldmsp{#1}}\fi}
\def\msb#1{\ifmmode
           \oldmsb{#1}
         \else \math{\oldmsb{#1}}\fi}
\def\supon{\catcode`\^=7}
\def\supoff{\catcode`\^=12}
\def\subon{\catcode`\_=8}
\def\suboff{\catcode`\_=12}
\def\supsubon{\supon \subon}
\def\supsuboff{\supoff \suboff}

\newcommand\actcharon{\catcode`\~=13}
\newcommand\actcharoff{\catcode`\~=12}

\newcommand\paramon{\catcode`\#=6}
\newcommand\paramoff{\catcode`\#=12}

\comment{
\let\oldlt=<
\let\oldgt=>

}

\comment{And now to turn us totally on and off...}

\newcommand\reservedcharson{ \commenton  \mathshifton  \atabon  \supsubon 
                             \actcharon  \paramon}

\newcommand\reservedcharsoff{\commentoff \mathshiftoff \ataboff \supsuboff 
                             \actcharoff \paramoff}

\newcommand\nojoe[1]{\reservedcharson #1 \reservedcharsoff}

\catcode`@=12
\reservedcharsoff

\comment{End of joe.tex character unreservations.}

\reservedcharson
\comment{ Must have ONLY ONE of these... trust these macros, they work
for CV:

for papers, etc:
}

\reservedcharsoff

\reservedcharson
\comment{Use this around text changed in each revision:}

\comment{Put the next one of these at the top of the paper file after
  each revision so that only the current changes to be reviewed show.

}
\reservedcharsoff

\comment{
There may be lots more header in the document, so turn all the
active characters on.  
We (and bibtex) use the active character ~ in names and a few other
places, and we use \sim rather than ~ when wishing for that %
symbol, so turn the active character on before the \begin{document}
even if the others are off with:
\actcharon
}
\reservedcharson

\comment{useful in revisions...move to top-aasjournal?
\usepackage[normalem]{ulem}
}

\comment{\watermark{draft}}
\received{2019 November 10}
\revised{2021 April 21}
\accepted{TBD}
\submitjournal{\textit{The Planetary Science  Journal}\comment{, in preparation. DRAFT of {\today} {\now}.}}

\comment{Paper-specific macros}
\newcommand\transit{\texttt{transit}}
\newcommand\MCcubed{\textsc{MC3}}
\newcommand\BART{\textsc{BART}}
\newcommand\BARTTest{\textsc{BARTTest}}
\newcommand\miniRT{\texttt{miniRT}}
\newcommand\TEA{\textsc{TEA}}
\newcommand\repack{\textsc{repack}}
\newcommand\CHIMERA{\textsc{CHIMERA}}
\newcommand\NEMESIS{\textsc{NEMESIS}}
\newcommand\TauREx{\textsc{Tau-ReX}}
\newcommand\rjupmean{\math{R\sub{\rm J,mean}}}

\comment{Programs are in \tt font: transit, miniRT.  They get caps at starts of sentences, but possessives are in regular font: ``\texttt{transit}'s''.
Packages are in \sc font, written as: {\BART}, {\BARTTest}, {\TEA}.  Yes, {\BART} and {\TEA} are also programs, but their package nature supersedes that.}

\shorttitle{BART I: Design, Tests, and HD 189733 b}
\shortauthors{Harrington \textit{et al.}}

\reservedcharson
\begin{document}
\reservedcharsoff
\comment{\input top-aasjournal}

\title{An Open-Source Bayesian Atmospheric Radiative Transfer (BART)
Code: I.\ Design, Tests, and Application to Exoplanet HD 189733 b}

\newcommand\affilucf{Planetary Sciences Group,
Department of Physics,
University of Central Florida,
Orlando, FL 32816-2385, USA}

\newcommand\affilucffsi{Florida Space Institute,
University of Central Florida,
Orlando, FL 32826-0650, USA}

\newcommand\affilaas{Space Research Institute,
Austrian Academy of Sciences,
Schmiedlstrasse 6,
A-8042 Graz, Austria}

\newcommand\affilnyuadph{Department of Physics,
New York University Abu Dhabi,
PO Box 129188 Abu Dhabi, UAE}

\newcommand\affilnyuadcss{Center for Space Science,
NYUAD Institute,
New York University Abu Dhabi,
PO Box 129188, Abu Dhabi, UAE}

\newcommand\affilnyuadcap{Center for Astro, Particle and Planetary Physics (CAP\sp{3}),
New York University Abu Dhabi,
PO Box 129188, Abu Dhabi, UAE}

\newcommand\affiluchile{Departamento de Astronomia,
Universidad de Chile,
Camino del Observatorio,
1515 Las Condes,
Santiago, Chile}

\newcommand\affilumich{Department of Astronomy, 
University of Michigan,
1085 S.\ University Ave.,
Ann Arbor, MI 48109, USA}

\newcommand\affilprinceton{Department of Astrophysical Sciences,
Princeton University,
Princeton, NJ 08544, USA}

\newcommand\affiloxford{Department of Physics,
University of Oxford,
Oxford OX1 3PU,
United Kingdom}

\newcommand\affilcornell{Center for Astrophysics and Planetary Science,
Space Sciences Building,
Cornell University,
Ithaca, NY 14853-6801, USA}

\author[0000-0002-8955-8531]{Joseph Harrington}
\affiliation{\affilucf}
\affiliation{\affilucffsi}

\author[0000-0002-9338-8600]{Michael D.\ Himes}
\affiliation{\affilucf}

\author[0000-0002-1347-2600]{Patricio E.\ Cubillos}
\affiliation{\affilaas}
\affiliation{\affilucf}

\author[0000-0002-0769-9614]{Jasmina Blecic}
\affiliation{\affilnyuadph}
\affiliation{\affilnyuadcap}
\affiliation{\affilucf}

\author{Patricio M.\ Rojo}
\affiliation{\affiluchile}

\author[0000-0002-8211-6538]{Ryan C.\ Challener}
\affiliation{\affilucf}
\affiliation{\affilumich}

\author{Nate B.\ Lust}
\affiliation{\affilucf}
\affiliation{\affilprinceton}

\author{M.\ Oliver Bowman}
\affiliation{\affilucf}

\author[0000-0002-3173-1637]{Sarah D.\ Blumenthal}
\affiliation{\affilucf}
\affiliation{\affiloxford}

\author[0000-0002-4989-6501]{Ian Dobbs-Dixon}
\affiliation{\affilnyuadph}
\affiliation{\affilnyuadcap}
\affiliation{\affilnyuadcss}

\author{Andrew S.\ D.\ Foster}
\affiliation{\affilucf}
\affiliation{\affilcornell}

\author{Austin J.\ Foster}
\affiliation{\affilucf}

\author[0000-0002-1024-9681]{M.\ R.\ Green}
\affiliation{\affilucf}

\author[0000-0003-4692-4607]{Thomas J.\ Loredo}
\affiliation{\affilcornell}

\author[0000-0002-8307-144X]{Kathleen J.\ McIntyre}
\affiliation{\affilucf}

\author{Madison M.\ Stemm}
\affiliation{\affilucf}

\author[0000-0002-1024-9681]{David C.\ Wright}
\affiliation{\affilucf}

\correspondingauthor{Joseph Harrington}
\email{jh@physics.ucf.edu}

\begin{abstract}

We present the open-source Bayesian Atmospheric Radiative Transfer ({\BART}) retrieval package, which produces estimates and uncertainties for an atmosphere's thermal profile and chemical abundances from observations.
\edit1{Several {\BART} components are also stand-alone packages, including the parallel} Multi-Core Markov-chain Monte Carlo (\MCcubed), which implements several Bayesian samplers; a line-by-line radiative-transfer model, {\transit}; a code that calculates Thermochemical Equilibrium Abundances, {\TEA}; and a test suite for verifying radiative-transfer and retrieval codes\edit1{, {\BARTTest}.
The} codes are in Python \edit1{and C}.
{\BART} and {\TEA} are under a Reproducible Research (RR) license, which requires reviewed-paper authors to publish a compendium of all inputs, codes, and outputs supporting the paper's scientific claims.
{\BART} and {\TEA} produce the \edit1{compendium's} content.
\edit1{Otherwise}, these codes are under permissive open-source terms, as are {\MCcubed} and {\BARTTest}, for any purpose.
This paper presents an overview of the code, {\BARTTest}, and an application to eclipse data for exoplanet HD 189733 b.
\edit1{Appendices address RR methodology for accelerating science, a reporting checklist for retrieval papers, the spectral resolution required for synthetic tests, and a derivation of} the effective sample size required to estimate any Bayesian posterior distribution to a given precision, which determines how many iterations to run.
Paper II, by \citeauthor{CubillosEtal2021psjBART2}, presents the underlying radiative-transfer scheme and an application to transit data for exoplanet HAT-P-11b.
Paper III, by \citeauthor{BlecicEtal2021psjBART3}, discusses the initialization and post-processing routines, with an application to eclipse data for exoplanet WASP-43b.
We invite the community to use and improve {\BART} and its components at http://GitHub.com/ExOSPORTS/BART/.

\end{abstract}


\section{INTRODUCTION}
\label{introduction}

Fitting an atmospheric spectrum model to remotely sensed data is the main way we learn the composition and thermal structure of planetary atmospheres, and the only technique viable for exoplanets for the forseeable future.
Inferring, or ``retrieving,'' gas properties from spectral data dates to 1859 \citep{KirchhoffBunsen1860AdPchemanalspec, Becker2017bookUnravelStarlight}.\comment{can't find this one but it's out there: Kirchhoff1859HDnatSonnenspek}
The line-by-line approach to calculating an atmospheric spectrum requires a computer, and dates to the early 1970s or before \citep{AndersonEtal1994spieHistHITRAN}.

Atmospheric retrievals for solar-system planets typically fit, either with a minimizer or by eye, a synthetic spectrum to high-resolution data (\math{\Delta\lambda/\lambda\sim} 1000 -- 100,000, where \math{\lambda} is wavelength) with signal-to-noise ratio (S/N) > 100 per wavelength channel.
In contrast, exoplanet data often have S/N \sim10 per point, and may have just a few points representing bandpasses larger than 1 {\micron}.
A simple atmospheric model with just 9 free parameters (four for chemical abundances and five for the temperature structure) might fit six or even fewer broadband photometric fluxes observed at different times and loosely called a spectrum.
The question then becomes, what kinds of atmospheres are consistent with the data, and how confidently?
The stakes for answering this question reliably are high, as the search for life \edit1{outside our solar system} will likely have its first potential positive result in spectroscopic data, and those data will be as relatively crude to the problem of life detection as the exoplanet example above was to the first exoplanet atmospheric characterizations.

Thus, it is not surprising that atmospheric radiative transfer (RT) met Bayesian inference.
Early grid-sampling approaches for exoplanet retrievals \citep{MadhusudhanSeager2009apjAtmRetrMeth} led quickly to Markov-Chain Monte Carlo (MCMC, \citealp{MadhusudhanSeager2010apjThermalHotJup}).
\citet{MadhusudhanEtal2011natWASP12batm} published the first fully Bayesian approach, with integrals over the posterior distribution to estimate the relative probabilities of two classes of models\edit1{,} C-rich and O-rich.

There are now \edit1{over} a dozen such codes, with more in development.
\citet{Madhusudhan2018bookAtmRetrExo} provides a review, and his Table 1 lists current codes, including references.
Retrieval is computationally expensive, involving a full RT calculation per Bayesian iteration.
The codes differ in their statistical approaches, with newer MCMC sampling algorithms reducing the number of iterations required to converge from tens of millions to hundreds of thousands, and Bayesian nested sampling showing promise of even faster computation, at least for models with fewer than 10 parameters \citep{Buchner2016statcomputNestedSamplingTests}.
Most of the codes use parallel processing, with Graphical Processing Units (GPUs) coming into play as well \citep{MalikEtal2017ajHELIOS, MalikEtal2018asclHELIOS, ZhangEtal2019paspPLATON}.
The latter makes sufficient simplifying assumptions (e.g., an isothermal atmosphere in chemical equilibrium) to run in minutes to an hour, much faster than the more-general approaches of most codes.
With the inclusion of high-temperature opacities, including those for ions present at such temperatures, this approach may be suitable for transits of very hot planets.

Machine learning promises dramatic speedups.
\citet{Waldmann2016apjDreamingAtmospheres} uses machine learning to identify the molecules of interest, which can otherwise take several full retrieval runs.
\citet{MarquezNeilaEtal2018natHELA} and \citet{ZingalesWaldmann2018arxivExoGAN} have both utilized machine learning as a surrogate for retrieval modeling and found comparable results, while reducing computing time from hundreds of CPU hours to seconds and minutes, respectively, although the match of posterior distributions was not perfect (a well known issue in machine learning).

Different retrieval codes address different observing geometries, depths in the atmosphere, and spectral regions.
Mid-infrared (mid-IR) data taken around secondary eclipse, for example, include both the star's spectrum and a planetary spectrum deriving primarily from the atmosphere's own dayside thermal emission and subsequent self-absorption.
They yield the planet-to-star flux ratio, derived from the disappearance of the planetary spectrum during the eclipse.
The reflected stellar component in the mid-IR is weak, due both to the rapidly falling stellar emission at these wavelengths for all types of stars and to low planetary albedoes in the mid-IR.
In the near-IR and optical, the stellar emission and planetary albedo are both larger, and the planetary emission decreases, leading to a mostly reflective planetary spectrum in the optical, except possibly for the hottest planets.

However, the flux ratio to the stellar spectrum, and thus data quality, is quite poor in the optical, by comparison to longer wavelengths.
For example, the Spitzer Space Telescope detected the first exoplanet-derived photons in eclipse observations for TrES-1 \citep{CharbonneauEtal2005apjTrES-1} and HD 209458b \citep{DemingEtal2005natHD209}.
For TrES-1, the 4.5 and 8 {\microns} planet-star flux ratios were 0.094
The optical reflected ratio of Jupiter at opposition (just before eclipse)\comment{albedo 0.538 \citep{MallamaEtal2017IcarJupAlbedo}} is about 4\tttt{-9}.
A hot-Jupiter planet with a 5
The main natural noise source in all such observations is the Poisson statistics of the unwanted stellar photons.

Phase-curve data sense varying degrees of both the dayside and the nightside, with no reflected component from the latter.
This allows signal disentanglement to relatively narrow planetary sectors, which can be identified with longitudes, \edit1{if the sub-observer latitude is low}, as is generally presumed for close-in, transiting planets.

In transit, the planetary atmosphere modulates the small component of the stellar radiation that passes through a significant atmospheric path length around the planetary limb.
It is the only geometry with no reflected component.
Disentangling what may be quite different atmospheric compositions and cloud structures at the morning, evening, and polar limb sectors is a challenge.
In addition, Rayleigh scattering and, for small planets, refraction \citep{BetremieuxSwain2018SmallPlanetLimbRef} around the planetary limb on the side away from the star complicate attempts to resolve the event in time to separate the different limb sectors, as does the unknown pole orientation.

However, resolving eclipses in time to locate the signal sources on the planetary dayside has had initial success for at least one planet.
\citet{MajeauEtal2012apjl2DMapHD189} and \citet{deWitEtal2012aapMappingHD189} both analyzed multiple Spitzer 8 {\micron} eclipses of HD 189733 b and localized the peak emission.
HD 189733 b provides the best eclipse signal of any planet, by far.
The promise of this approach, with much more and better data and combined with the retrieval technique that is the subject of this paper, is a full 3D reconstruction of the atmosphere's thermal and condensation structure, and at least 2D recovery of the chemical composition.
Doing so independently for multiple eclipses would yield a time-resolved view of an exoplanet's atmosphere, and is the holy grail of exoplanet atmospheric retrieval.

Given the demand for the method, the high impact of some of the application papers (e.g., \citealp{LineEtal2014apjRetrievalCO, MadhusudhanEtal2011natWASP12batm}), and the prospect that this might be the method that discovers the effects of life on another planet, one might wonder why there were only three established implementations in the first five years after the method matured \citep{MadhusudhanSeager2009apjAtmRetrMeth, MadhusudhanSeager2010apjThermalHotJup, MadhusudhanEtal2011natWASP12batm, BennekeSeager2012apjRetrieval, LineEtal2013apjRetrieval1}.

The answer, of course, is that the calculation is challenging to implement.
RT depends on molecular line lists containing billions of lines that change shape with temperature and pressure, and Markov chains require thousands to millions of function evaluations.
So, both memory and processor demands are high.
Implementing the calculation requires expertise in both atmospheric RT and Bayesian statistics, two disciplines rarely taught to the same people before about 2015.
The coding effort itself is involved, with practical application requiring parallel programming and attention to both memory and computational efficiency, again at a level not typically taught to scientists.

Demands on the calculation are evolving.
For example, \citet{OreshenkoEtal2017apjlWASP12b}, noting that WASP-12b is hot enough for thermochemical equilibrium to dominate, retrieve atomic abundances and compute the approximate molecular chemistry at each step.
\citet{FraineEtal2014NatHATP11b} add an arbitrary shift in the spectrum between data from different instruments.
Extensive code alterations require good software design and change management, including version control, if the code is not to become tangled and unusable over time, yet, historically, few astronomers have been trained in software engineering.

Further, the science itself poses challenges.
With the data quality now available for exoplanets, solutions can be degenerate or indeterminate \citep{DemingSeager2017jgrpIllRealAtmExo}.
Data are still noisy, sparse, coarse in wavelength space, and spatially unresolved.
Non-simultaneous observations introduce biases (e.g., stars vary over time).
The 3D atmospheric dynamics are unknown, and are difficult to infer.
The addition of even modest amounts of data can substantially change a solution, leading one to wonder whether even that solution is valid.
Nonetheless, Bayesian retrieval is the state of the art, and can at least be used to determine what future observations and physical data will most constrain an atmospheric model.

In this paper and companion works by \citet[\edit1{hereafter BART2  [in review, reviewer should have a copy]}]{CubillosEtal2021psjBART2}
and \citet[\edit1{hereafter BART3 [in review, reviewer should have a copy]}]{BlecicEtal2021psjBART3}, we describe our open-source
retrieval code, called Bayesian Atmospheric Radiative Transfer ({\BART}).
Our code was designed from the outset to be used and extended by
others.
A hybrid of user-friendly Python and fast C, there are both
user and developer documents in addition to these technical papers.
The code has been tested against both hypothetical problems with known answers and against the results of other codes on real data.

Exoplanet characterization results are famously difficult to reproduce, due to the brevity and sometimes the outright inaccuracy of published reports.
As analyses become increasingly complex, describing the hundreds of decisions in them becomes impossible in a paper of reasonable length, and would be uninteresting to most readers.
Further, it is frequently impossible to know which decisions are important.
We recently discovered significant differences between two supposedly identical calculations, one in Python and one in the Interactive Data Language, due to differences in their native median and standard deviation routines.
With some calculations (such as retrieval) taking years to implement correctly, even with a published description in hand, checking or building upon the work of others becomes impractical without access to the source code, input data, and configuration information used to support a scientific claim.

The Reproducible Research (RR) movement promotes the rapid advance of science by requiring that these be included in a compendium published with each paper.
Producing the compendium should be easy, if codes are written to support it and if researchers keep good records.
While such sharing undoubtedly accelerates science, it carries a burden and some risks for the researcher, particularly if done openly before publication.

In our development of a new kind of software license for science, we attempt to raise awareness of the problems that proprietary computer codes create with respect to research reproducibility and efficiency, and that open development creates with respect to quality, priority, and credit \citep{NASEM2018NAPNASAOpenSource}.
By restricting others' use before publication, and requiring citation and the full disclosure of code and data that support published scientific claims, our Reproducible Research Software License resolves these issues.

This paper gives an overview of {\BART}'s design goals and our approaches in achieving them, describes its high-level design, \edit1{and} discusses performance optimizations\edit1{ and the convergence of Bayesian samplers.
It} presents the {\BARTTest} test suite and its application to {\BART}, demonstrating {\BART} on data for HD 189733 b.
\edit1{Appendices} discuss reproducible research and how {\BART}'s license promotes better science, provide a checklist for users and reviewers of items to include in published \edit1{retrieval} reports, \edit1{derive the number of Monte Carlo iterations required to determine Bayesian credible regions to a given precision, and address the resolution required in synthetic retrieval spectra.}

\edit1{The paper series} include\edit1{s} full derivations of some well known equations because, in writing {\BART}, we found disagreements or incompatible conventions in the literature, equations expressed as proportionalities, and other impediments to implementing a quantitative code correctly.
Likewise, we explain some topics in more detail than needed merely to justify our claims to existing experts, \edit1{in hope that} {\BART} will be used by students (and senior researchers!)\ not already expert in all of the disparate topics involved.
Our goal is not merely to present a code, but to improve the practice and understanding of retrieval.

\edit1{Two companion papers describe {\BART}'s calculations and present additional applications.}  \edit1{\citet[BART2]{CubillosEtal2021psjBART2}}
present {\transit}, the RT code originally written by \citet{Rojo2006PhD} and heavily modified for use in {\BART}, and its application to eclipse and transit geometry.
The paper applies {\BART} to transit data for HAT-P-11b.
\edit1{\citet[BART3]{BlecicEtal2021psjBART3}} describe the inputs, user configuration, and initialization; two modules from the optimization loop, the atmospheric profile generator and the spectrum integrator; and the outputs and post-processing routines, including the calculation of contribution functions.
The paper applies {\BART} to eclipse data for WASP-43b.
Prior papers describe {\BART}'s Thermochemical Equilibrium Abundances ({\TEA}, \citealp{BlecicEtal2016apjsTEA}) code, used in the initialization, and {\BART}'s Markov Chain Monte-Carlo statistical package \citep[{\MCcubed},][]{CubillosEtal2017apjRednoise}, which drives the main loop.

\section{GOALS AND IMPLEMENTATION APPROACH}
\label{sec:goals}

{\BART}'s main goal is to infer atmospheric properties from \edit1{particularly} noisy observations, including data from multiple observatories and, eventually, simultaneous fits to different types of observations (e.g., eclipses, transits, and phase curves).
A second goal is to be a platform for comparing different approaches to the retrieval problem, to assess the effects of different inputs (such as different line lists for the same molecule), and to compare multiple algorithms.
Third, it should be flexible enough to implement future calculations, whatever they may be.
Rather than just contributing yet another algorithm into the comparison, {\BART} aims for the flexibility to implement any \edit1{applicable physics} and be a venue for conducting comparisons by varying only one part of the calculation at a time.

Thus, we designed a modular, object-oriented framework capable of such flexibility, and implemented a straightforward, one-dimensional (1D) retrieval scheme with a modest variety of thermal and chemical profiles for the initial\edit1{-}release described here and in the two companion papers.
The framework allows {\BART} to incorporate the full panoply of physical, chemical, and biological effects on planetary observations, as well as a variety of statistical and observational approaches, as code modules contributed by interested experts.

\edit1{Also, we have endeavored to give the user control over every physically relevant parameter of the calculation.
RT output-grid resolution and starting point, parameter ranges, Voigt-profile resolution and extent, number of Voigt profiles in each dimension of the Voigt-profile grid, reference pressure, layer pressures, limiting optical depth, and many others are all user choices, with reasonable defaults set for hot-Jupiter exoplanets.}

The version of {\BART} described here models the atmosphere as a 1D vertical grid of temperatures and chemical abundances.
{\BART}'s design makes it possible to implement a 3D atmospheric grid, including wavelength-dependent scattering by clouds and hazes, or to feed the RT calculation with data from global circulation or active chemical models.

To enable such contributions, we have adopted an open-source development model, provided documents for both users and developers, and implemented a variety of tests.  The tests verify the code's correctness and make it more difficult for future modifications to introduce bugs.

Since the portions most likely to be altered are also those least requiring computational speed, most {\BART} modules are written in Python and its NumPy array mathematics package, which are popular and free.
They expose the planet's physical description to the user, allowing user-written routines to alter or add to the description before the RT calculation.
The routines that process the \edit1{modeled} spectra afterward are likewise exposed.
\edit1{We supply both the source distribution and pre-compiled binaries in containers that make installation singularly easy.}

\section{HIGH-LEVEL DESIGN}
\label{sec:design}

\begin{figure}[thb]
\centering
\includegraphics[width=\linewidth, clip]{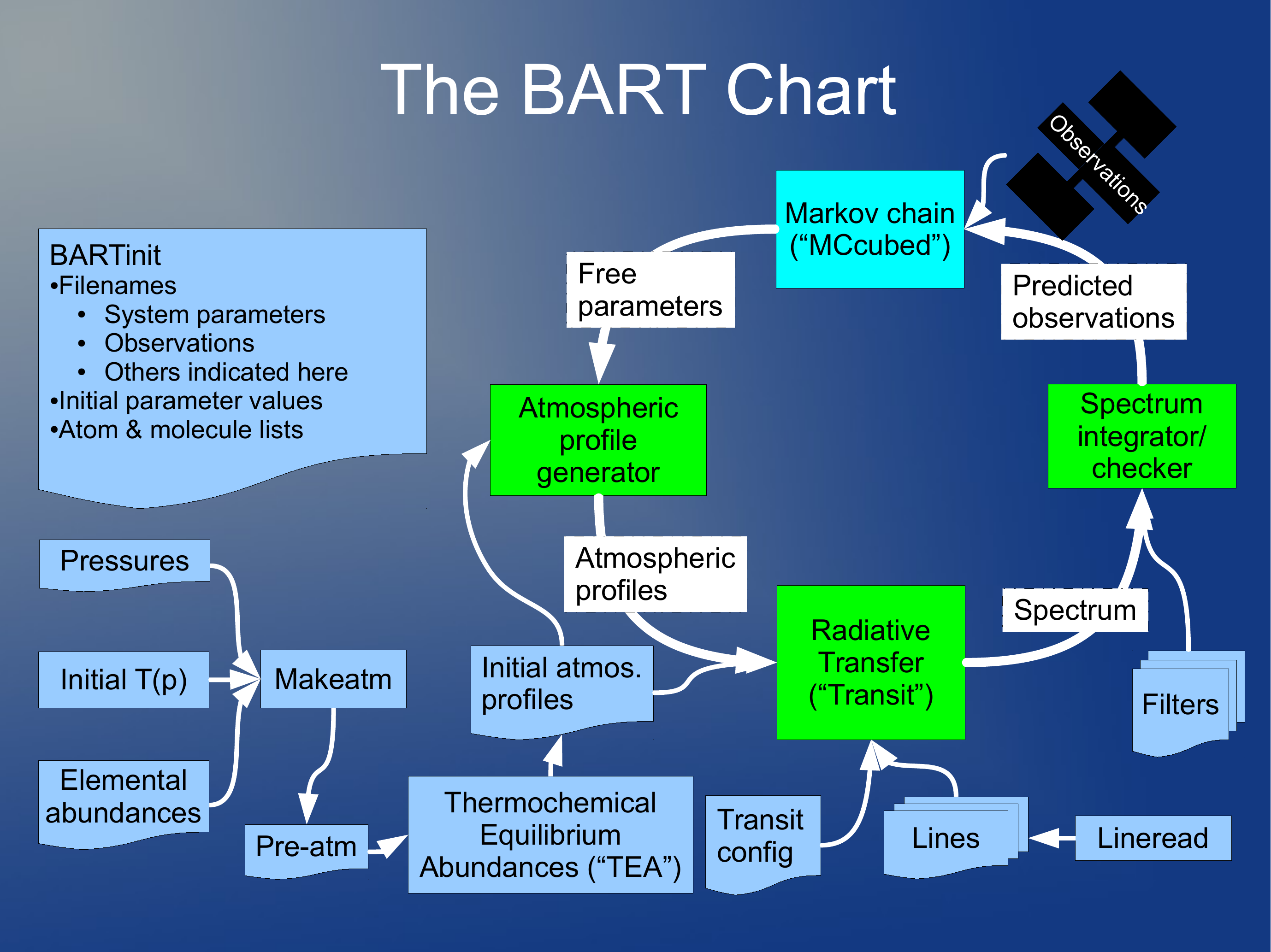}
\caption{The {\BART} Chart.
In the main optimization loop (thick white arrows in upper right), the statistics package ({\MCcubed}, aqua box, top) feeds sets of free parameters to the RT machinery (\edit1{BART2 and BART3; }green boxes are code modules\edit1{;} white boxes name the information being transferred).
Blue boxes \edit1{are initialization and support (BART3).  Straight bottoms} indicate routines\edit1{, while wavy bottoms indicate input from static data sources (files).  The black silhouetted space telescope represents the observations}.
See Section \ref{sec:design}.}
\label{fig:BARTchart}
\end{figure}

Fundamentally, {\BART} is an optimizer and parameter-space explorer.
That is, it finds the class of model atmospheres that produce the simulated spectroscopic observations that are most similar to some measured spectrum, within its uncertainties.
This is true even for problems with fewer measurements than free model parameters, although in this case a single best fit cannot be determined.
Thus, the entire RT apparatus is merely an elaborate function being fit to data.
Setting up (reasonably) self-consistent initial conditions, providing the initial data for the RT calculation, and producing plots and tabulated spectra are substantial tasks that surround the fitting procedure.

Figure \ref{fig:BARTchart} presents the execution flow of {\BART}.
The optimization loop is in the upper right (connected by thick arrows).
The initialization and support routines surround the main loop (blue boxes).
{\BART} requires numerous inputs, including planetary system parameters, the observations being modeled, initial values and limits of free model parameters, line lists, filter or spectrometer bandpass functions, the list of atmospheric layers in the model, elemental abundances for the initial atmosphere, and the list of chemical species to model.
In the chart, many of these appear as blue ``data files'' with wavy bottoms.

The initialization \edit1{(BART3)} evaluates the starting temperature-pressure profile, \math{T(p)}, in each layer.
It uses {\BART}'s Thermochemical Equilibrium Abundances ({\TEA}) code \citep{BlecicEtal2016apjsTEA} to calculate molecular abundances in each layer.
The user can choose to use these profiles, an arbitrary set of abundance profiles, or constant vertical abundances.

At the top of the main loop, our optimization and statistics package, Multi-Core Markov-Chain Monte Carlo ({\MCcubed}, aqua box in Figure \ref{fig:BARTchart}, \citealp{CubillosEtal2017apjRednoise}, see below) produces sets of atmospheric parameters from its walk through the parameter space.
The three green boxes at the bottom of the optimization loop are together the function being fit to the data.
An atmospheric profile generator turns sets of parameters from {\MCcubed} into a grid of temperatures and chemical abundances.
The {\transit} RT code uses that grid to calculate an emission spectrum or transmission spectrum modulation for the model atmosphere.
The spectrum integrator/checker produces per-channel measurement predictions from the spectrum (BART3).
{\MCcubed} then calculates {\chisq} and iterates.

{\MCcubed} implements both a simple Metropolis-Hastings sampler and the differential-evolution Markov-chain algorithms of \citet{Braak2006DifferentialEvolution} and \citet[added to {\MCcubed} since its publication]{Braak2008SnookerDEMC}.
Each MCMC chain has one {\transit} program attached to it, just one of which is illustrated in Figure \ref{fig:BARTchart}.

The profile generator \edit1{(BART3)} parameterizes \math{T(p)} several different ways.  Currently, it scales the abundance profiles it receives from the initialization by a constant parameter per profile.
With better data than currently available for exoplanets, the profile generator could apply abundance profiles with free parameters, as it now does for thermal profiles.
As very hot planets are thought always to be in thermochemical equilibrium, a faster TEA could enable recalculating equilibrium abundances at each iteration, and using atomic abundances as free parameters (e.g., \citealp{OreshenkoEtal2017apjlWASP12b}).

\edit1{BART2} details {\transit}, {\BART}'s RT code.
Initially a PhD dissertation project under JH \citep{Rojo2006PhD}, it implemented the tangent geometry of a transiting exoplanet observation.
We modified it to handle multiple similar calculations without restarting and re-initializing, feeding it different model atmospheres via the Message Passing Interface (MPI).
We also implemented the emergent-ray geometry of an eclipse observation, and optimized many of its calculations, either removing them from {\transit}'s main loop or pre-calculating them before the {\BART} run entirely.
\texttt{Transit} implements a gray cloud \edit1{deck and Rayleigh scattering; p}arameterized scattering is a planned extension.
Even with \edit1{observations} much coarser than the width of a line, the non-linearity of the radiation integral requires reasonably accurate and highly resolved line shapes, sampling \edit1{those opacities} onto a\edit1{n} output grid \edit1{that resolves the lines} in the RT forward model, and then integration over observational bandpasses to compare to data.

The RT is complex and must execute fast, so {\transit} is a C program, written in an object-oriented style, and wrapped for Python with the Simplified Wrapper Interface Generator (SWIG).
It was also designed to be altered, although with necessarily greater effort than the Python in which the rest of {\BART} is implemented.
It stores the large, static, pre-calculated opacity table in shared memory, to avoid duplicating it for each MCMC chain.

\texttt{Transit} accepts an atmospheric description that specifies the temperature and molecular composition (including isotopologues) at each of a given set of atmospheric layers.
It either computes the opacities on the spot for a given atmospheric model or interpolates from a pre-calculated table.
For forward models, it can use either approach, but retrieval requires the pre-calculated grid, for efficiency.
The 4-dimensional grid is calculated for each molecule on the wavenumber grid of the output spectrum and over the range of temperatures and pressures present in the atmosphere.
One stand-alone run of {\transit} pre-computes this grid, which can be used for many related calculations by {\BART}, or for stand-alone {\transit} runs.

The output of {\transit} is either a transmission modulation spectrum or an emission spectrum.
The spectrum integrator integrates these over filter or spectrometer bandpass transmission functions, for comparison to data by {\MCcubed}.

Once {\MCcubed} produces a sufficient number of samples (Section \ref{sec:conv}), it calculates relevant statistics and produces plots for the spectrum and all the individual and pairwise parameter histograms by marginalizing the posterior distribution.

The post-processing routines (\edit1{BART3}) identify the best-fit {\transit} run, calculate contribution functions for each bandpass, calculate the credible regions of the thermal profile, and make retrieval-specific plots.  Test and real examples, with sample inputs and outputs, appear below.

\section{OPTIMIZING THE CALCULATION}

The original \citet{Rojo2006PhD} {\transit} is a single-pass RT program that takes minutes to hours to run, depending on the number of lines, layers, and wavelength channels.
Optimization is thus critical for a Bayesian approach involving \ttt{5}--\ttt{7} iterations.
Most of the calculation involves reading the line lists and processing the data into opacities in each layer and at each wavelength.
These became {\transit}'s first pass, which produces an opacity file useful for many runs.
This pass runs before {\BART}, as a stand-alone program, and can take several hours on its own \edit1{(days} if using modern \ttt{9+}-line lists for multiple molecules \edit1{without pre-processing)}, so even this initialization required optimization.

The second pass, used in {\BART}, reads that file, receives a set of atmospheric profiles, and produces either an emission spectrum or a transmission spectrum modulation.
This pass must run in under a few seconds, regardless of the number of lines.

Here are our most significant optimizations over a single-pass RT solver \edit1{(we refer to BART2 and BART3 for details of the calculations)}:

\begin{enumerate}
\item Define separate initialization and spectrum calculation steps.
\item Put a loop around the spectrum calculation, removing as much as
possible from the loop.
\item Precalculate and save the opacity table, so the RT just interpolates the opacities in a given atmosphere.
\item Precalculate a 3D table of Voigt profiles, covering the range of expected Doppler and Lorentz widths over a wavenumber range reaching well into the distribution's broad tails.
\item Sum \edit1{the strengths of lines} with the same profiles \edit1{in the 3D Voigt table} before broadening.
\item Set a user-defined line-strength lower bound, relative to the strongest line at each temperature--pressure combination in the opacity table.
Ignore lines below this threshold.
\item Only calculate to a user-defined optical depth \edit1{at each} wavenumber.
\item Use long-lived, initialized instances of {\transit}'s second pass (``workers'' or ``servers'') that receive atmospheric profiles from the Bayesian sampler and communicate spectra to them, to avoid the overhead of restarting them.
\item Use shared memory, so all MCMC workers access one copy of the opacity table.
\item Use reliable Bayesian samplers that require few separate workers and converge quickly.
\end{enumerate}

For items 3 and 4, {\transit} computes the Voigt profiles in the 3D table at very high resolution \edit1{(thousands of samples per output grid interval) and aligned to the output grid}.
When \edit1{filling} the opacity array, \edit1{for each line, it shifts the appropriate profile by an integer number of profile samples to best match the line peak and samples its values at the output grid wavenumbers.
Due to the nonlinearity of the RT calculation, this produces more consistent and accurate spectra than averaging opacity over bins.
Appendix \edit1{\ref{ap:synthreterrors}}, BART2, and our tests against spectra calculated with other methods in Section \ref{sec:compbarstow} demonstrate the robustness of this approach.
Users must select output grids that sample a few times per line width or risk missing entire lines, although sampling other lines at their peaks mitigates this somewhat.}

The final item also bears some discussion.
{\BART} currently offers three \edit1{Bayesian} samplers.
Since the parameter space is rarely Gaussian, and often shows nonlinear parameter correlations, the Metropolis-Hastings \edit1{(MH)} algorithm converges slowly, and often not at all.
The \citet{Braak2006DifferentialEvolution} differential-evolution Monte Carlo algorithm (DEMC) \edit1{requires} twice as many chains as free parameters.
Although it generally converges much faster and more reliably than MH, DEMC still sometimes has a few chains that refuse to converge for strongly correlated parameter posteriors.
The \citet{Braak2008SnookerDEMC} Snooker DEMC algorithm (DEMCzs) converges well with as few as three chains, and typically converges faster with more (total samples matter, not just samples per chain).  
It converges reliably in our tests.

We attempted to parallelize \texttt{pylineread}, the program that ingests line lists of various formats, but found that it ran in the same time, regardless of the number of cores.
The limiting factor is the time to read and write the files, which parallelization cannot straightforwardly improve (using fast storage would help).

\section{BAYESIAN SAMPLERS AND CONVERGENCE}
\label{sec:conv}

\edit1{Although written with respect to {\BART}, this section, as with much of this paper, applies to all Bayesian retrievals, and in this case all Monte Carlo Bayesian analyses, whether using MCMC or another sampler.}

{\BART} is fundamentally a model --- the atmospheric profile generator, {\transit}, and the spectrum integrator, taken as a unit --- being compared to data by a Bayesian parameter-space explorer.
As such, it has all the benefits and liabilities of the Bayesian approach.

Our Monte Carlo convergence goals are to forget the initial conditions and to produce enough samples to estimate posterior summaries\edit1{,} such as credible regions\edit1{,} with sufficient precision.  {\BART} uses two separate approaches to accomplish these tasks.

First we ask, how many iterations convince us that the parameter-space sample resembles the unknowable posterior distribution well enough to begin sampling from it, i.e., that the initial conditions are forgotten?
This is a subtle topic in the literature, with no perfect answer.
We have implemented one of the best-known tests, that of \citet[GR,][]{GelmanRubin1992}, requiring that all parameters achieve a GR statistic within 1
Unfortunately, simply discarding the run prior to GR convergence violates the Markov assumption, introducing bias into the results.
BART thus allows the user to specify a number of burn-in steps to ignore and computes the GR statistic \edit1{for the kept samples as a convergence diagnostic}.

Second, we ask how many steps (MCMC iterations) are needed \edit1{to} find credible regions to a given accuracy.
In Appendix \edit1{\ref{ap:MCMCsteps}}, we show that the effective sample size (ESS) needed to ensure that a credible region contains estimated probability \math{\hat{C}} with accuracy \math{s\sub{\hat{C}}} is
\begin{equation}
{\rm ESS} \approx \frac{\hat{C}(1-\hat{C})}{s\sbp{\hat{C}}{2}}.
\end{equation}
In most MCMC samplers, steps are small and thus the samples are correlated.
For {\BART}, there may be \ttt{2}--\ttt{4} (or more) steps per effectively independent sample (SPEIS), depending on the sampler and the problem, and the number can vary a great deal between the parameters.
There are many methods to estimate SPEIS.
BART does it for each parameter by summing the autocorrelation function from zero lag to some small, positive threshhold of autocorrelation value, and using the largest value among the parameters.
The number of steps required is the product of SPEIS and ESS.
For example, if one calculates a 95.45\% credible region with ESS \math{\approx} 1700, the 1\math{\sigma} uncertainty on the probability content of this credible region is 0.5\%.
If SPEIS=1000, one must run 1.7\tttt{6} iterations.

\edit1{{\BART} runs a user-specified number of iterations.
Good MCMC practice is to do trial runs to determine SPEIS and the GR convergence length, calculate the required ESS, and run a number of iterations larger than the sum of the required steps and GR convergence length and many times the GR convergence length.}
\edit1{
{\BART} will discard a user-specified number of burn-in steps, which can be set to the GR convergence length calculated from trial runs.
Every 10
After the burn-in steps, {\BART} also calculates GR convergence when saving.
At the end of the run, it computes SPEIS and ESS for that run.
W}e determine \math{s\sub{\hat{C}}} for a given \math{\hat{C}} \edit1{(usually 68.27

It is possible for runs started in a high-probability region not to reach GR convergence in practical run lengths (e.g., \ttt{7} iterations, which might take many days on 10 cores, depending on many factors).
Often, one can solve such problems by choice of sampler, parameters, and priors.

A parameter-space sampler seeks the most probable location and explores around it to ascertain the credible region.
What if there is no information in one or more parameters?
That is, what if the posterior histogram for that parameter is basically flat, or flat up/down to some level, and then zero?
The sampler will run back and forth over the flat range for that parameter while trying different combinations of the remaining parameters, searching for some improvement in {\chisq}.
For many samplers, this introduces a high computational cost.

The first thing to do in such cases is to widen the allowed range of all such parameters dramatically, to ensure that good fits are not excluded by the range.
If a good fit appears in the new range, subsequent runs might use a narrower range that includes it.

If no good fits emerge, the two cases are still informative, but in a negative way.
Fairly flat posterior histograms suggest that the data may have nothing to say about that parameter.
One should eliminate the parameter, rerun the fit, and compare Bayes factors (or approximations like the Bayesian Information Criterion) to see whether the parameter is useful.
If not, reports should say that the data were uninformative regarding that parameter.
An example is an atmospheric fit including CH\sub{4} in a spectral region where CH\sub{4} has very low opacity, such that no matter how high the abundance, there would be no measurable effect on the spectrum.

Posterior histograms that are flat up/down to a cutoff indicate limits.
In our HD 189733 b retrieval (Section \ref{sec:hd189}), the spectral range includes CO lines, but none were detected.
The cutoff gives the maximum allowed abundance.
In the case of absorption, a lower limit with a flat region at higher values indicates a saturated spectrum.
Increasing the abundance changes nothing significant in the region with data; all lines/bands are already saturated.
In such cases, after exploring a wide abundance range, it is legitimate to reduce the range to a small region containing the cutoff, and report the limit.
This may speed up convergence in the final runs, depending on the sampler.

As a final check on run quality, trace plots (parameter value {\em vs.}\ iteration number) should show each parameter running over the entire allowed parameter range many times, and not sticking too long in one region, nor avoiding any.

We may hope, at some future point, to automate the selection of parameters to include in a run.
While this would certainly make things easier for the user, there is always the risk that peculiar posteriors might fool an algorithm into improperly keeping or removing a parameter.
For the time being, then, the user must be as diligent in checking the posterior plots, trace plots, and run time as in selecting the input settings, choosing the line lists, and even analyzing the input data.
{\BART} is not intended as a black box.

\section{TESTS and {\BARTTest}}
\label{sec:tests}

\edit1{At the level required for most retrievals, calculating RT is sufficiently complicated that one cannot verify the correctness of efficient codes by inspection.}  Codes calculating it are, therefore, subject to numerous types of errors (``bugs''), including subtle changes of values that produce plots that look correct, but are wrong.
We have thus developed {\BARTTest}, an independent package of quantitative and qualitative tests for both RT and retrieval codes.
This section presents an initial set of tests, using {\transit} and {\BART} as the first test subjects.

{\BARTTest} has four kinds of tests: analytic RT, comparison RT, synthetic retrieval, and real-data retrieval.
The analytic and synthetic tests quantitatively assess correctness against known results.
Several tests developed to catch file-reading and data-combining bugs also appear in the analytic group.
The comparison and real-data retrieval tests compare complex, real-world calculations among multiple codes.
The reliability of these tests depends on the strength of the consensus result.
The analytic RT tests can be useful in diagnosing differences between results if a comparison test does not match the consensus.

Given the number of variables and setup parameters, this text focuses on general description and important points.
The {\transit} and {\BART} configuration files for each test appear in the {\BARTTest} package as a reference for users to configure their own RT and retrieval codes to run these tests.
The version of {\BARTTest} described here appears in the compendium.
\textbf{These detailed configurations and their meanings in our codes are the official versions of these tests,} not the high-level descriptions in this paper.
Others performing these tests should thus configure their codes to mimic the configurations, including the specific line lists, wavelengths computed, layer boundaries, thermal profile, included species, etc.
For retrievals, it is important to use the same observations and uncertainties, even if better results appear in the literature in the future.

We encourage RT- and retrieval-code authors to validate their codes with {\BARTTest}'s analytic tests, to contribute their results to build the consensus for comparison tests, and to add new tests, especially to handle calculations not yet found in {\transit} or {\BART}, such as those involving hazes and clouds.
For example, in Section \edit1{\ref{sec:compbarstow}}, we compare {\transit} to the \edit1{hot-Jupiter cases of \citet{BarstowEtal2020mnrasRetrievalComparison}}.
Results and new tests may be submitted via pull requests at the code's development repository on GitHub.

Table \ref{tbl:tests} summarizes the tests.  Subsequent subsections expand on some of them and define terminology in the table.

\atabon\begin{longtable*}{p{2.3cm}p{4.5cm}p{4.5cm}p{5.9cm}}
\caption{Summary of Tests\label{tbl:tests}}
\\
\hline\hline Name & Purpose & Atmospheric Composition & Notes \\
\hline
\multicolumn{4}{c}{\textbf{Radiative-Transfer Analytic Tests}}\\
\hline
\texttt{f01oneline}
& Location, width, shape, and strength of a single, known line.
& One layer of LG1 at \sim40 mbar, rest CG1 (see Table \ref{tbl:gases}).
& Calculating the line shape in transmission is nontrivial, so that case is not tested.
\\
\texttt{f02fewline}
& Combination of multiple, separate lines from one molecule.
& Same as \texttt{f01oneline}, but with LG2.
& 
\\
\texttt{f03multiline} 
& Combination of lines from multiple molecules and line lists.
& Three layers each have a different gas with three lines (LG2, LG3, LG4). Others have a gas with no lines (CG1).
& 
\\
\texttt{f04broadening}
& Broadening line shape in opacity, isolating the broadening calculation from the radiation integral.
& One layer of LG1, rest CG3.
& This test produces an opacity table sampled every 0.005 cm\sp{\math{-1}} over a short range, in addition to an intensity spectrum.
All other tests produce only intensity spectra.
\\
\texttt{f05abundance}
& (Near) linear relationship between abundance and line depth at low optical depth.
& Based on \texttt{f01oneline}, but abundance varies in 10 steps uniformly from 10\sp{\math{-4}} -- 10\sp{\math{-3}}, with CG1 filling in.
& LG1 trades off against CG1 to keep line broadening constant.
Line is optically thin for near-linear absorption increase.
\\
\texttt{f06blending}
& Line blending from different molecules in the same layer.
& The \sim9 mbar layer has 1
& Two lines in LG1 and LG2 are \sim0.04 cm\sp{-1} apart at \sim2.29 {\microns}.
The wavenumber sampling interval is 0.005 cm\sp{\math{-1}}.
\\
\texttt{f07multicia}
& Multiple CIA sources, similar to \texttt{f03multiline}.
& Uniform 85
CIA line lists.
& One of the following: No CIAs, H\sub{2}-He CIAs, both H\sub{2}-H\sub{2} and H\sub{2}-He CIAs (some codes may also require LG1).
\\
\texttt{f08isothermal}
& Background emission and emission-absorption cancellation for the
isothermal case.
& Uniform composition of 60\% H\sub{2}, 10\% each CO, CO\sub{2}, CH\sub{4}, \& H\sub{2}O, using their full line lists, but no CIAs.  Constant background and atmosphere temperatures.
& Full line lists for all species.  Result is  a Planck spectrum.  Uses just \citet{SharpBurrows2007apjsAMOOpac} gases, as some codes only have these, but users may configure many more.  Not offered in transmission, as result is not a Planck spectrum.
\\
\hline
\multicolumn{4}{c}{\textbf{Radiative-Transfer Comparison Tests}}\\
\hline
\texttt{c01hjcleariso}
& Forward model of cloudless HD 189733 b-like planet.
& Isothermal \math{T(p)} profile in emission \& transmission. Mean temperature \sim1100 K. Full line lists for CH\sub{4}, CO, CO\sub{2}, H\sub{2}O, NH\sub{3}, and H\sub{2}. H\sub{2}-H\sub{2} \& H\sub{2}-He CIAs.
& Test of all code features on realistic cases.  Validated by comparison to others.\\
\texttt{c02hjclearnoinv}
& Forward model of cloudless HD 189733 b-like planet.
& Same as \texttt{c01hjcleariso}, except noninverted \math{T(p)} profile in emission \& transmission.
& Same.\\
\texttt{c03hjclearinv}
& Forward model of cloudless HD 189733 b-like planet.
& Same as \texttt{c01hjcleariso}, except inverted \math{T(p)} profile in emission \& transmission.
& Same.
\\
\edit1{\texttt{c04hjcleariso\-BarstowEtal}}
& \edit1{Forward model of cloudless HD 189733 b-like planet.}
& \edit1{Follows models of \citet{BarstowEtal2020mnrasRetrievalComparison}.  Includes some CO-only models (1.0 \rsun, 1.0 \mjup, 1.0 \rjupmean\sp{*}, 0.85 H\sub{2}:0.15 He.  10 ppmv CO at 1500 K; 100 ppmv CO at 1000 K and 1500 K) and Model 0 (0.781 \rsun, 1.162 \mjup, 1.138 \rjupmean\sp{*}, 0.85 H\sub{2}:0.15 He, 1500 K, 300 ppmv H\sub{2}O, 350 ppmv CO).}
& \edit1{Same.}
\\
\edit1{\texttt{c05hjcloudiso\-BarstowEtal}}
& \edit1{Forward model of cloudy HD 189733 b-like planet.}
& \edit1{Follows Model 1 of \citet{BarstowEtal2020mnrasRetrievalComparison} (0.781 \rsun, 1.162 \mjup, 1.138 \rjupmean\sp{*}, 0.85 H\sub{2}:0.15 He, 1500 K, 300 ppmv H\sub{2}O, 350 ppmv CO, clouddeck at 10 mbar).}
& \edit1{Same.}
\\
\hline
\hline
Name
& Purpose
& Data
& Notes\\
\hline
\multicolumn{4}{c}{\textbf{Retrieval Synthetic Tests}}\\
\hline
\texttt{s01hjcleariso}
& Can we retrieve what we put in?
& Model from \texttt{c01hjcleariso}.
 \\
\texttt{s02hjclearnoinv}
& Same.
& Model from \texttt{c02hjclearnoinv}.
 \\
\texttt{s03hjclearinv}
& Same.
& Model from \texttt{c03hjclearinv}.
 \\
\edit1{\texttt{s04hjcleariso\-BarstowEtal}}
& \edit1{Same.}
& \edit1{Model 0 from \texttt{c04hjclearisoBarstowEtal}.}
 \\
 \edit1{\texttt{s05hjcloudiso\-BarstowEtal}}
& \edit1{Same.}
& \edit1{Model 1 from \texttt{c05hjcloudisoBarstowEtal}.}
 \\
\hline
\multicolumn{4}{c}{\textbf{Retrieval Real-Data Test}}\\
\hline
\texttt{r01hd189733b}
& Reality check
& Photometry: {\em Spitzer} IRAC channels 1-4, IRS 16 {\microns}, MIPS 24 {\microns}.
Spectra: {\em Spitzer} IRS, {\em HST} NICMOS G206 grism.\sp{1}
& Multiple reductions of these data exist.  Tests must use the same eclipse depths and uncertainties as {\BARTTest} to be accurate comparisons.\\
\hline
\multicolumn{4}{l}{\parbox{\textwidth}{\sp{1} IRAC is the InfraRed Array Camera.  IRS is the InfraRed Spectrograph.  MIPS is the Multiband Imaging Photometer for Spitzer.  \textit{HST} is the \textit{Hubble Space Telescope}.  NICMOS is the Near Infrared Camera and Multi-Object Spectrograph.}}\\
\multicolumn{4}{l}{\parbox{\textwidth}{\edit1{\sp{*} \citet{BarstowEtal2020mnrasRetrievalComparison} reports the planetary radii in terms of the volumetric mean radius of Jupiter (69,911 km), rather than the IAU-defined value of \rjup (71,492 km, \citealp{IAU2016apjNominalValues}).  For codes that use the IAU value, the radii must be converted accordingly.}}}\\
\end{longtable*}\ataboff

\subsection{Analytic RT Tests}

Our simple RT code, {\miniRT}, performs {\BARTTest}'s analytic calculations for eclipse geometry (Equation (14)--(18) of \edit1{BART2}).
Written in Python, its goal is verifiability by inspection, not efficiency.
It follows how a human thinks about RT.
Additional {\BARTTest} routines accept the output of the code being tested in human-readable form, compare these to the output from {\miniRT}, and make comparison plots.

Many tests use a set of fictional gases with line lists constructed to facilitate the tests (see Table \ref{tbl:gases}).
Otherwise, these gases behave like the common molecules given in the table.
To make mathematical confirmation straightforward, there is no continuum opacity in most tests, and there are just a few lines in each list.

\begin{table}[ht]
\centering
\caption{Test Line Lists and Fictitious Test Gases}
\label{tbl:gases}
\atabon\begin{tabular}{llll}
\hline\hline
Name\sp{1} & Like\sp{2} & # of lines & \math{\lambda}\\
\hline
CG1 & H\sub{2}  & 0 & \\
CG2 & He        & 0 & \\
CG3 & N\sub{2}  & 0 & \\
LG1 & H\sub{2}O & 1 & 2.28919 \\
LG2 & CH\sub{4} & 3 & 2.28921, 2.15, 3.20\\
LG3 & CO        & 3 & 2.38, 2.50, 2.54\\
LG4 & CO\sub{2} & 3 & 2.86, 3.02, 3.78\\
\hline
\multicolumn{4}{l}{\sp{1} CG = Clear Gas, LG = Line Gas.}\\
\multicolumn{4}{l}{\parbox{3in}{\sp{2} Properties (mass, isotopes, etc.)\ same as this molecule, except line list.}}
\end{tabular}\ataboff
\end{table}

All tests have an emission (eclipse geometry) version.
{\BARTTest} includes transmission (transit geometry) cases only where it makes sense.
For example, in \texttt{f01oneline}, the emission case is simply the
product of the line strength, a Voigt curve, density, and layer thickness, which verifies by inspection.
The transmission case entails calculating the slanted path length through the single layer of LG1 of rays at multiple altitudes, calculating the optical depth and transmission as above, multiplying by \math{2\pi r}, where \math{r} is layer distance from the planet center, and integrating over altitude.
Further, the slant path may hit the layer twice or only partially.
This loses the inspection-level simplicity of the emission case and combines these calculations with the Voigt function, making the test less diagnostic.
Broadening is the same calculation for eclipse and transit geometries, so well-written codes will have one routine for it and will not require two tests.
The transmission case for test \texttt{f05abundance} would require assessing a linear change in \math{\tau} = 1 altitude, where \math{\tau} is optical depth, which has the issues outlined above and also requires many tightly spaced layers.

Next, we expand on selected tests.

\subsubsection{\texttt{f04broadening}: Line Broadening}

At least theoretically, spectral lines broaden into Voigt profiles, \math{V}.
These are the convolution of Gaussian (\math{G}) and Lorentzian (\math{L}) functions.
The Gaussian derives from Doppler broadening due to the Maxwell-Boltzmann distribution of velocities in a gas.
The Lorentzian derives from Heisenberg uncertainty in the transition energy due to short state lifetimes, especially at high pressures and temperatures.
As a function of wavenumber, \math{\nu}:
\atabon\begin{eqnarray}
G(\nu;\sigma) & = & \frac{1}{\sigma\sqrt{2\pi}}e\sp{-\nu\sp{2}/2\sigma\sp{2}},\\
L(\nu;\gamma) & = & \frac{\gamma}{\pi(\nu\sp{2}+\gamma\sp{2})},
\end{eqnarray}\ataboff
where \math{\sigma} and \math{\gamma} are the Gaussian width and Lorentzian half width, respectively (see \edit1{BART2} for a more detailed description).

\begin{figure}[tb]
\centering
\includegraphics[width=\linewidth, clip]{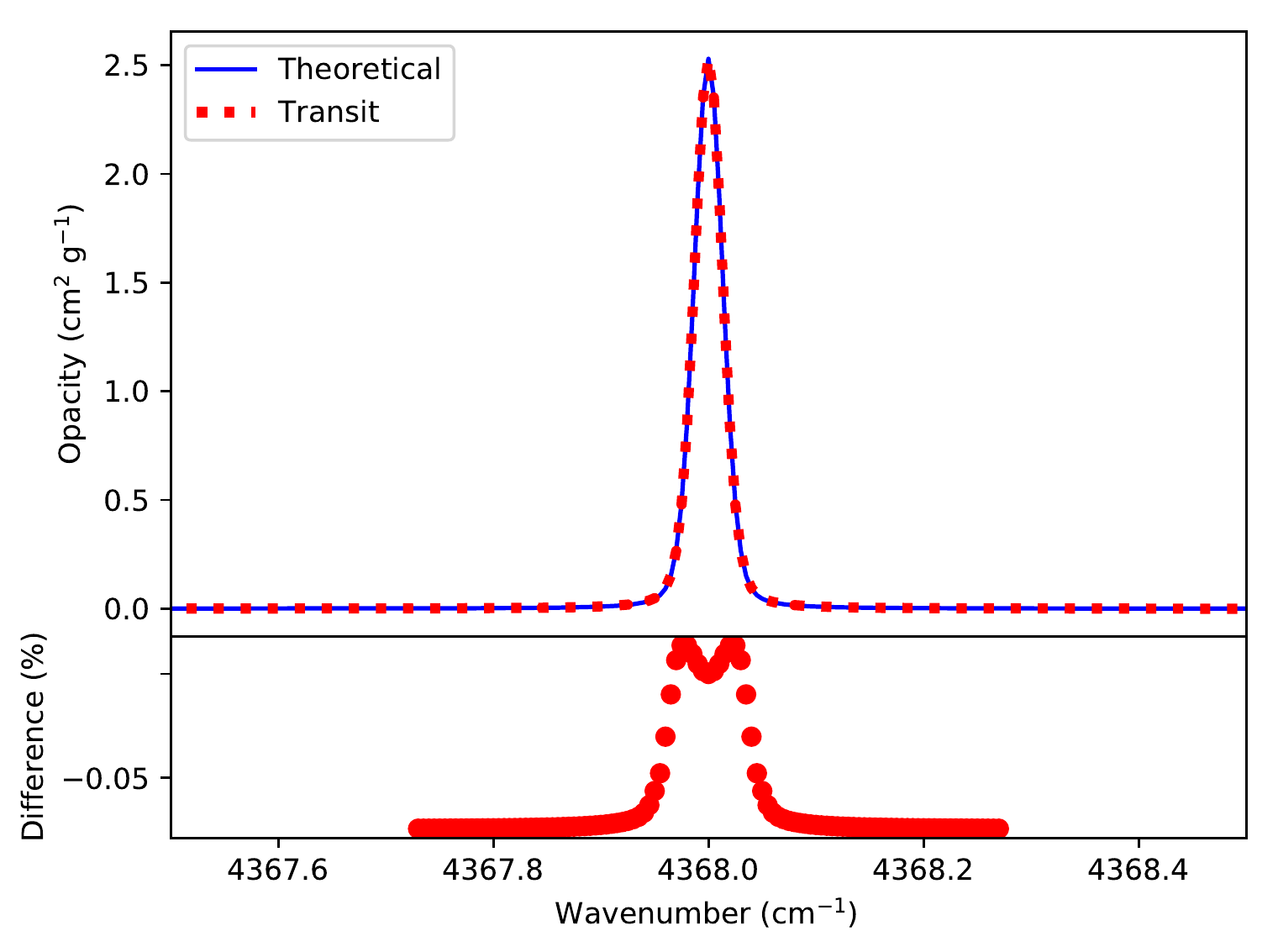}
\caption{Comparison of {\transit} line shape to {\miniRT}'s Faddeeva function for that case \edit1{(test \texttt{f04broadening})}.  Here and throughout, the axes (horizontal, in this case) may state a constant offset, for clarity.
\label{fig:voigtcomp}}
\end{figure}

The Voigt profile can be constructed from the Faddeeva function:
\atabon\begin{eqnarray}
w(z) & = & \frac{i}{\pi}
  \int\sbp{-\infty}{\infty}
    \frac{e\sp{-\eta\sp{2}}}{z-\eta}{\rm d}\eta\\
z    & = & \frac{\nu+i\gamma}{\sigma\sqrt{2}}\\
\eta & = & \frac{\nu\sp{\prime}}{\sigma\sqrt{2}}\\
w(z) & = & \frac{i}{\pi}
  \int\sbp{-\infty}{\infty}
    \frac{e\sp{\frac{-{\nu\sp{\prime}}\sp{2}}{2\sigma\sp{2}}}}
         {\left(\frac{\nu+i\gamma}{\sigma\sqrt{2}}-
          \frac{\nu\sp{\prime}}{\sigma\sqrt{2}}\right)}
    \frac{{\rm d}\nu\sp{\prime}}{\sigma\sqrt{2}}\\
     & = & \frac{i}{\pi}
  \int\sbp{-\infty}{\infty}
    \frac{e\sp{\frac{-{\nu\sp{\prime}}\sp{2}}{2\sigma\sp{2}}}
          (\nu-i\gamma-\nu\sp{\prime})}
         {(\nu-\nu\sp{\prime})\sp{2}+\gamma\sp{2}}
    {\rm d}\nu\sp{\prime}
\end{eqnarray}\ataboff
Taking the real part of the Faddeeva function, one obtains the Voigt function:
\atabon\begin{eqnarray}
\frac{{\rm Re}[w(z)]}{\sigma\sqrt{2\pi}} 
  & = & \frac{\gamma}{\sigma\sqrt{2\pi\sp{3}}}
    \int\sbp{-\infty}{\infty}
    \frac{e\sp{\frac{-{\nu\sp{\prime}}\sp{2}}{2\sigma\sp{2}}}}
         {(\nu-\nu\sp{\prime})\sp{2}+\gamma\sp{2}}
    {\rm d}\nu\sp{\prime}\\
     & = & \int\sbp{-\infty}{\infty}
  G(\nu\sp{\prime};\sigma)
  L(\nu-\nu\sp{\prime};\gamma)
  {\rm d}\nu\sp{\prime} \\
  & = & V(\nu; \sigma, \gamma)
\end{eqnarray}\ataboff

SciPy \citep{VirtanenEtal2020NatMethSciPy1} offers a Python binding to a fast C implementation of \math{w(z)}, \texttt{scipy.special.wofz()}.
{\BARTTest} uses this binding to compare to {\transit}, which uses an approximation of this function, described by \citet{PierluissiEtal1977jqsrtVoigt}.

\edit1{
Before building an opacity table, {\transit} creates a pre-calculated table of Voigt profiles for a range of \math{\sigma} and \math{\gamma} values.  
When broadening each molecular line, {\transit} uses the profile with the closest parameters.  
This approach avoids needing to calculate a Voigt profile for each line, which becomes computationally expensive for extensive line lists (e.g., ExoMol).
For each molecule, {\transit} computes the opacity table at the pressures and wavenumbers specified in the atmospheric configuration file and over a grid of temperatures.
}
\comment{Not citing https://ui.adsabs.harvard.edu/abs/2020JQSRT.25507228T/abstract as we cite the papers for the specific line lists we use.}

When calculating an emission or transmission spectrum for a planet, it interpolates in the opacity table (if specified) to the temperature of each atmospheric layer.
These approximations sharply reduce {\transit}'s run time.
The \texttt{f04broadening} test assesses their combined effect on accuracy.
This test uses the default temperature sampling interval (100 K) and a grid of Voigt profiles over 60 \math{\sigma} and 60 \math{\gamma} values.
The atmospheric layer containing the line-producing species has a temperature of 1442.58 K and a pressure of 0.33516 bar.
With a difference of {\textgreater}42 K to the closest temperature in the opacity grid, this considers a case with (almost) the largest possible interpolation error.
\texttt{Transit} differs from {\BARTTest} by {\textless}0.1\% (Figure \ref{fig:voigtcomp}).

At extremely high spectral resolution or at long wavelengths, one must configure {\transit}'s pre-calculated Voigt table to ensure sufficient accuracy, which can be assessed by running a modified version of this test (set the resolution/wavelength range, move the fake line to that location).

\subsubsection{\texttt{f05abundance}: Varying Abundance}

\begin{figure}[b]
\centering
\includegraphics[width=\linewidth, clip]{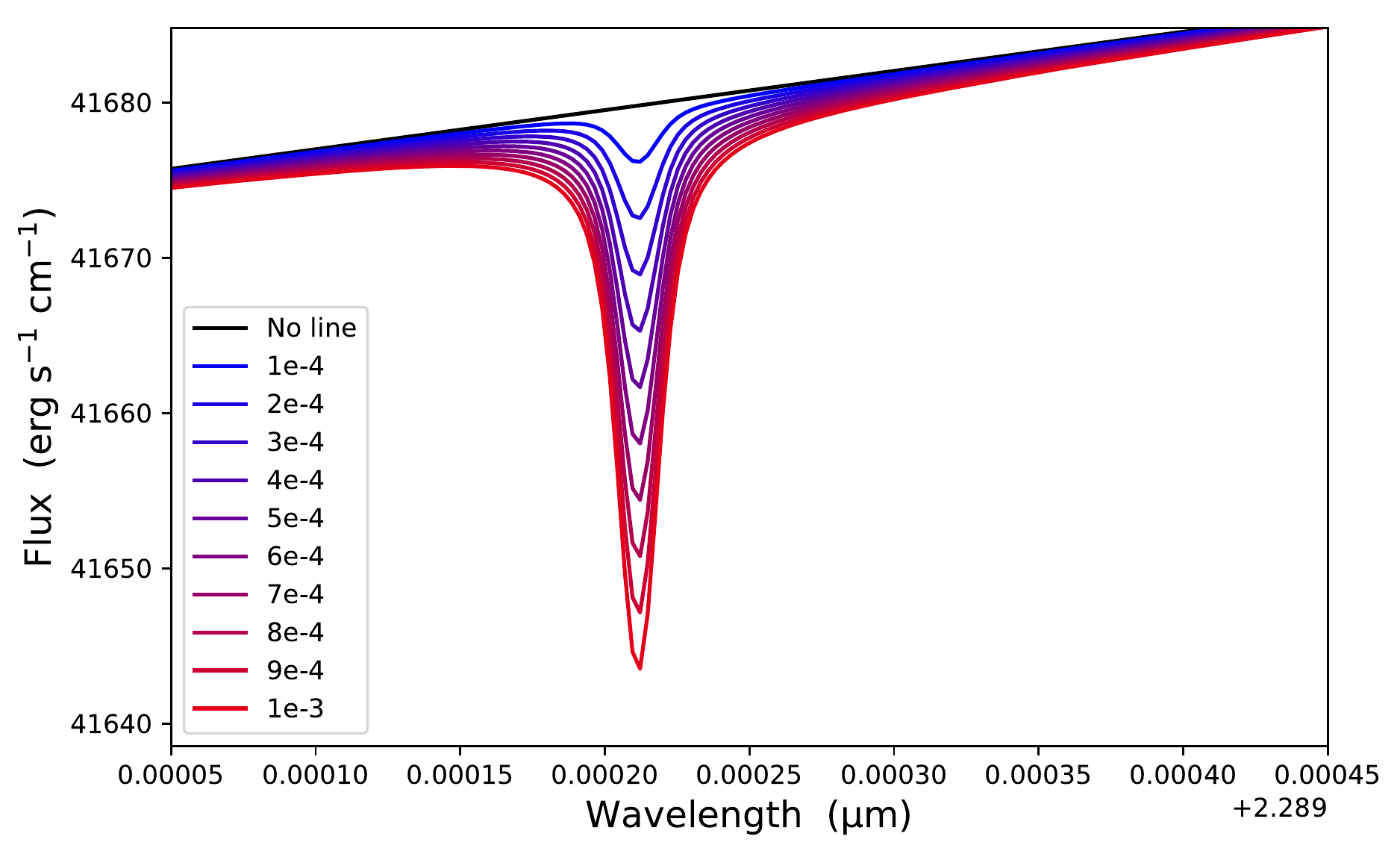}
\caption{Spectra produced by {\transit} for the eleven cases \edit1{in the \texttt{f05abundance} test}: the reference spectrum with no line, and the ten spectra with 0.01\% to 0.1\% LG1 in steps of 0.01\%.
}
\label{fig:abundance}
\end{figure}

This tripwire test relies on the property that, in the optically thin regime (optical depth \math{\tau \ll 1}), the fraction of the interior (blackbody) radiation absorbed, \math{A}, scales nearly linearly with the optical depth, since the linear term dominates in its Taylor-series expansion,
\begin{equation}
  A = 1 - \frac{F}{F\sub{0}} = 1 - e\sp{-\tau} \approx 1 - \sum\sbp{n=0}{\infty} \frac{(-\tau)\sp{n}}{n!}
  \approx \tau - \frac{\tau\sp{2}}{2} + \frac{\tau\sp{3}}{6} - \ldots,
\end{equation}
where \math{F} is transmitted flux and \math{F\sub{0}} is interior flux.
\math{\tau} is linearly dependent on the density, which is proportional to the abundance.  For two spectra of a gas with lower and higher abundance, each frequency channel should closely obey
\begin{equation}
  f = \frac{F\sub{0} - F\sub{h}}{F\sub{0} - F\sub{l}},
\end{equation}
where \math{F\sub{0}} is the spectral flux without any lines,
\math{F\sub{l}} is the flux in the low-abundance spectrum, 
\math{F\sub{h}} is the flux in the high-abundance spectrum, 
and \math{f} is the abundance ratio.

We test this using a non-inverted atmospheric model uniformly composed of 0.00\% - 0.10\% LG1, with the remainder CG1 (any CG species will do).
The abundance of LG1 varies in steps of 0.01\%.
The 0.00 abundance case is \math{F\sub{0}}, the planetary interior blackbody without the LG1 spectral line.
Taking \math{F\sub{l}} as the 0.01\% abundance spectrum and starting \math{F\sub{h}} with the 0.02\% case, \math{f} takes on the integers 2--10.
Figure \ref{fig:abundance} shows the results of {\transit}.
Table \ref{tbl:abundance} shows the output of {\BARTTest}.

\begin{table}[ht]
\centering
\caption{Varying Abundance Test}
\label{tbl:abundance}
\atabon\begin{tabular}{lll}
\hline\hline
Abundance & \math{f}\sp{1} {\em vs.} 0.01\% case\\
\hline
0.02\% & 1.999784\\
0.03\% & 2.999535\\
0.04\% & 3.999005\\
0.05\% & 4.998531\\
0.06\% & 5.997734\\
0.07\% & 6.996982\\
0.08\% & 7.995986\\
0.09\% & 8.994910\\
0.1\%  & 9.993651\\
\hline
\sp{1} Factor difference in line depth
\end{tabular}\ataboff
\end{table}

\subsubsection{\texttt{f08isothermal}: Isothermal Atmosphere}

This tripwire test recognizes that, without scattering, line emission and absorption are equal in any optically thick, isothermal gas mixture.
The mixture thus emits as a blackbody.
The atmosphere for \texttt{f08isothermal} has many molecules uniformly present in all layers and a full line list for all of the species present.
This should produce a blackbody emission spectrum peaking at the wavenumber corresponding to the atmospheric model's temperature, according to Wien's law.
Deviations may arise from approximations or precision issues in the code.
Figure \ref{fig:isothermal} shows the theoretical Planck function plotted over {\transit}'s result.

\begin{figure}[htb]
\centering
\includegraphics[width=\linewidth, clip]{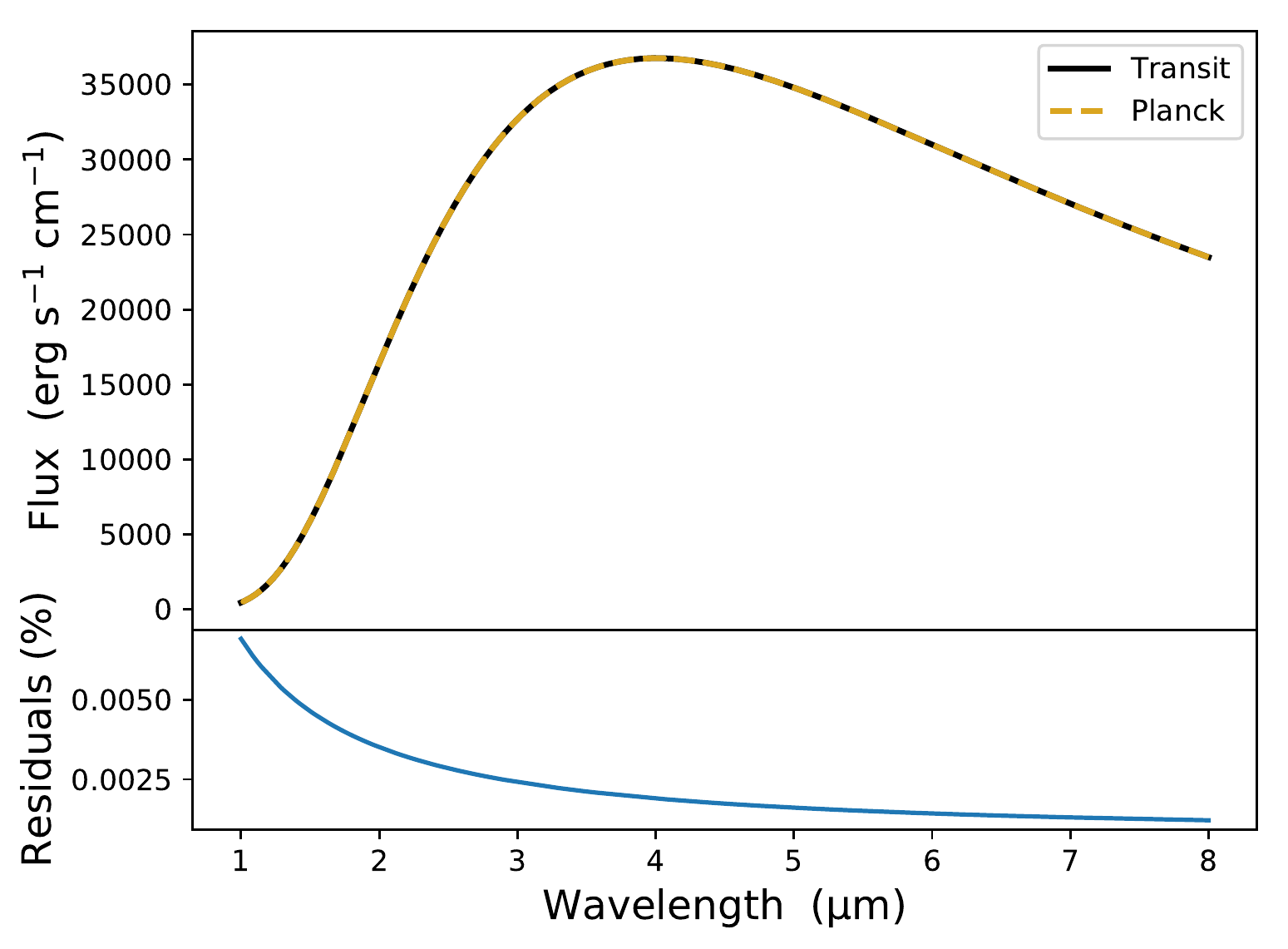}
\caption{\texttt{Transit}'s result for an isothermal atmosphere with theoretical Planck function overplotted \edit1{(test \texttt{f08isothermal})}.}
\label{fig:isothermal}
\end{figure}

\subsection{Comparison RT Test}
\label{sec:comparisontest}

To avoid aggregating working parts into an erroneous whole, one must validate the entire RT calculation.
As the complexity is too great to verify reliably by inspection, \edit1{\texttt{c01hjcleariso, c02hjclearnoinv,} and \texttt{c03hjclearinv} are} tests \edit1{inspired by} the HD 189733 system.
The strength of \edit1{these tests} rests on the number of participating codes, so we invite the community to perform these calculations in their own codes and to submit the results for inclusion in {\BARTTest}.

It is important to identify the sources of any differences without assuming that the tested codes are correct, as there is no assurance that several codes do not all share the same bug.
With such honest testing, as the group of tested codes grows, so the likelihood of a groupthink bug decreases.
The comparison of many codes implementing a single model also shows the range of outputs due to modeling approaches and assumptions, even with identical inputs.

Our model planet resembles HD 189733 b.
Tests must adopt the stellar radius of 0.756 solar radii, stellar temperature of 5000 K, planetary radius of 1.138 \rjup \citep{TorresEtal2008apjHD189733b}, and planetary gravity of 2182.73 cm s\sp{\math{-2}}, which corresponds to a mass of 1.14 \mjup.
The reference pressure of 0.1 bar corresponds to this planetary radius.
Tests must calculate spectra for the wavelength region between 1 and 11 {\microns}, as the most spectroscopically active species show features in this region.
Tests should calculate both emission intensity spectra (secondary eclipse) and transmission modulation spectra (\edit1{primary} transit).

\begin{figure}[ht]
\centering
\includegraphics[trim={0 1.8cm 0 0},width=\columnwidth]{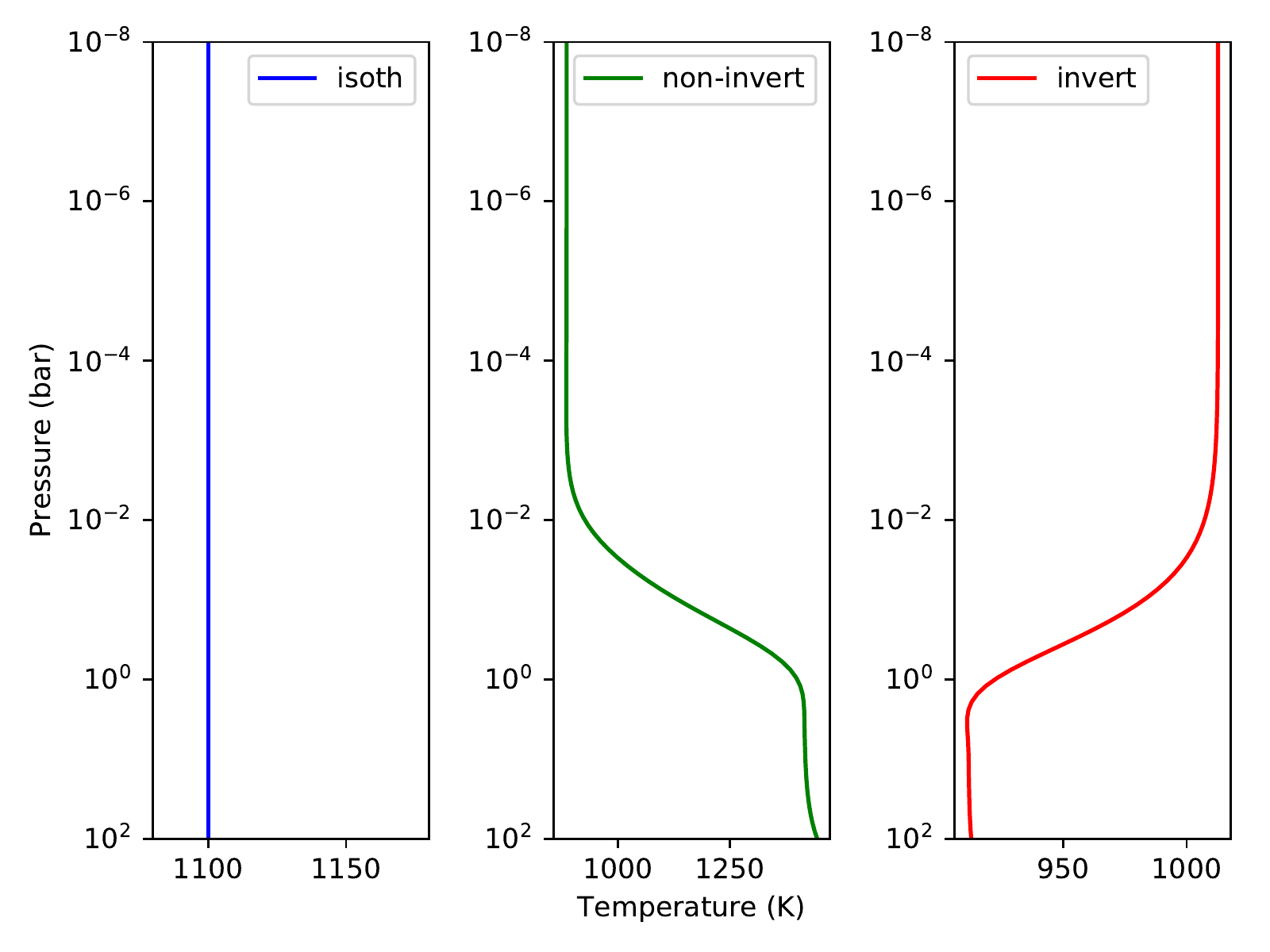}
\vspace{10pt}
\caption{Isothermal, non-inverted, and inverted \math{T(p)} profiles
used for the RT model tests \texttt{c01hjcleariso}, \texttt{c02hjclearnoinv}, \texttt{c03hjclearinv}\edit1{, and their associated retrieval tests (\texttt{s01hjcleariso}, \texttt{s02hjclearnoinv}, and \texttt{s03hjclearinv})}.
The inverted and non-inverted profiles use the \math{T(p)} parameterization of \citet{LineEtal2013apjRetrieval1}.
The profiles have effective temperatures generally around 1100 K, to resemble HD 189733 b.
\label{fig:PTs}}
\end{figure}

The common inputs are the isothermal, inverted, and
non-inverted \math{T(p)} profiles given in Figure \ref{fig:PTs}.
Profiles are close to an effective temperature \math{T\sub{\rm eff}} =
1100 K, assuming zero albedo and uniform day-night distribution.  They
derive from the temperature-parameterization model
of \citet{LineEtal2013apjRetrieval1}.

\subsubsection{\citet{BarstowEtal2020mnrasRetrievalComparison} Forward Models}
\label{sec:compbarstow}

\edit1{
To compare {\BART} with additional peer-reviewed codes, we emulate some of the setups described in \citet{BarstowEtal2020mnrasRetrievalComparison}, which were executed using the {\NEMESIS} \citep{IrwinEtal2008jqsrtNEMESIS, LeeEtal2012MNRASHD189733b}, {\CHIMERA} \citep{LineEtal2013apjRetrieval1}, and {\TauREx} \citep{WaldmannEtal2015apjTauREx} codes.  
These codes utilize the correlated-\math{k} method, whereas {\transit} uses a line-by-line approach.
(If the user-selected output resolution is low enough to miss some lines, it is properly called ``line sampling'' rather than ``line-by-line'', but there is no difference in what the code does.
As this is a user choice, we call it ``line-by-line'', below.)
Specifically, we emulate the setups for the cloud-free Model 0, cloudy Model 1, and three of the CO-only cases.  
We summarize these setups in Table \ref{tbl:tests} (see \texttt{c04hjclearisoBarstowEtal} and \texttt{c05hjcloudisoBarstowEtal}), but we direct readers to \citet{BarstowEtal2020mnrasRetrievalComparison} for more detailed descriptions of the tests and to {\BARTTest} for the exact setups.
We note that \citet{BarstowEtal2020mnrasRetrievalComparison} report planetary radii in terms of Jupiter's mean volumetric radius (\rjupmean, 69,911 km), rather than the IAU-defined value of \rjup \citep[71,492 km,][]{IAU2016apjNominalValues}; not properly accounting for this will lead to a vertical offset in the transmission spectra.
}

\begin{figure*}[htp]
\includegraphics[width=0.49\textwidth, clip=True]{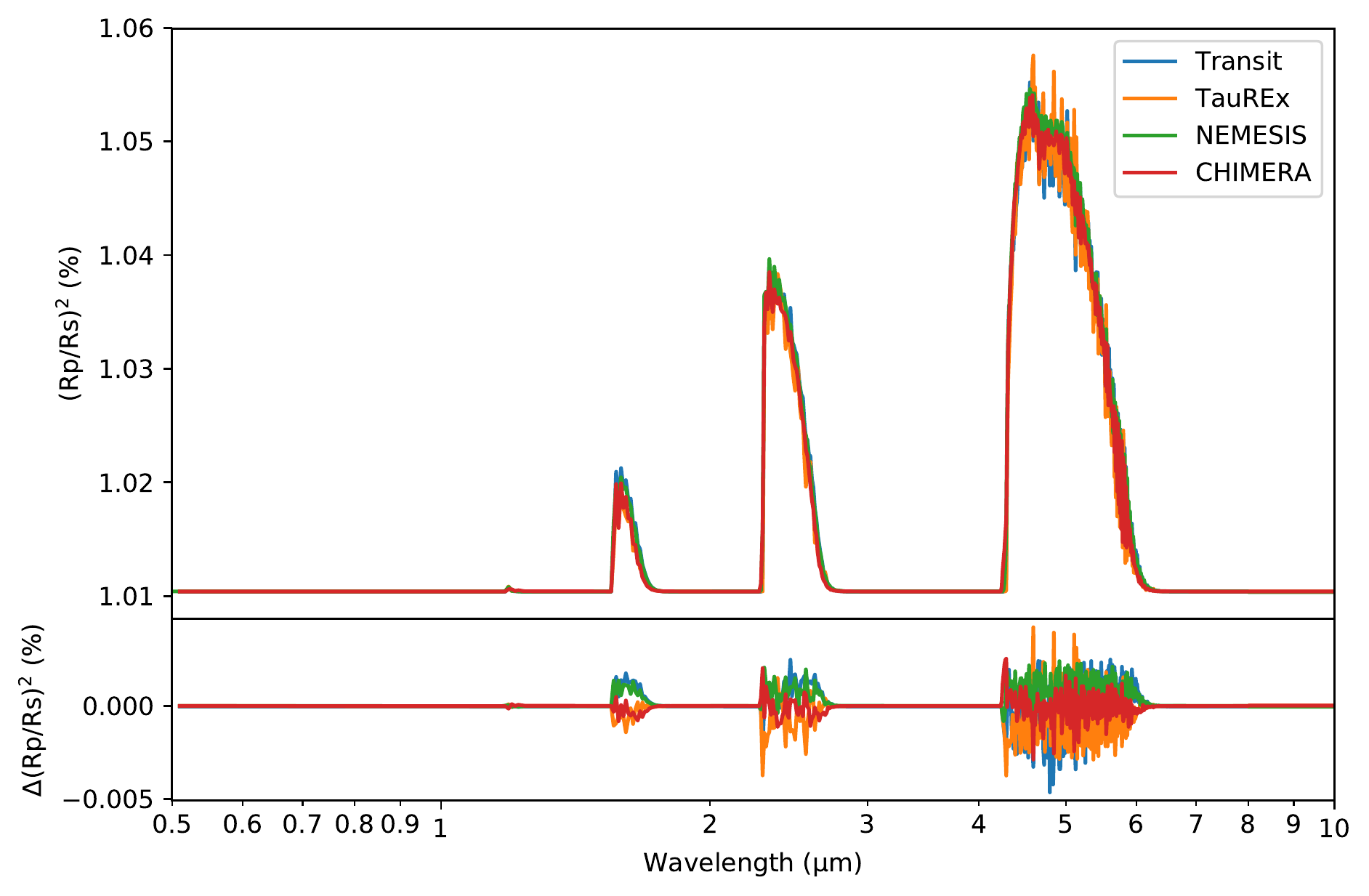}\hfill
\includegraphics[width=0.49\textwidth, clip=True]{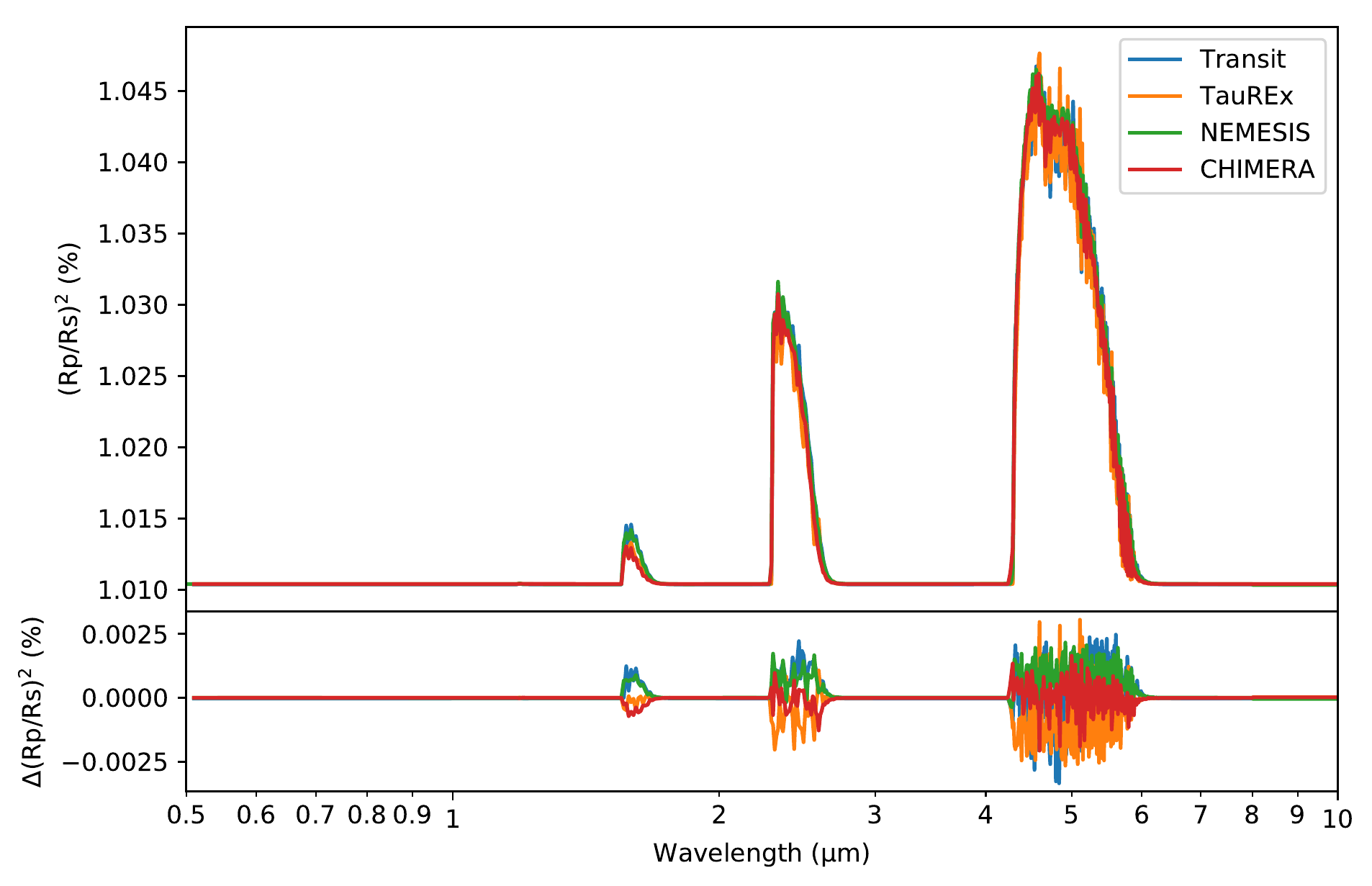}\\
\includegraphics[width=0.49\textwidth, clip=True]{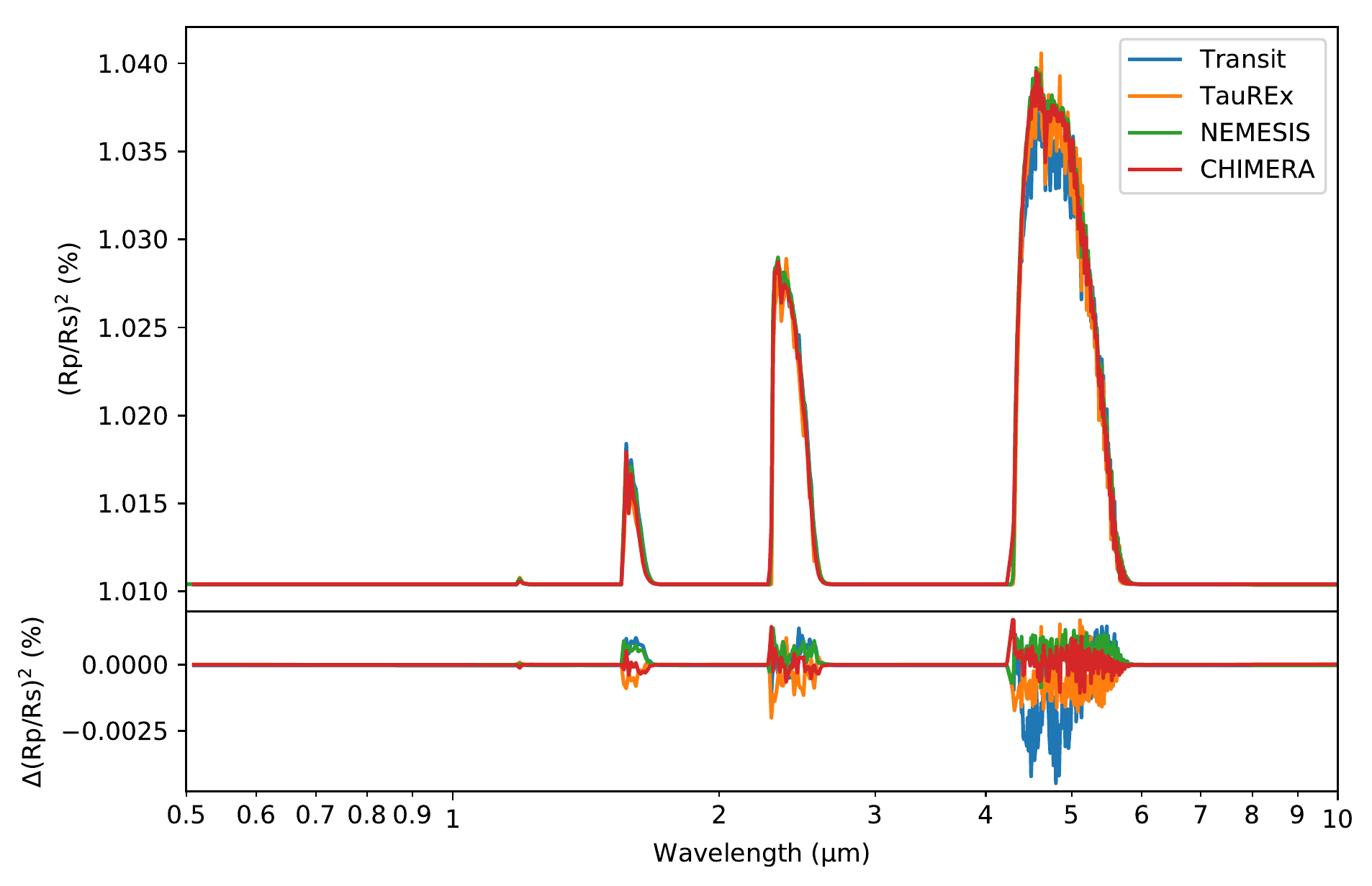}\hfill
\includegraphics[width=0.49\textwidth, clip=True]{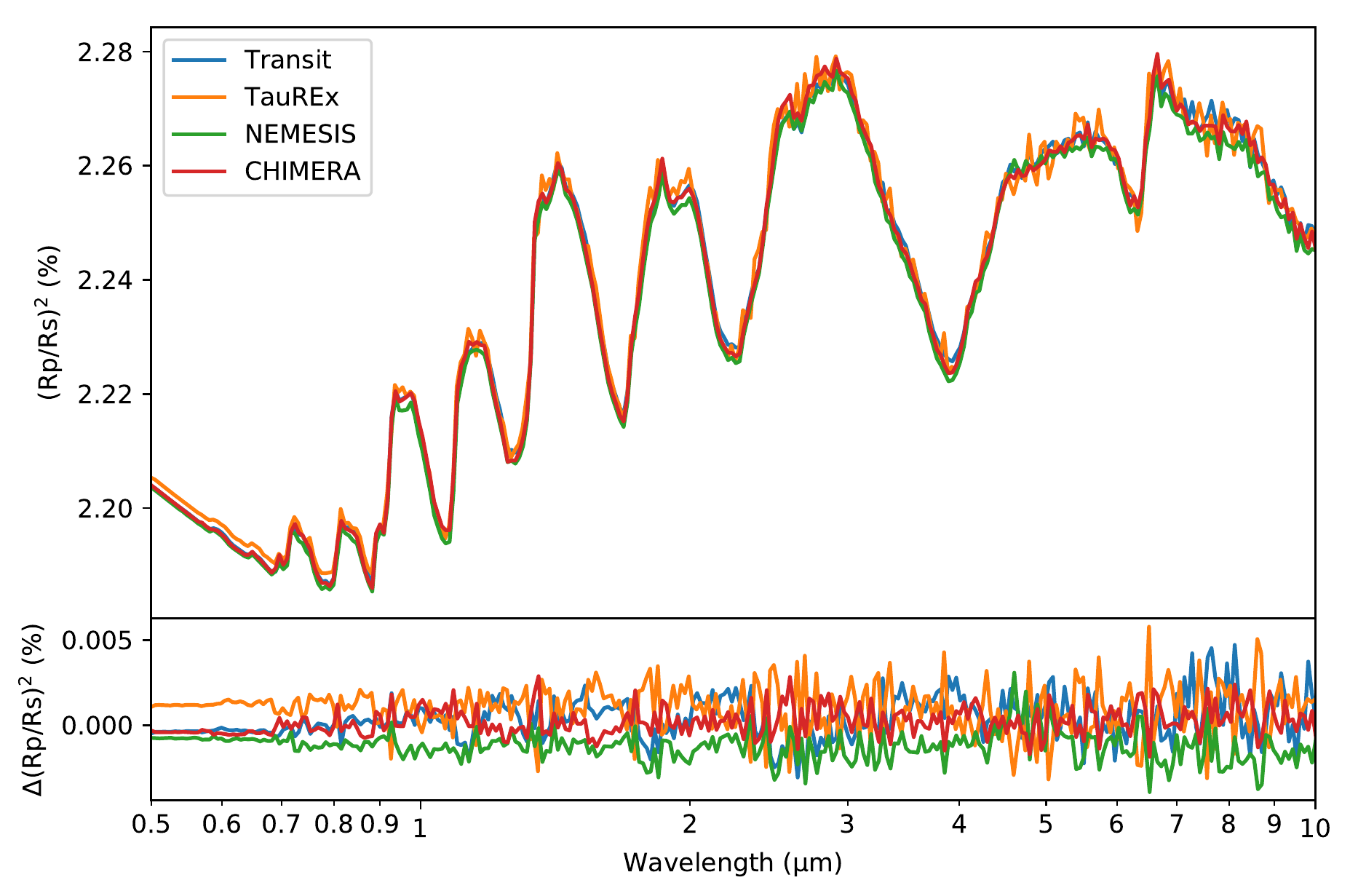}\\
\includegraphics[width=0.49\textwidth, clip=True]{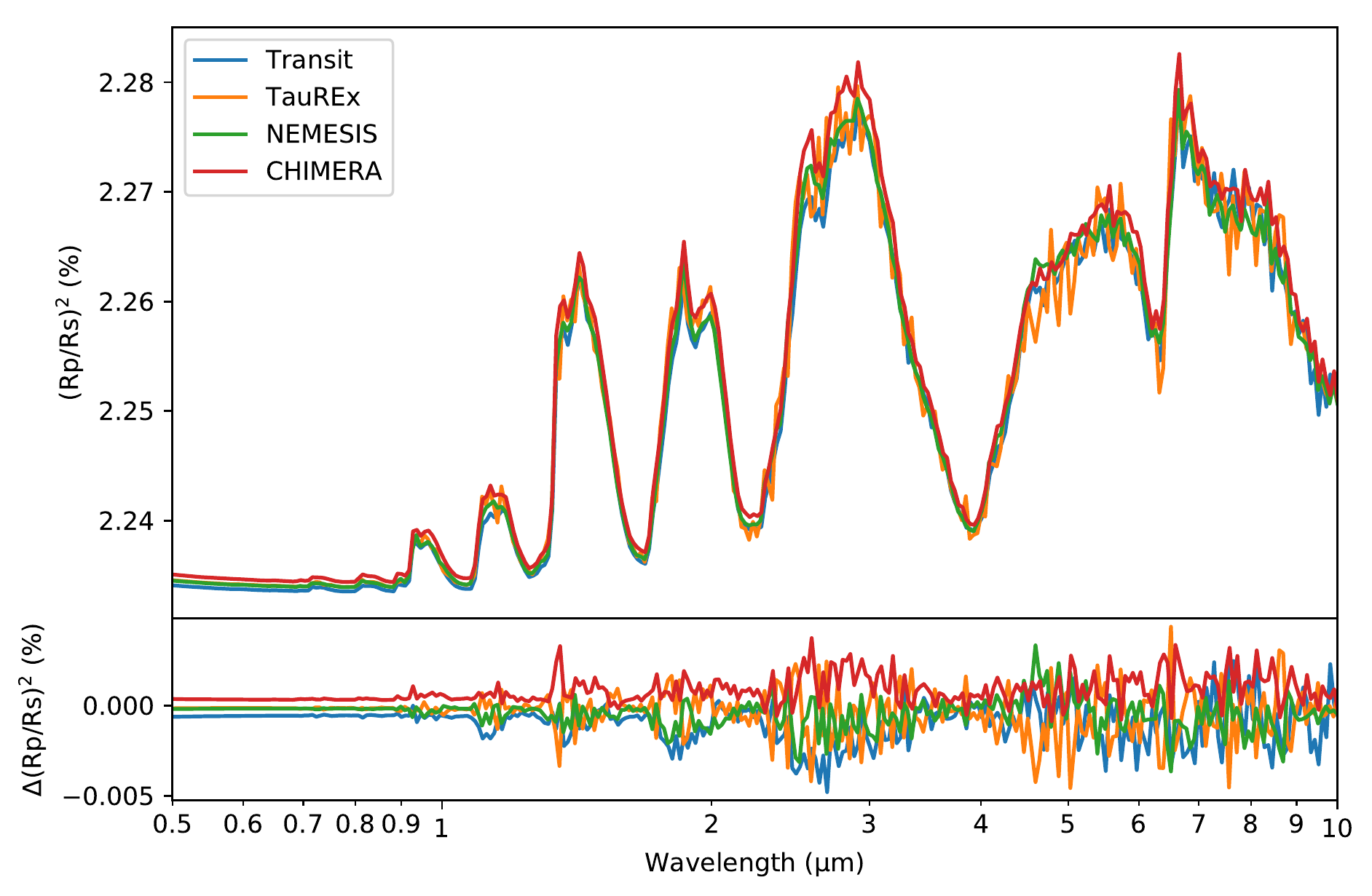}\hfill\\
\caption{Forward spectra for the \texttt{c04hjclearisoBarstowEtal} and \texttt{c05hjcloudisoBarstowEtal} tests: CO at 1500 K for \ttt{-4} mixing ratio (top left) and \ttt{-5} mixing ratio (top right), CO at 1000 K for \ttt{-4} mixing ratio (middle left), the cloudfree HD 189733 b-inspired Model 0 (middle right) and cloudy Model 1 (bottom left).  The plotted residuals are computed based on the average of the {\NEMESIS}, {\CHIMERA}, and {\TauREx} spectra, as in \citet{BarstowEtal2020mnrasRetrievalComparison}.
\label{fig:BarstowEtal-synth-spectra}}
\end{figure*}

\edit1{
Figure \ref{fig:BarstowEtal-synth-spectra} shows comparisons between the spectra produced by {\transit}, {\NEMESIS}, {\CHIMERA}, and {\TauREx}.  
As in \citet{BarstowEtal2020mnrasRetrievalComparison}, we bin the CO spectra to steps of 0.01 {\microns} and compute the residuals with respect to the average of the {\NEMESIS}, {\CHIMERA}, and {\TauREx} spectra; for the Model 0 and 1 cases, we bin according to {\CHIMERA}'s reported wavelengths.  
In general, there is close agreement between {\transit} and the other codes.
The differences are on the order of the differences between the other codes, despite {\transit}'s \edit1{opacity-sampling} approach.
}

\subsection{Synthetic Retrieval Tests}
\label{sec:synthrets}

To test {\BART} retrievals, we used the synthetic planets of the \texttt{c01hjcleariso}, \texttt{c02hjclearnoinv}, \texttt{c03hjclearinv}\edit1{, \texttt{c04hjclearisoBarstowEtal}, and \texttt{c05hjcloudisoBarstowEtal}} tests.
The tests are called \texttt{s01hjcleariso}, \texttt{s02hjclearnoinv}, \texttt{s03hjclearinv}\edit1{, \texttt{s04hjclearisoBarstowEtal}, and \texttt{s05hjcloudisoBarstowEtal}}.
We consider both eclipse and transit geometry for each atmospheric model\edit1{, except for the \citet{BarstowEtal2020mnrasRetrievalComparison} cases, which are only in transmission}.

\edit1{
To generate eclipse and transit depths for the \texttt{s01} -- \texttt{s03} cases, we use 47 channels spanning 2 -- 5 {\microns} that have perfect transmission over their spectral ranges.
We use relatively high-S/N synthetic data so that the retrieved credible regions will be relatively small, and thus more likely to expose small coding errors when compared to inputs.
We set uncertainties such that each channel has \math{S/N} = 50 for emission cases and 300 for transmission cases.
}
For stellar emission, we use a K2 solar abundance Kurucz stellar model \citep{CastelliKurucz2003iausATLAS9GridModelAtmospheres}.
\edit1{
For the \texttt{s04} and \texttt{s05} cases, we consider each of the simulated spectra by {\NEMESIS}, {\CHIMERA}, and {\TauREx} with noise levels of 60 parts per million.
}

We include opacities for the species as described in Section \ref{sec:comparisontest}.
The \edit1{\texttt{s01} -- \texttt{s03}} retrievals each have five parameters for the \math{T(p)} profile \citep{LineEtal2013apjRetrieval1}; five for the scaling factors of the log abundances of H\sub{2}O, CO, CO\sub{2}, CH\sub{4}, and NH\sub{3}; and, for the transmission cases, a parameter for the planetary radius at 0.1 bar.
\edit1{The \texttt{s04} and \texttt{s05} retrievals have free parameters for the isothermal temperature, the planetary radius at 10 bar, the log mixing ratios of H\sub{2}O and CO, and the pressure corresponding to an opaque cloudtop.}
\edit1{All cases feature uniform priors on the model parameters; parameters that are the logarithm of the true parameter therefore have log-uniform priors.}

{\BART}'s results \edit1{for the \texttt{s01} -- \texttt{s03}} retrievals are similar in some respects (Figures \ref{fig:synth-retrievals-spectra}, \ref{fig:synth-retrievals-ecl}, and \ref{fig:synth-retrievals-tra}).
The retrieved thermal profiles and molecular abundances generally match the inputs in the regions of the atmospheres probed by these synthetic observations (Figures \ref{fig:synth-retrievals-ecl} and \ref{fig:synth-retrievals-tra}, middle columns).
The lower atmospheres are generally poorly constrained, as the spectrum is \edit1{minimally} influenced by those pressure levels \edit1{at the wavelengths of the synthetic data}.
The emission cases provide better constraints than the transmission cases on the \math{T(p)} profile and abundances, as expected \citep{Griffith2014rsptaExoDegenreateSolutions, HengKitzmann2017mnrasTransmissionSpectraTheory, Madhusudhan2018bookAtmRetrExo}.

The isothermal emission case's retrieved abundances and \math{T(p)} profile demonstrate the inability to detect molecular features from an isothermal atmosphere.  
\edit1{In the region with sensitivity, the best-fit thermal profile is isothermal; the inversion seen in the explored thermal profiles corresponds to regions with negligible or no contribution to the spectrum.}
We allow for any \math{T(p)} profile rather than enforcing an isothermal condition \edit1{because} it would not be known {\em a priori}\/ whether the atmosphere were isothermal.
\edit1{The 1D marginalized posteriors for the molecular abundances are poorly constrained and tend to favor a log mixing ratio \textless -4, with significant probability for log mixing ratios \textless -8, consistent with a lack of spectral features for an isothermal atmosphere.}

Table \ref{tbl:retrievals-speis-ess} shows the SPEIS, ESS, and posterior accuracies for these retrievals.
\edit1{The large SPEIS (and small ESS) are due to a combination of factors.  We choose the highest SPEIS value among all chains and all parameters as a conservative estimate; the non-inverted eclipse case has a SPEIS \textgreater 15,000, with a median SPEIS of \math{\sim}2,730.  Compared to the \citet{BarstowEtal2020mnrasRetrievalComparison} cases and the HD 189733 b retrieval, these SPEIS values are significantly greater.  This may be related to a numerical effect seen in synthetic retrievals tests (see Appendix \ref{ap:synthreterrors}).}

\begin{figure*}[htp]
{\centering \textbf{\large{Transmission} \hspace{7cm} \large{Emission}}\par}
\includegraphics[width=0.49\textwidth, clip=True]{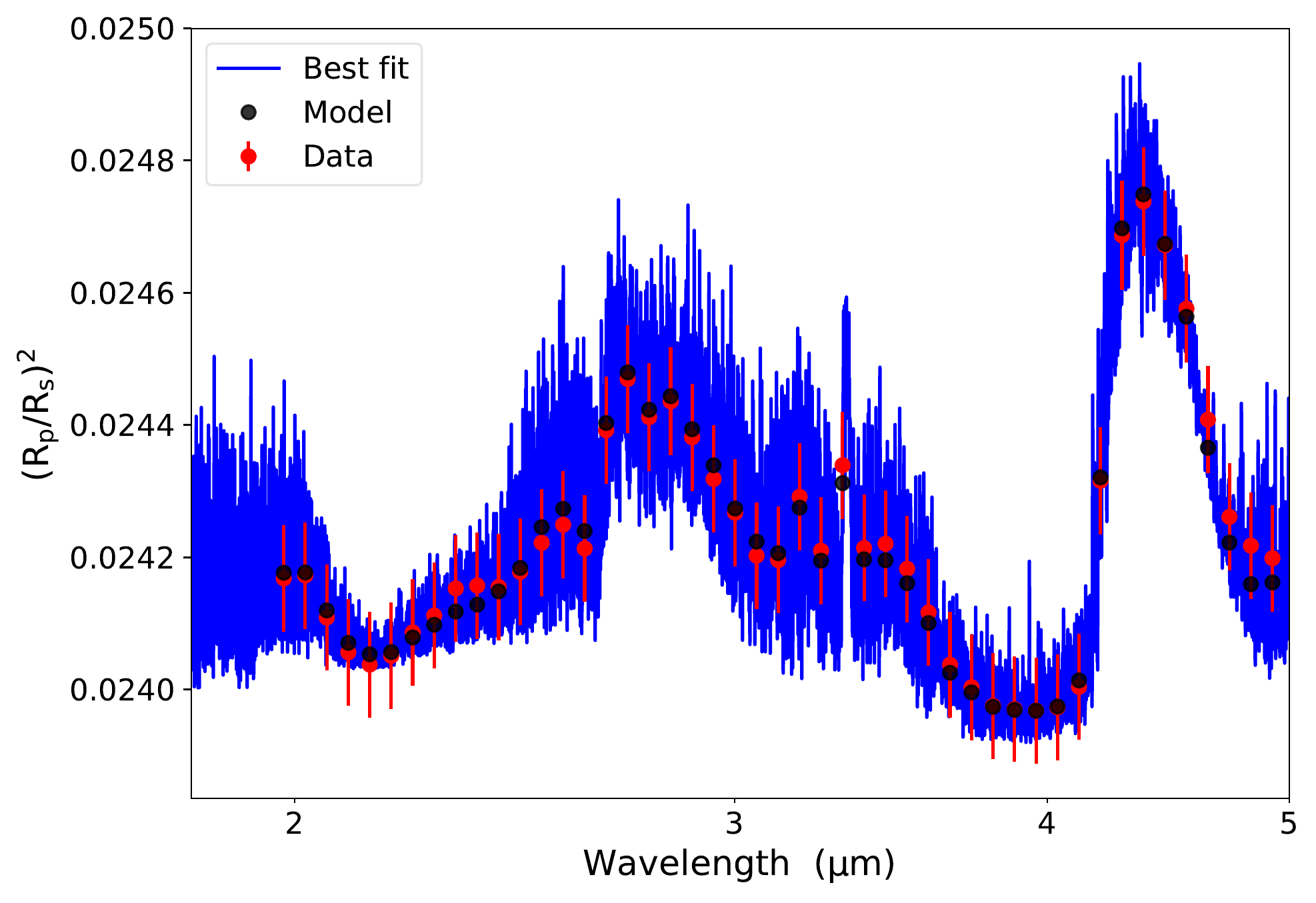}\hfill
\includegraphics[width=0.49\textwidth, clip=True]{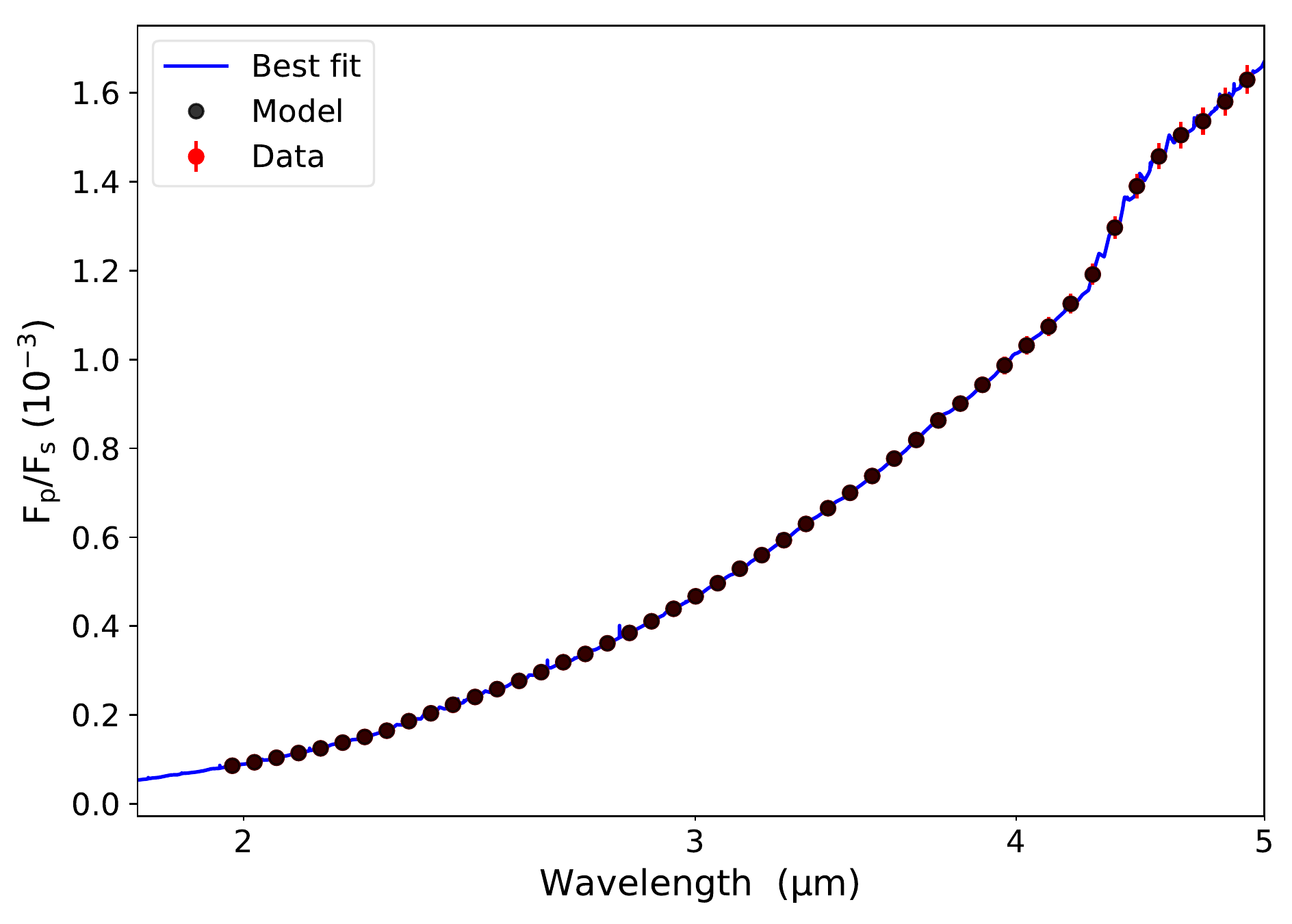}
\includegraphics[width=0.49\textwidth, clip=True]{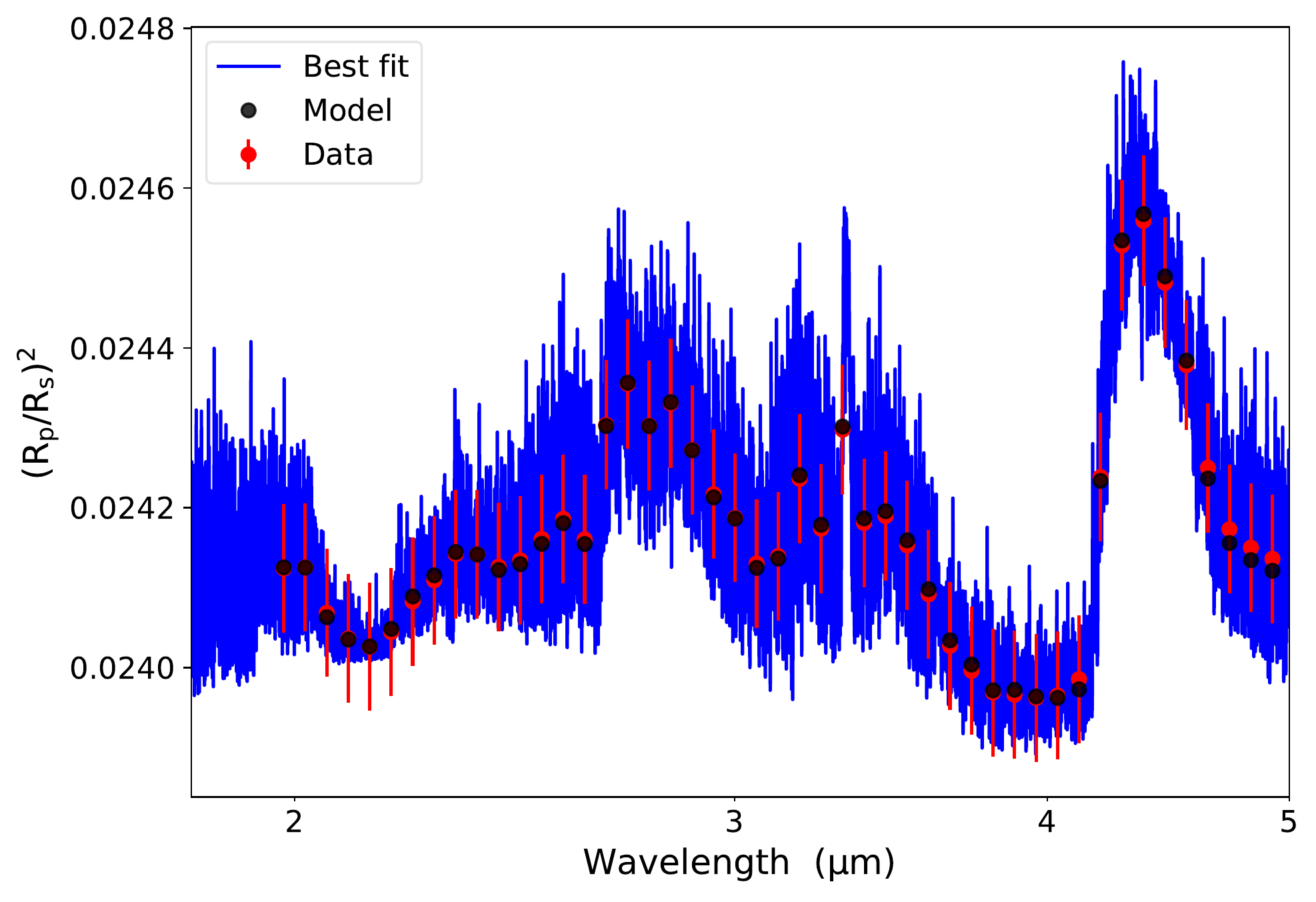}\hfill
\includegraphics[width=0.49\textwidth, clip=True]{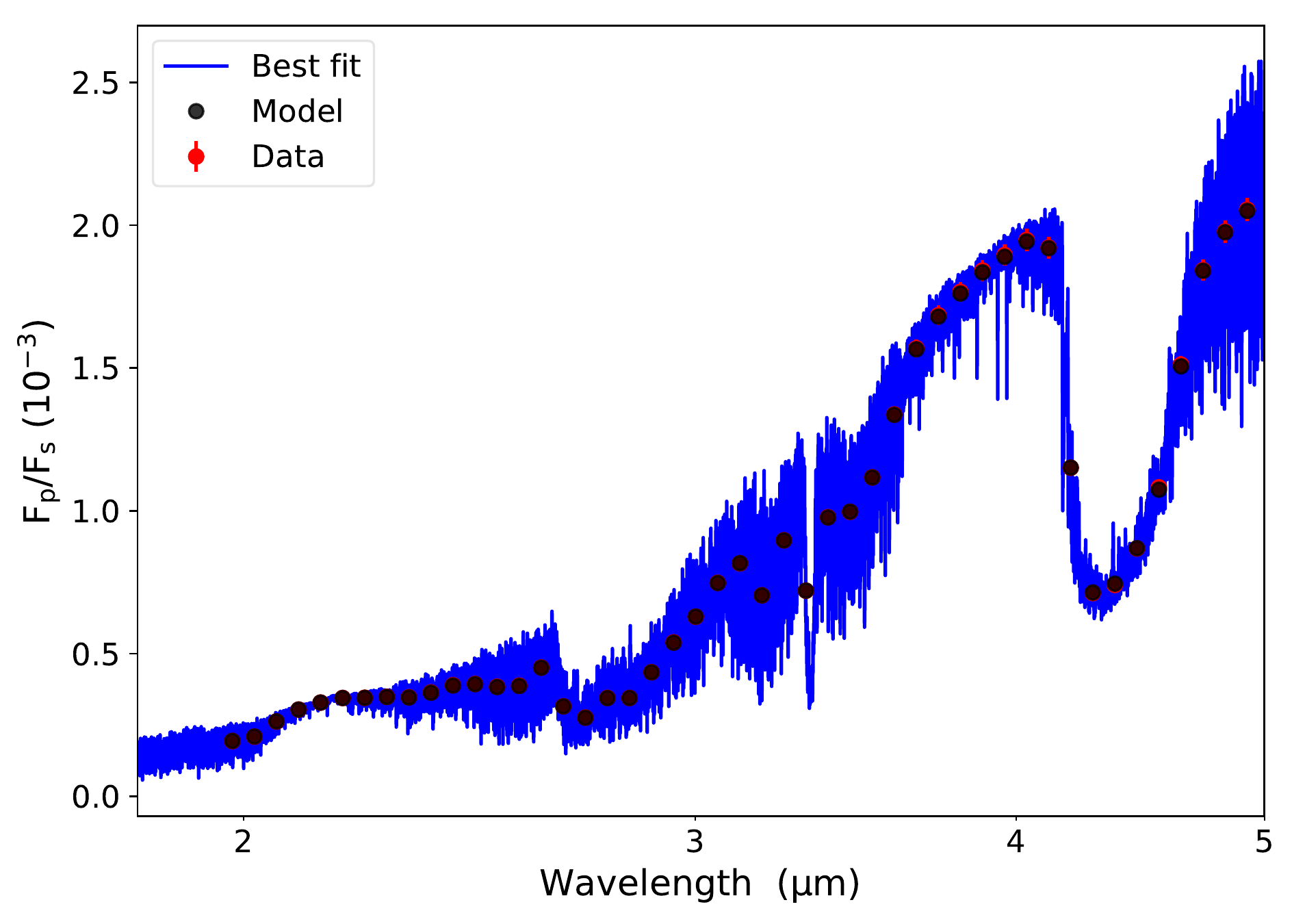}
\includegraphics[width=0.49\textwidth, clip=True]{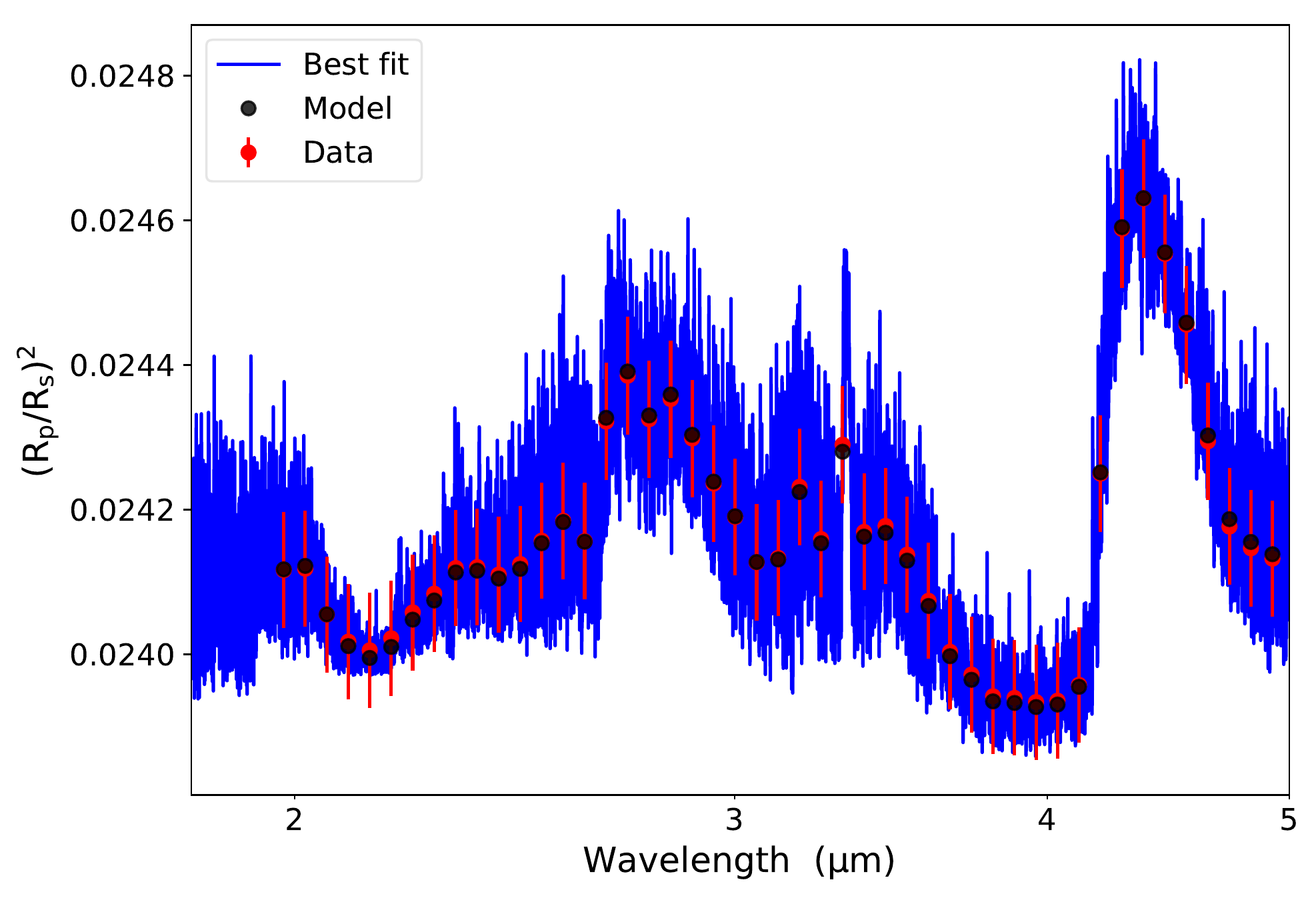}\hfill
\includegraphics[width=0.49\textwidth, clip=True]{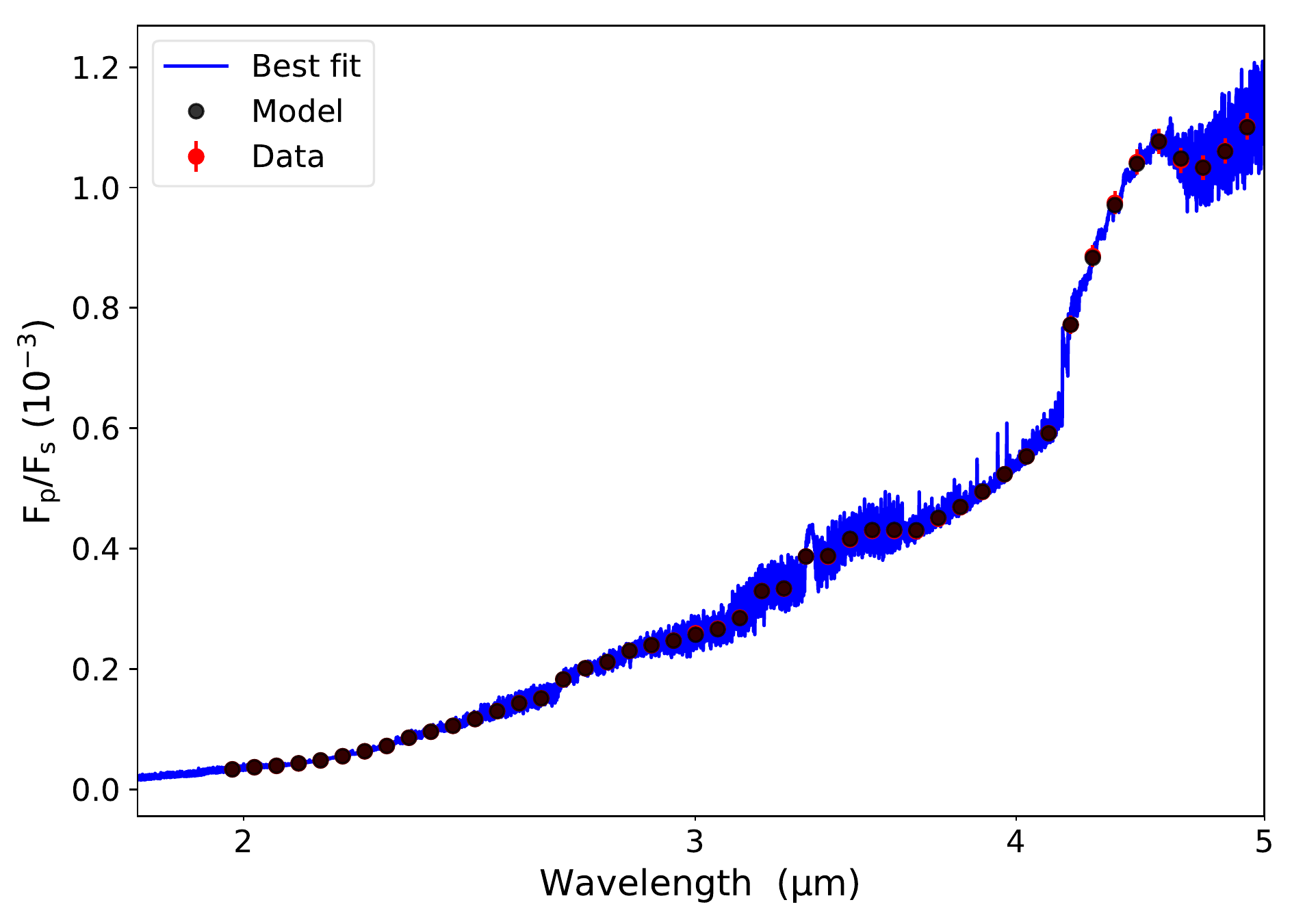}
\caption{Best-fit spectra for the six synthetic retrieval tests \edit1{of \texttt{s01hjcleariso}, \texttt{s02hjclearnoinv}, and \texttt{s03hjclearinv}}: isothermal (top row), non-inverted (middle row), and inverted (bottom row) atmospheres in transit (left column) and eclipse (right column) geometries.  \edit1{The transmission data look similar for all three cases as they only differ in thermal profiles, to which transit geometry is generally insensitive.}
\label{fig:synth-retrievals-spectra}}
\end{figure*}

\begin{figure*}[htp]
{\centering \textbf{\large{Emission}}\par\medskip}
\includegraphics[width=0.340\textwidth, clip=True]{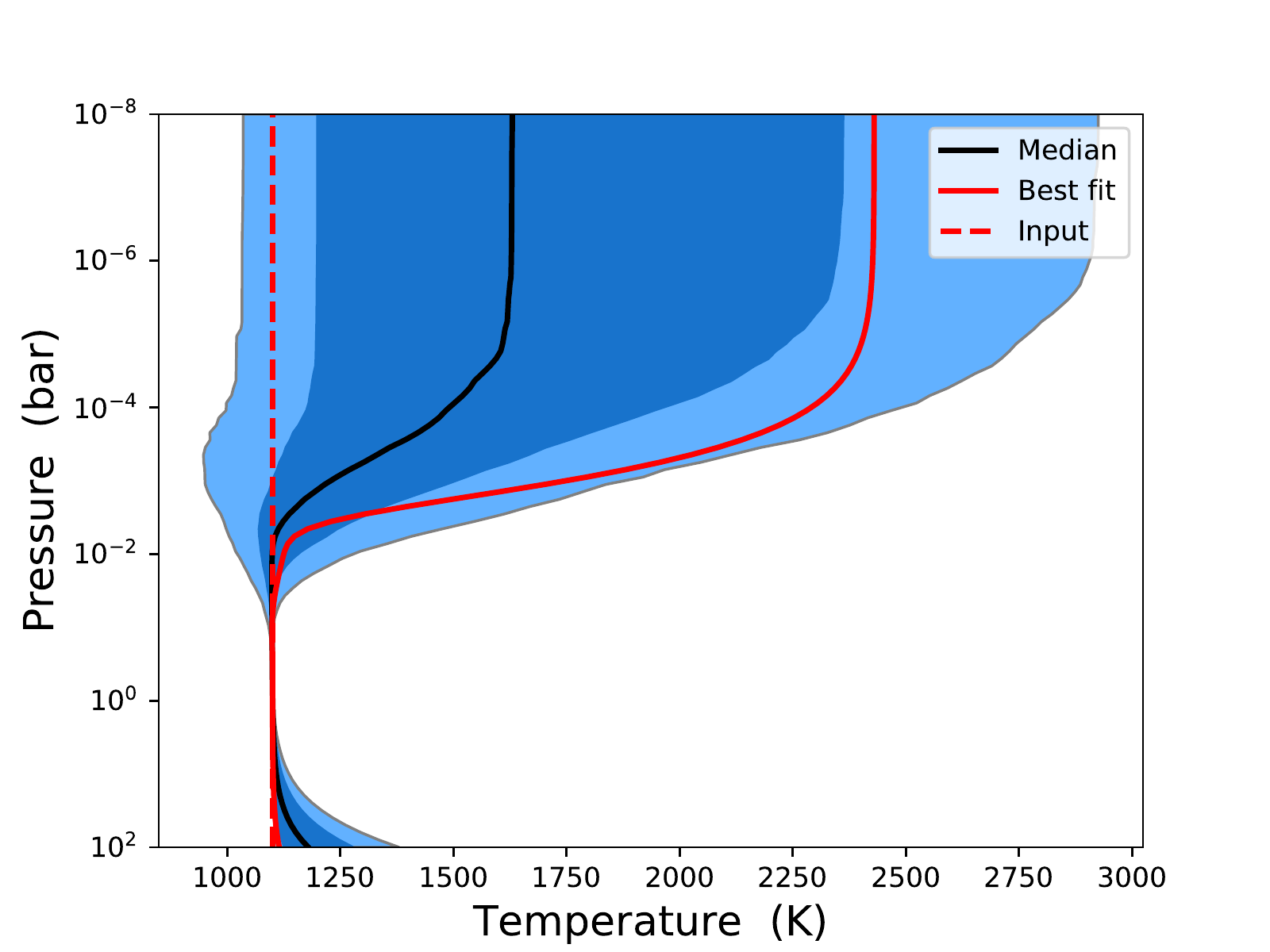}\hfill
\includegraphics[width=0.310\textwidth, clip=True]{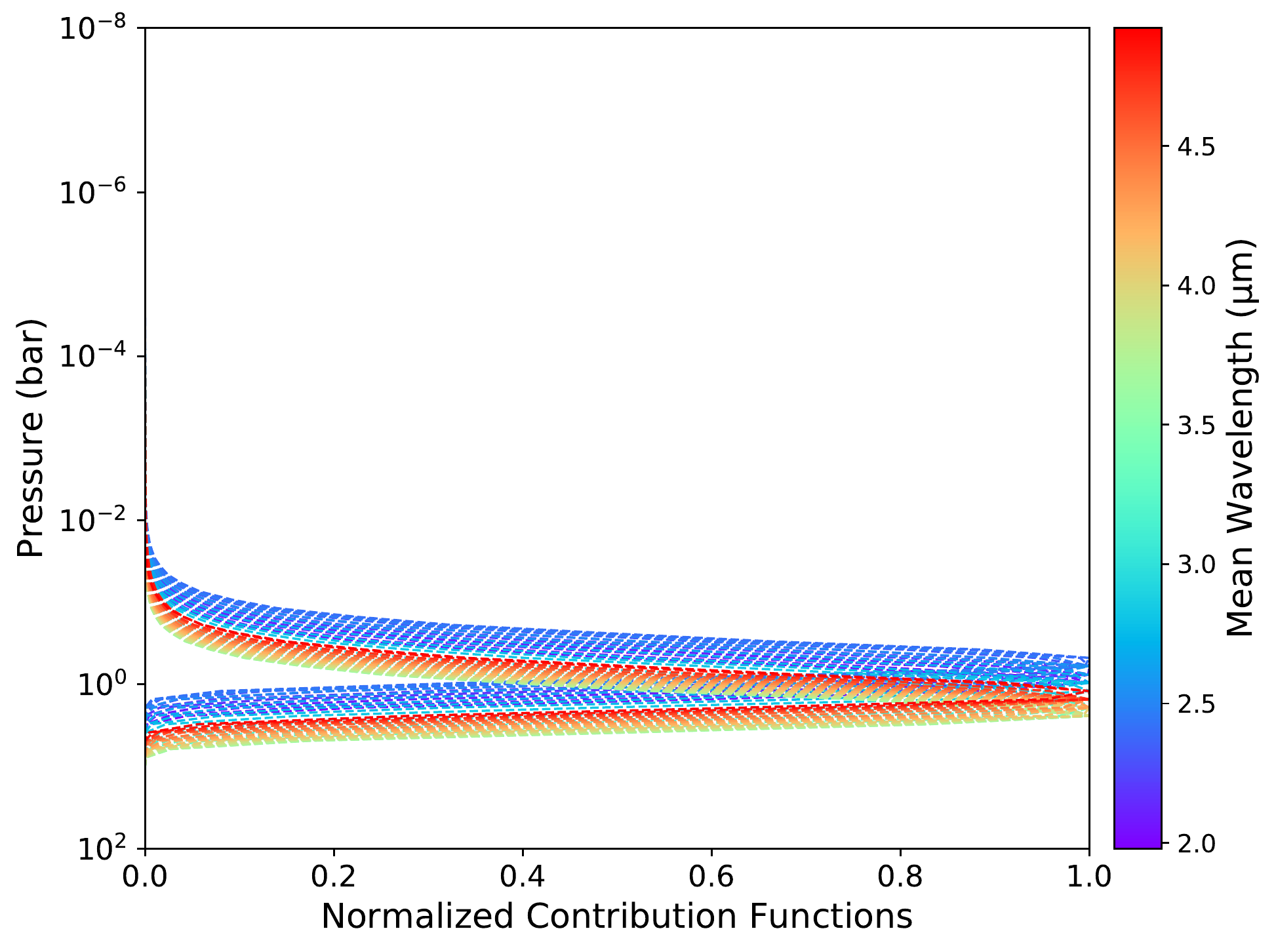}
\includegraphics[width=0.340\textwidth, clip=True]{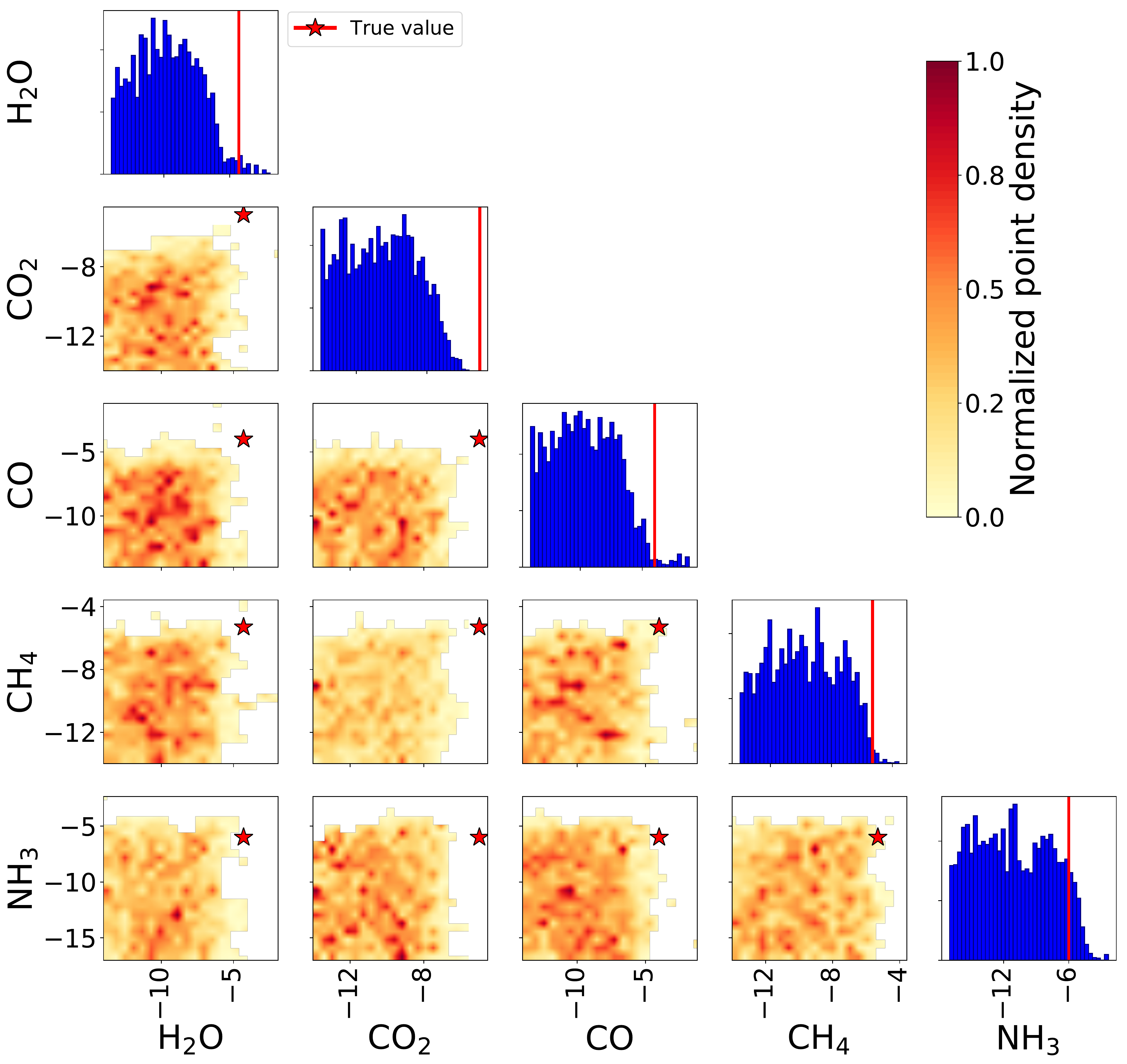}\newline
\includegraphics[width=0.340\textwidth, clip=True]{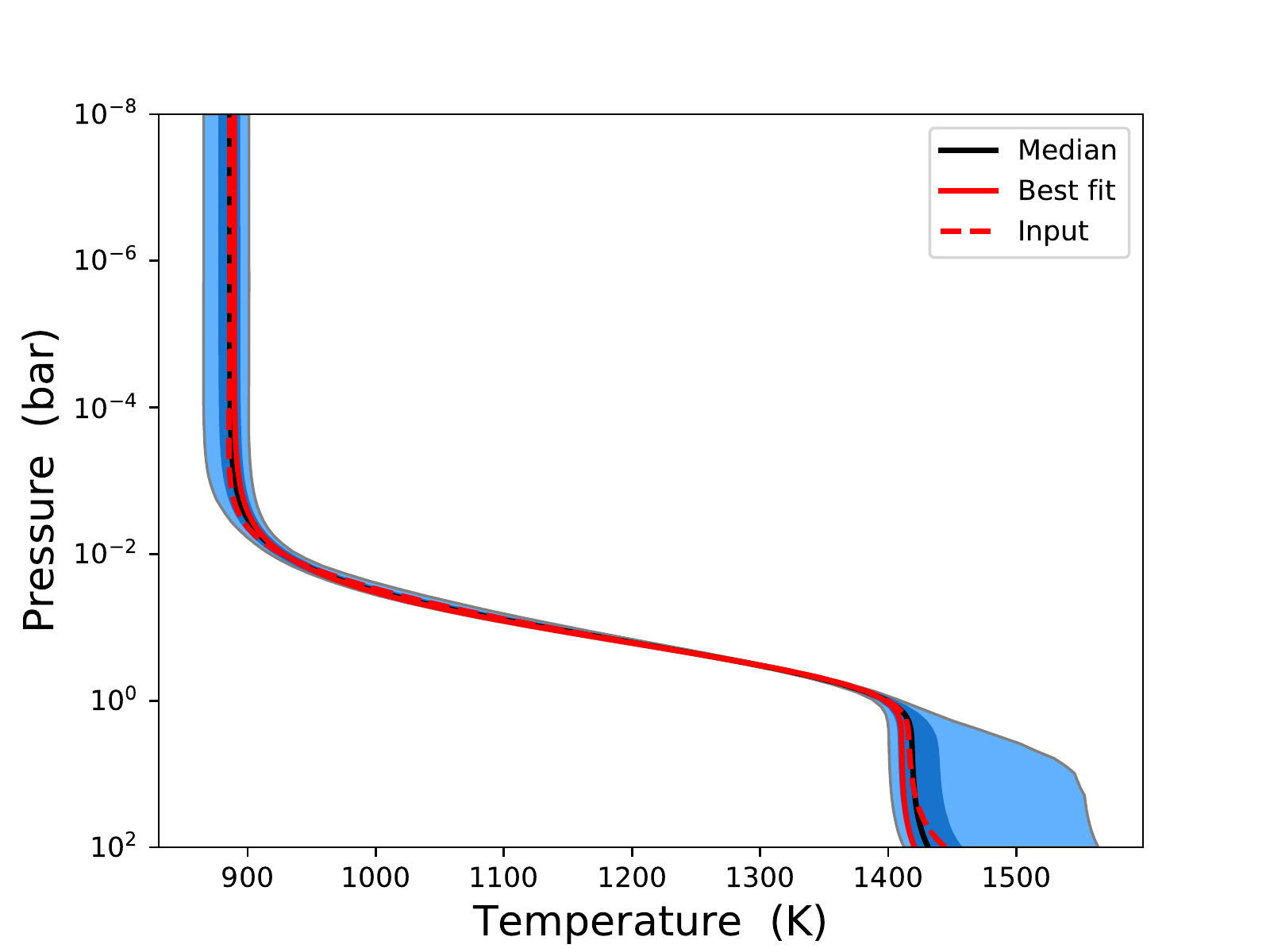}\hfill
\includegraphics[width=0.310\textwidth, clip=True]{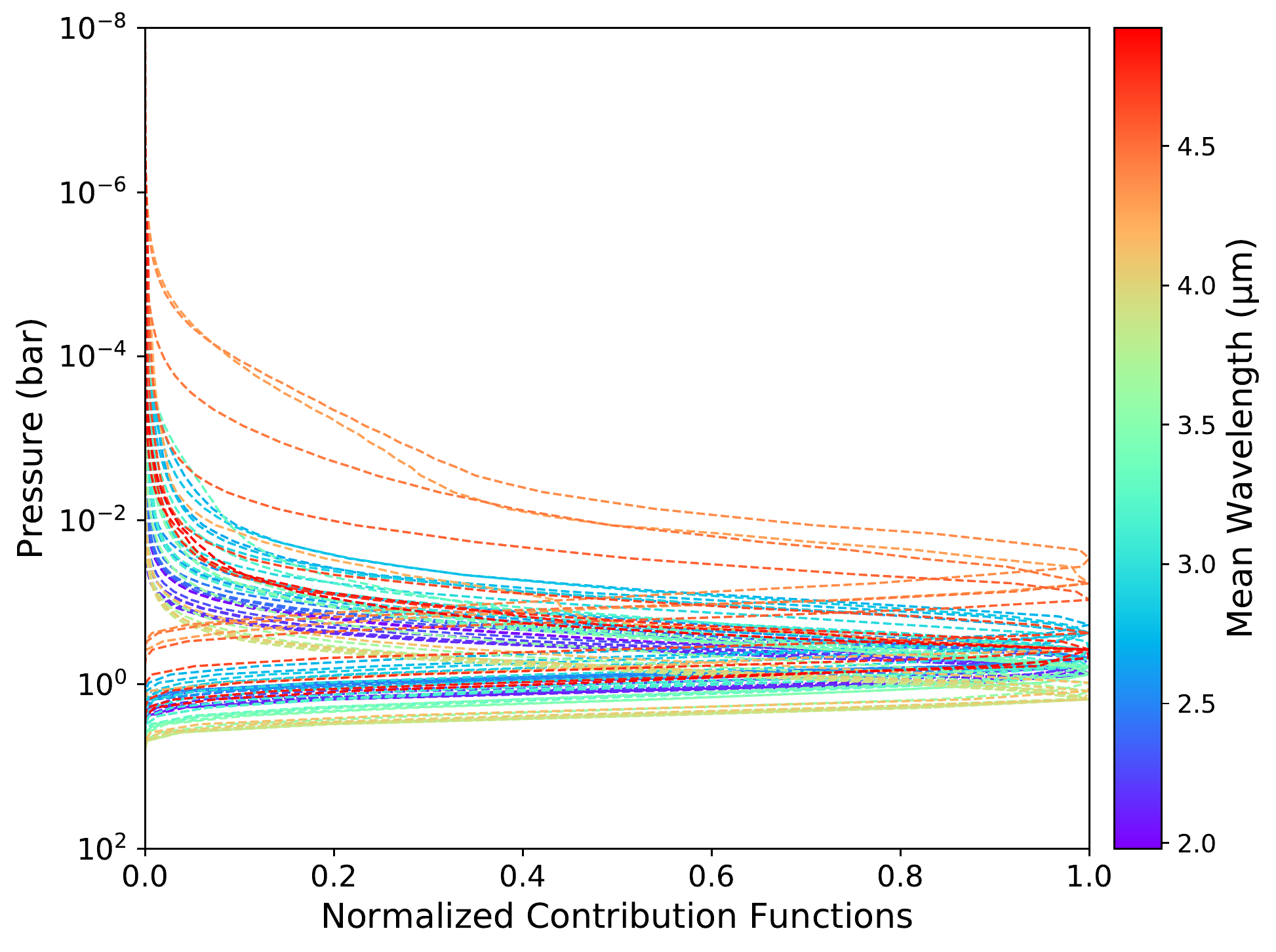}
\includegraphics[width=0.340\textwidth, clip=True]{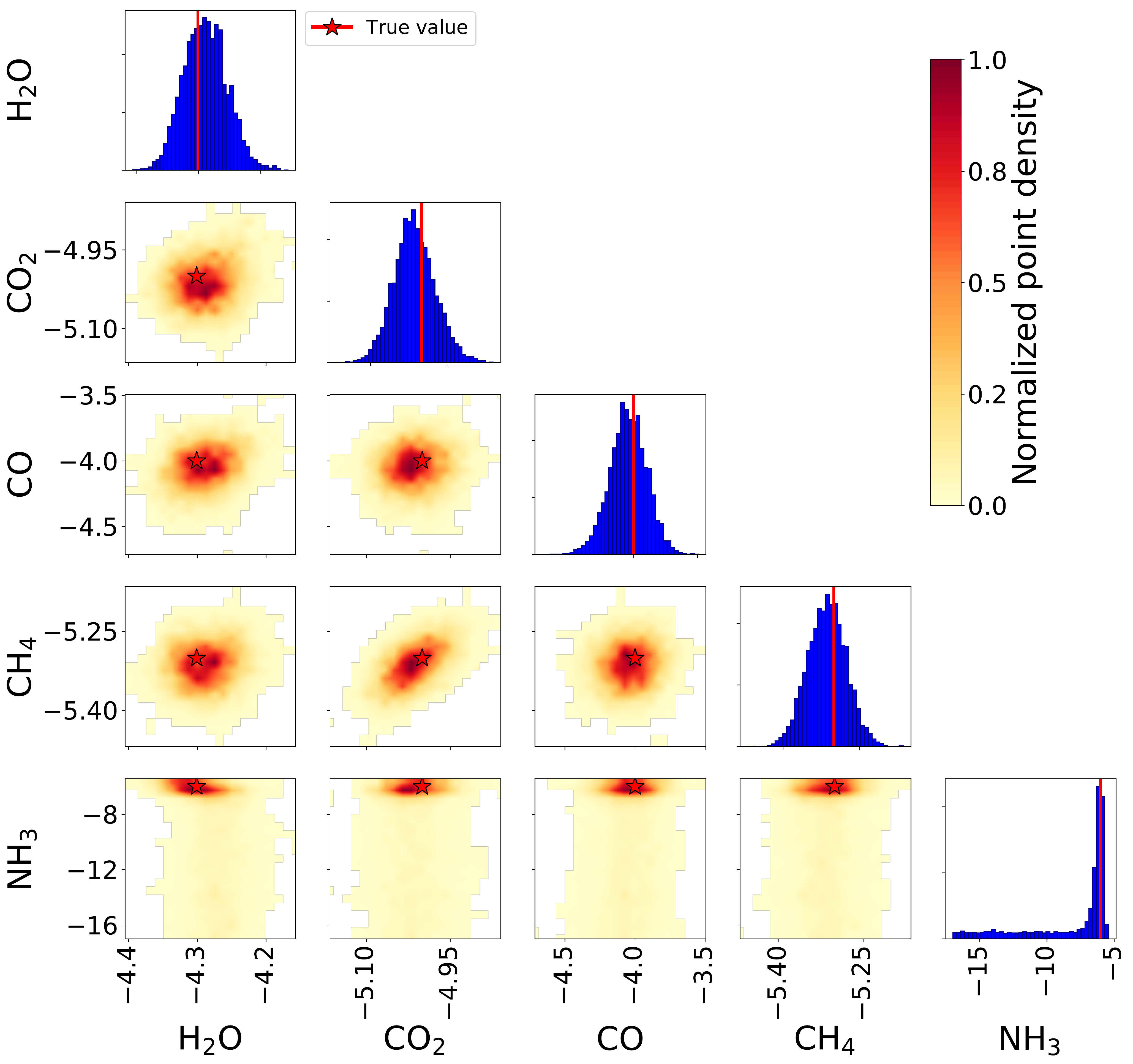}\newline
\includegraphics[width=0.340\textwidth, clip=True]{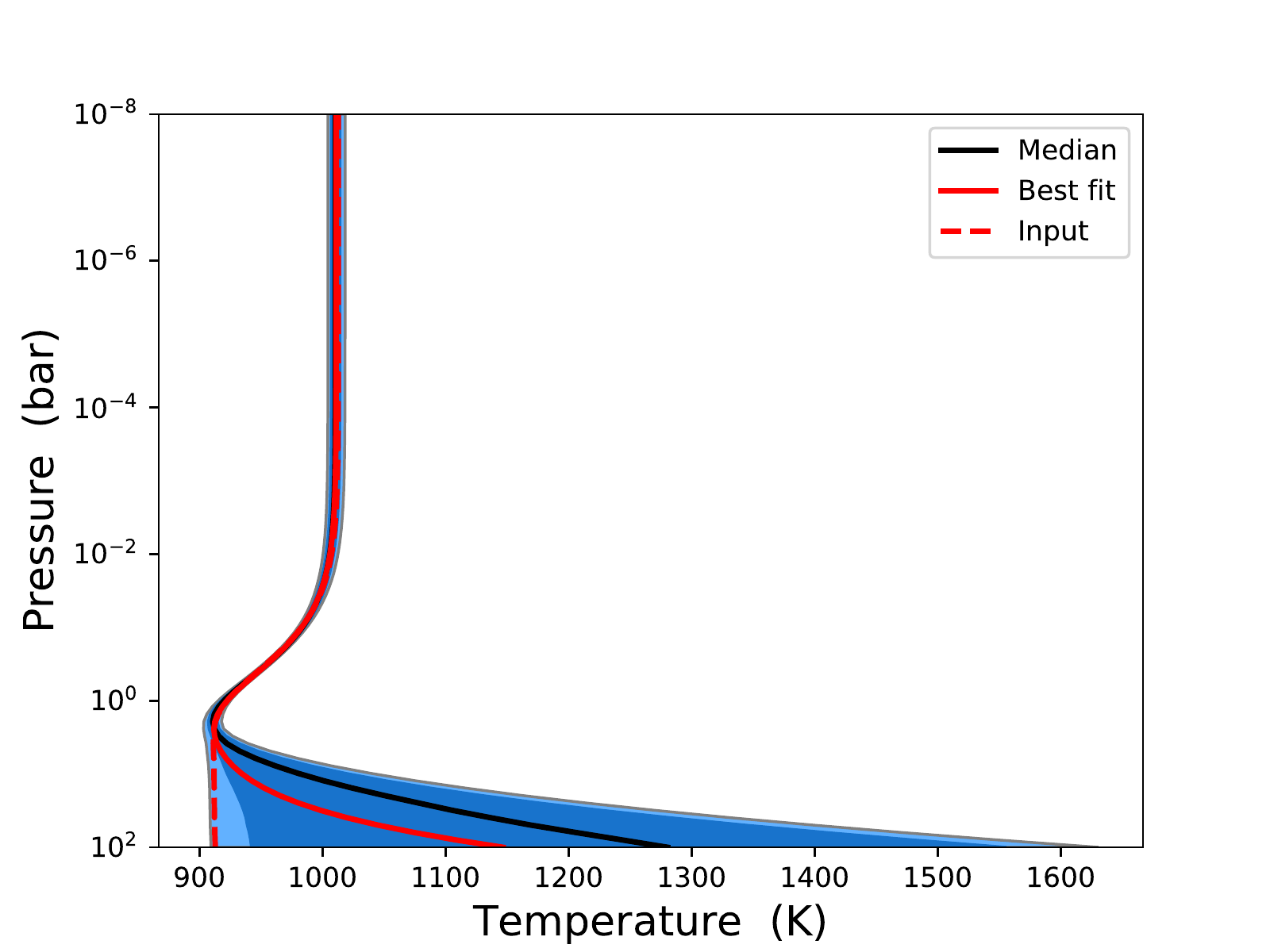}\hfill
\includegraphics[width=0.310\textwidth, clip=True]{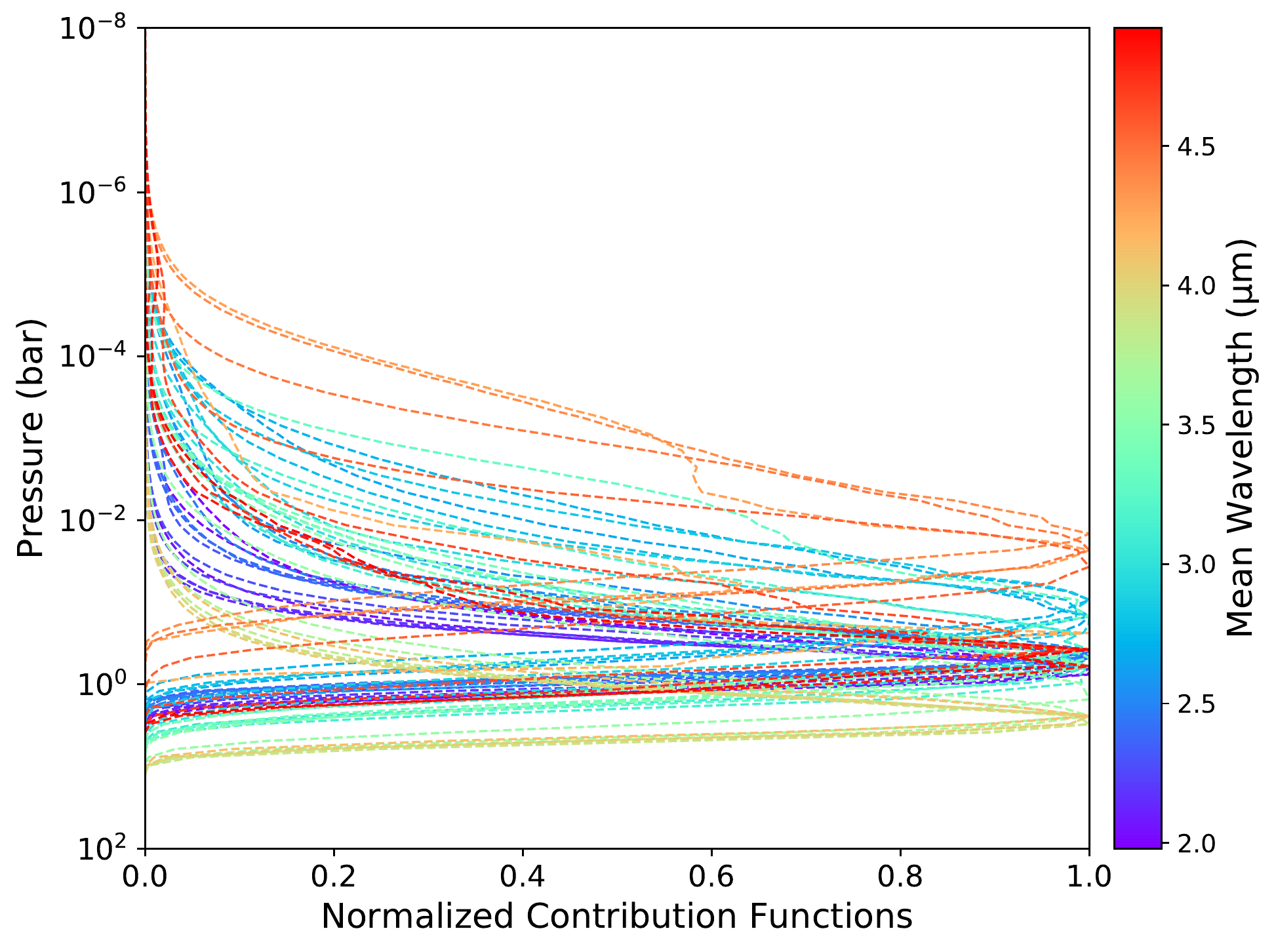}
\includegraphics[width=0.340\textwidth, clip=True]{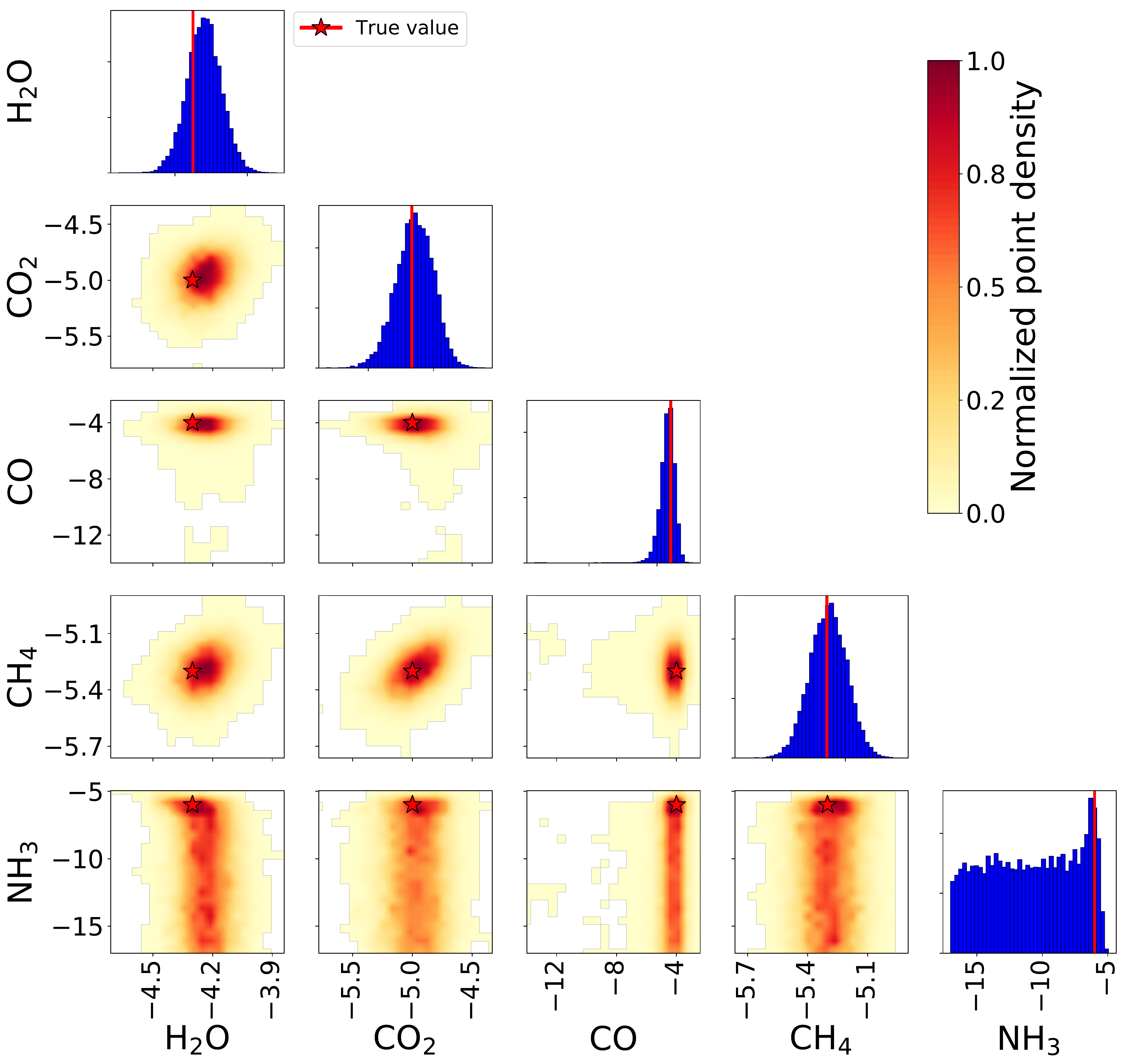}
\caption{Comparison between input and retrieved \math{T(p)} profiles (left column), the normalized contribution functions (middle column, \edit1{BART3}), and best fit molecular abundances (right column) for the isothermal (top row), non-inverted (middle row), and inverted (bottom row) emission cases \edit1{of the \texttt{s01hjcleariso}, \texttt{s02hjclearnoinv}, and \texttt{s03hjclearinv} tests}.
Dark- and light-blue shading in \math{T(p)} profile plots designate the 68.27\% and 95.45\% credible regions).
These regions and the median derive from all the fits on a per-pressure-level basis.
The\edit1{y} do not follow the functional form of the individual profiles.
Few, if any, individual \math{T(p)} profiles, including the best fit, stay confined to these regions, especially where the contribution functions indicate low sensitivity.
\edit1{BART accurately retrieves the \math{T(p)} profile and abundances wherever they contribute to the spectra.
However, isothermal, non-scattering emission spectra are blackbodies, insensitive to abundances.
Most MCMC-proposed spectra in those cases are not isothermal.
Those with detectable structure (high abundances) are rarely accepted, creating the artificially low upper limits.
This affects all Monte Carlo Bayesian retrievals.
}
\label{fig:synth-retrievals-ecl}}
\end{figure*}

\begin{figure*}[htp]
{\centering \textbf{\large{Transmission}}\par\medskip}
\includegraphics[width=0.340\textwidth, clip=True]{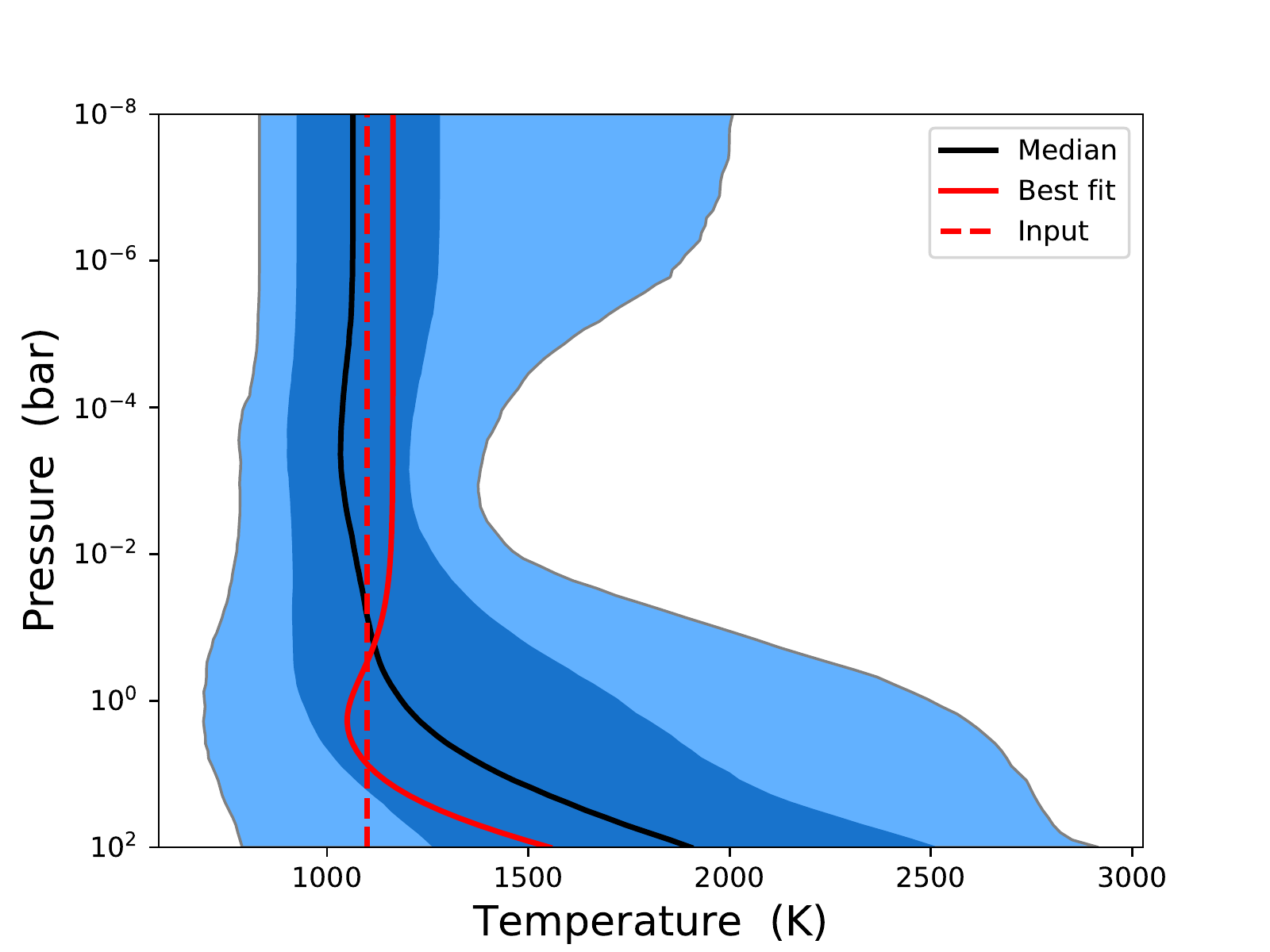}\hfill
\includegraphics[width=0.310\textwidth, clip=True]{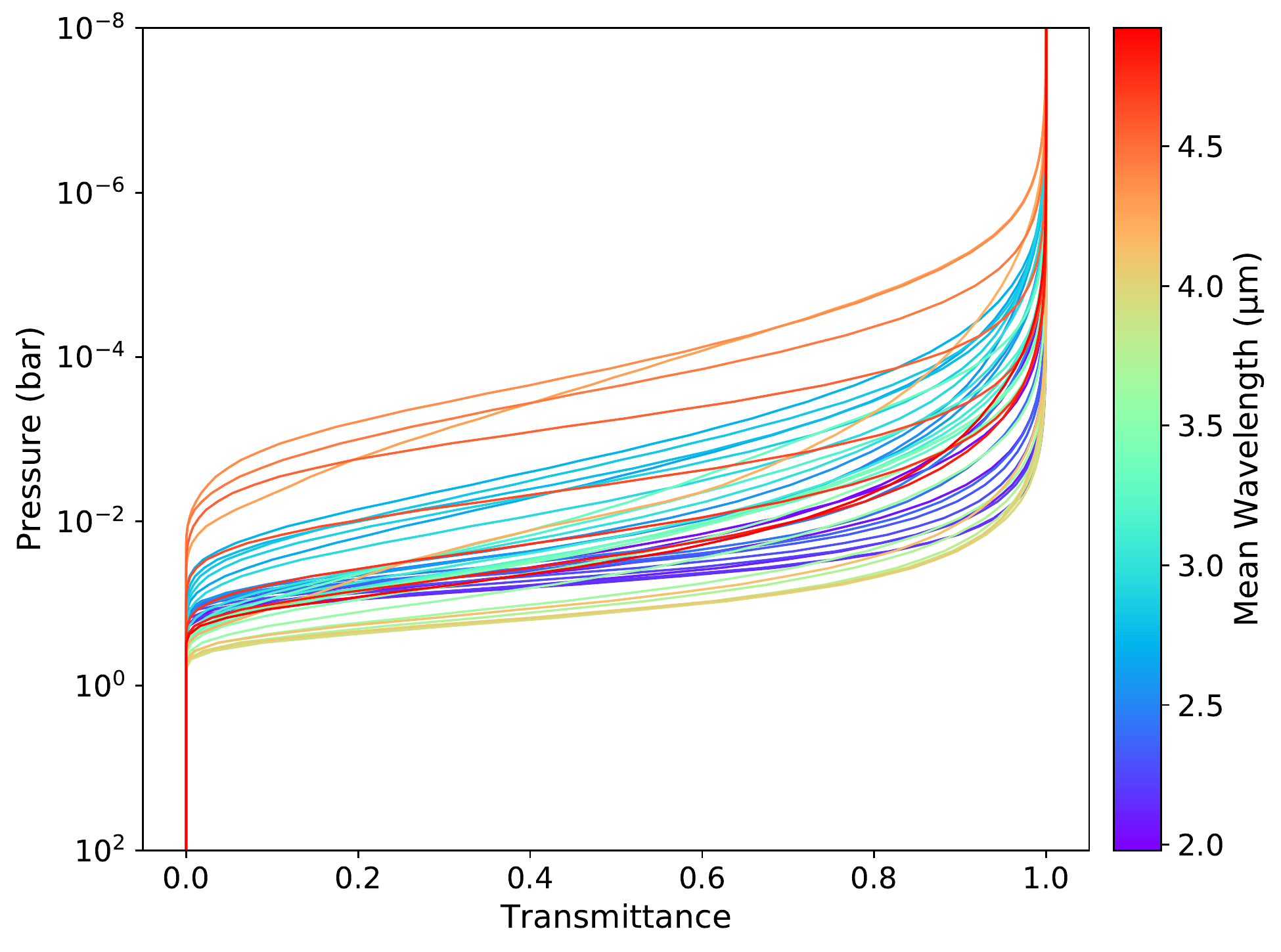}
\includegraphics[width=0.340\textwidth, clip=True]{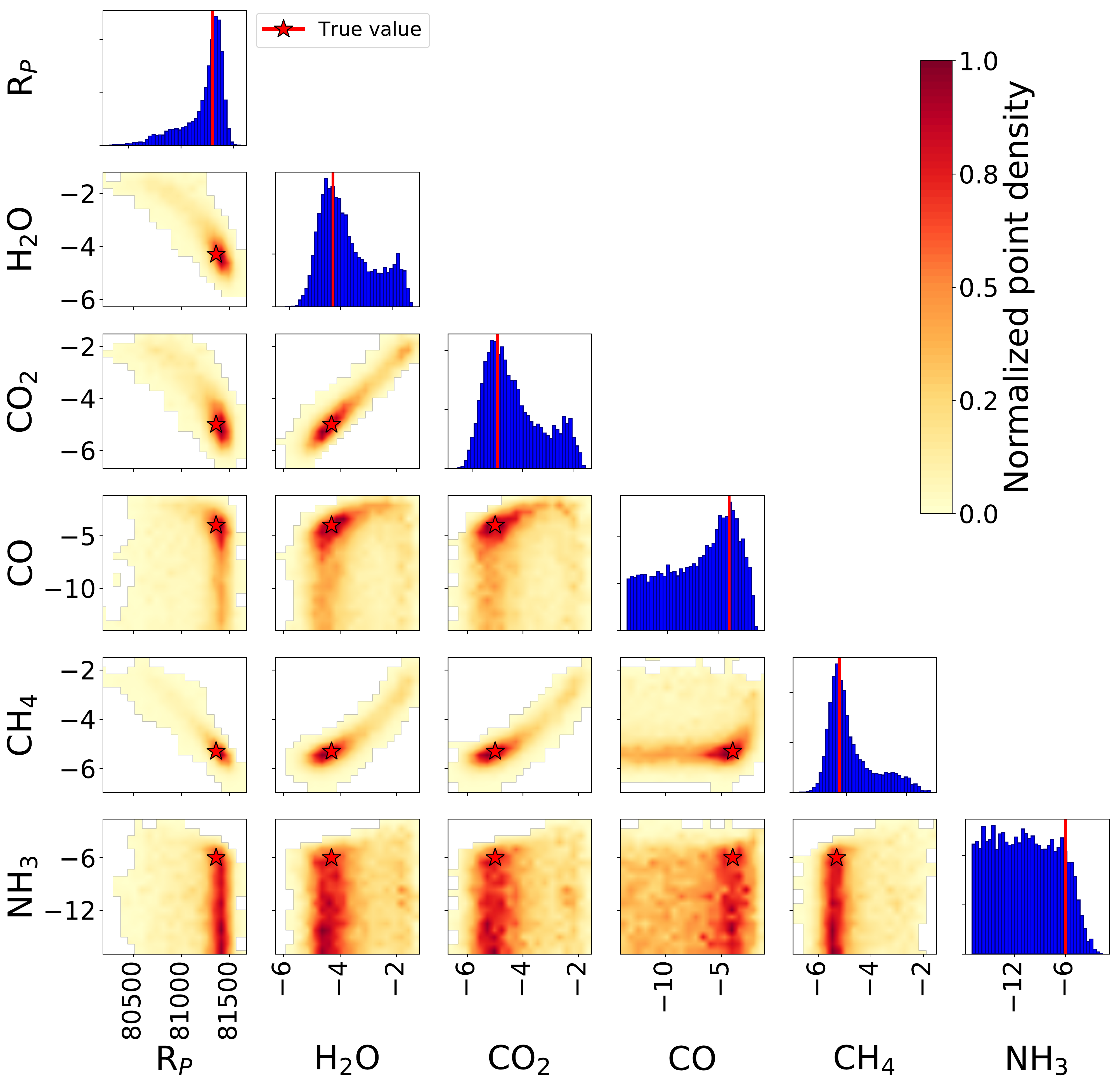}\newline
\includegraphics[width=0.340\textwidth, clip=True]{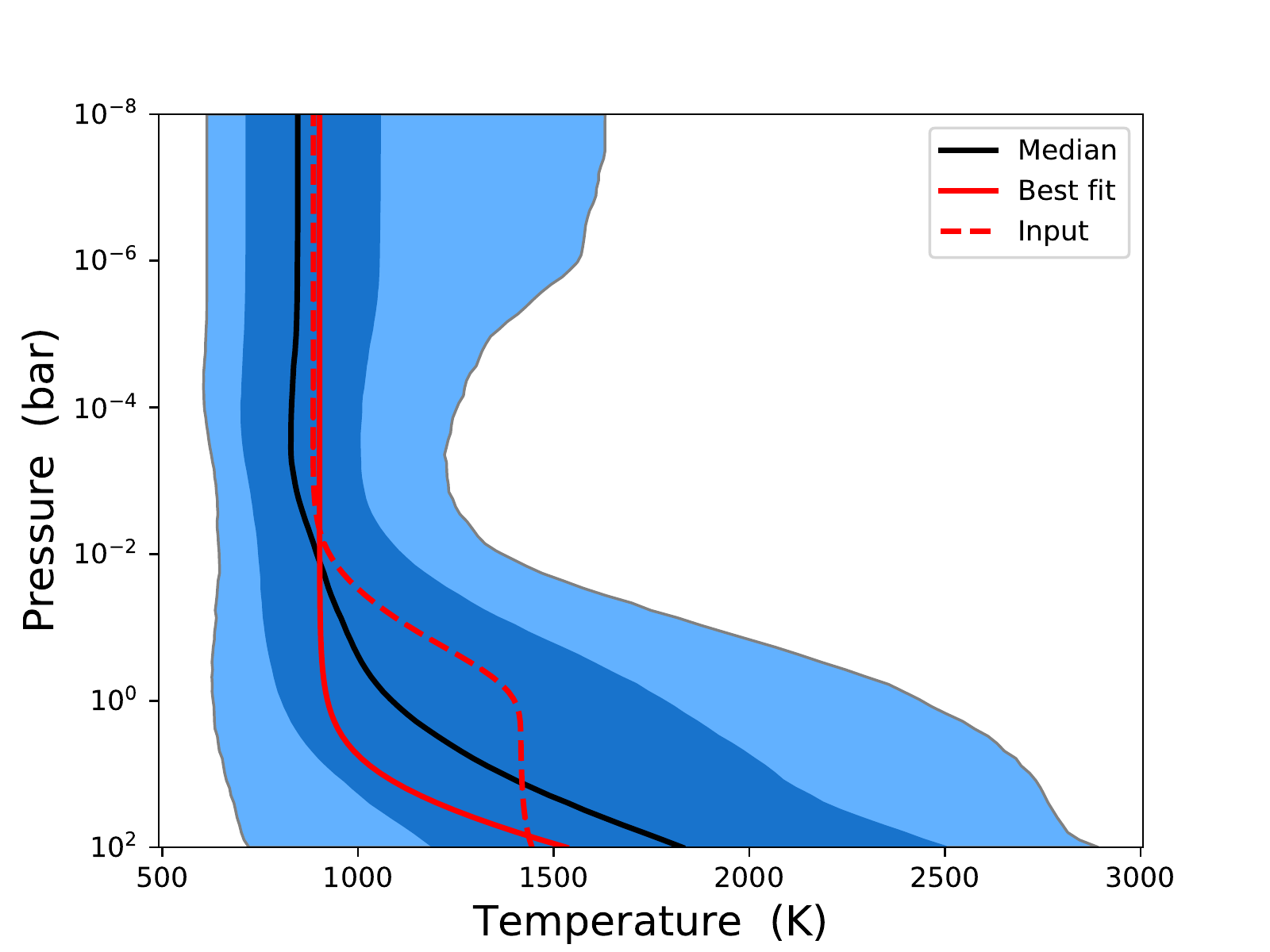}\hfill
\includegraphics[width=0.310\textwidth, clip=True]{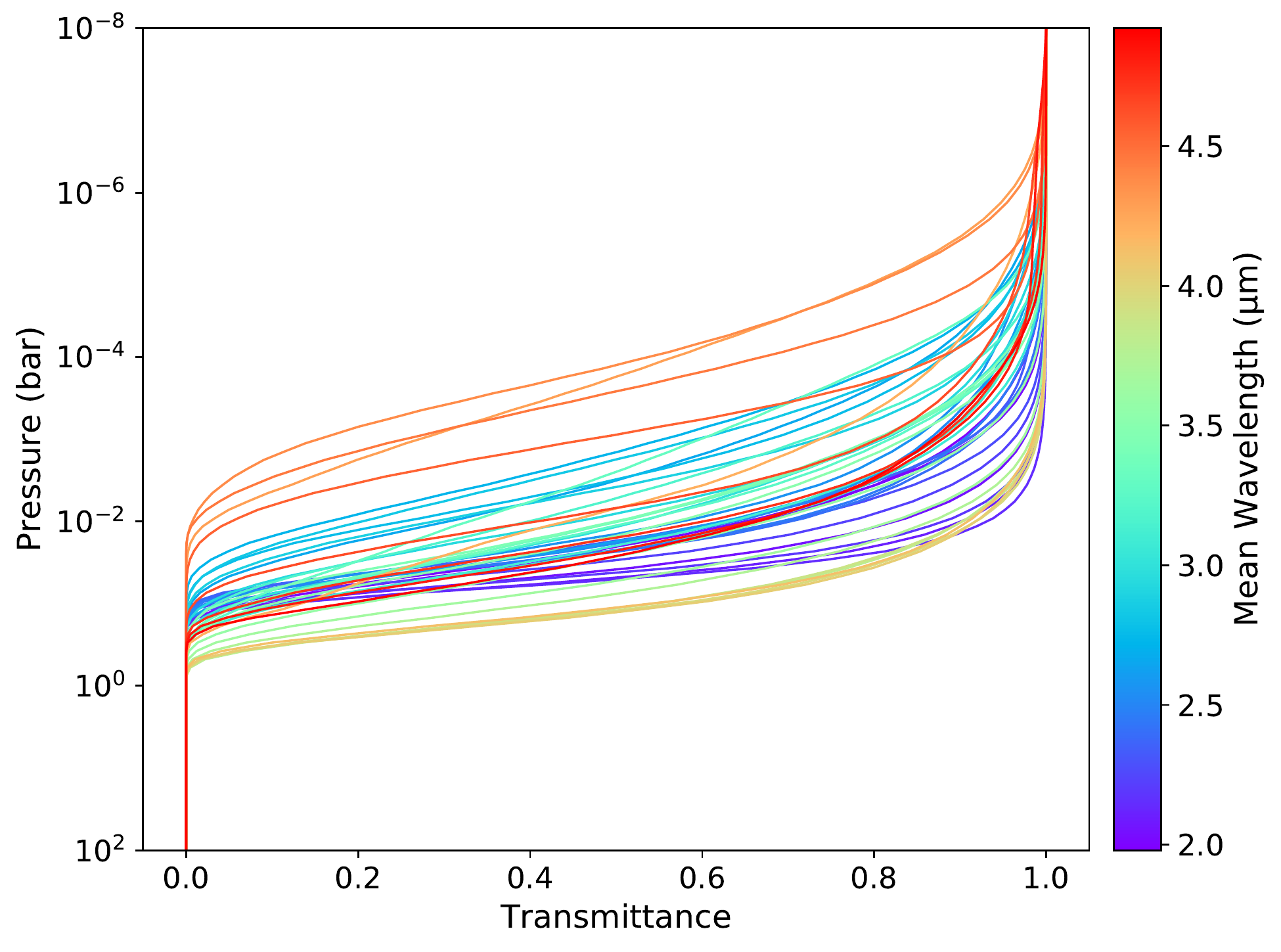}
\includegraphics[width=0.340\textwidth, clip=True]{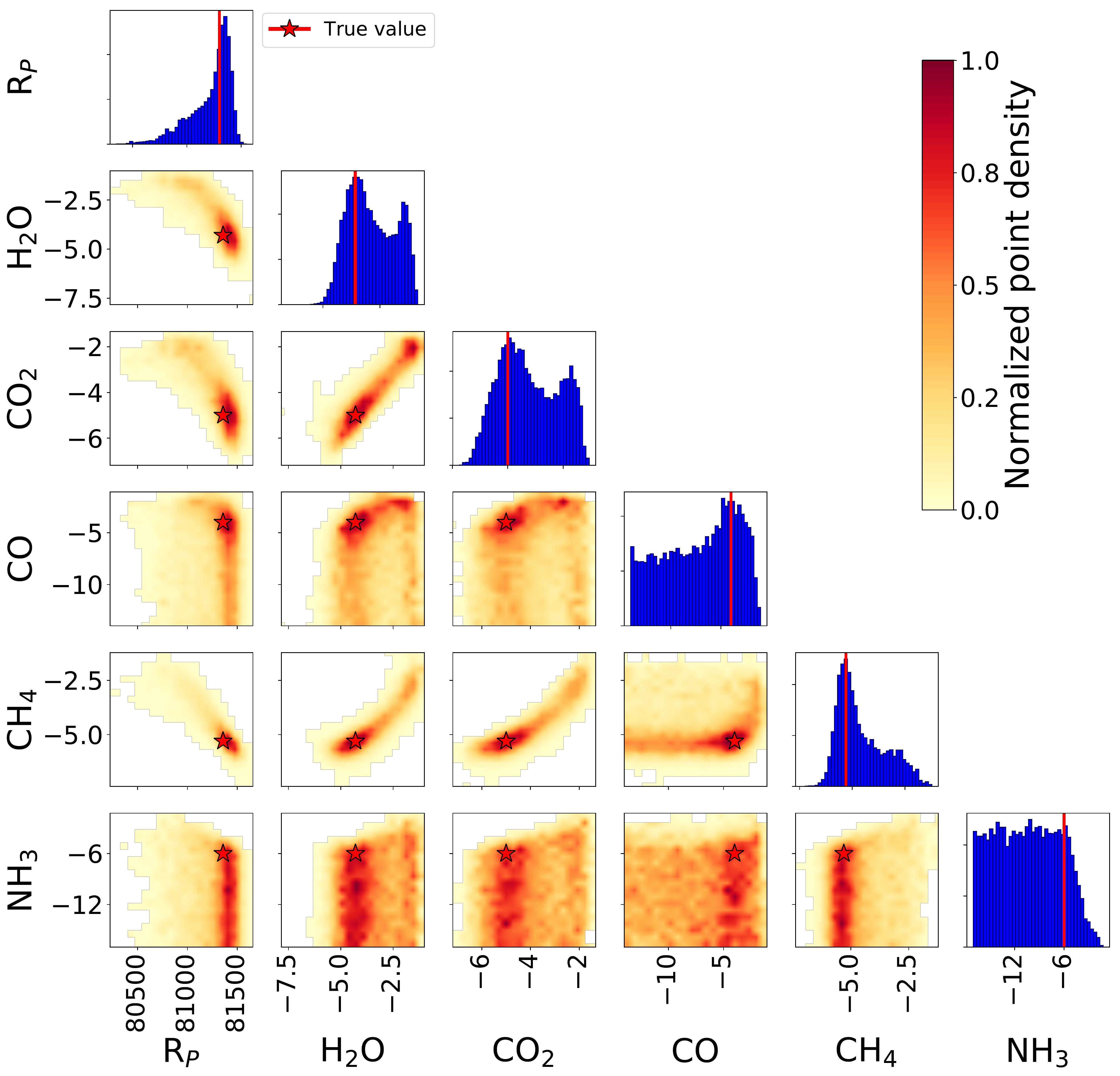}\newline
\includegraphics[width=0.340\textwidth, clip=True]{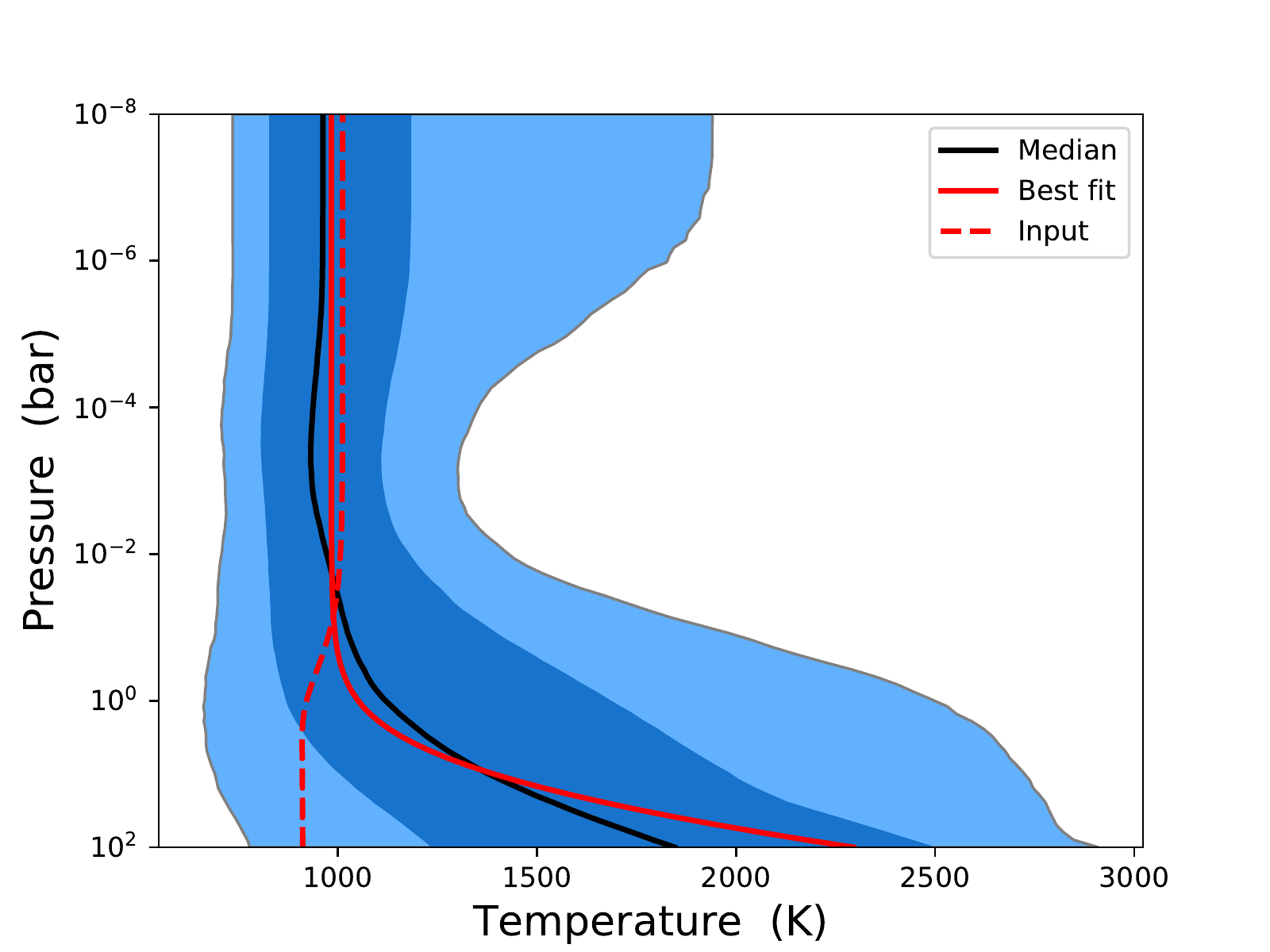}\hfill
\includegraphics[width=0.310\textwidth, clip=True]{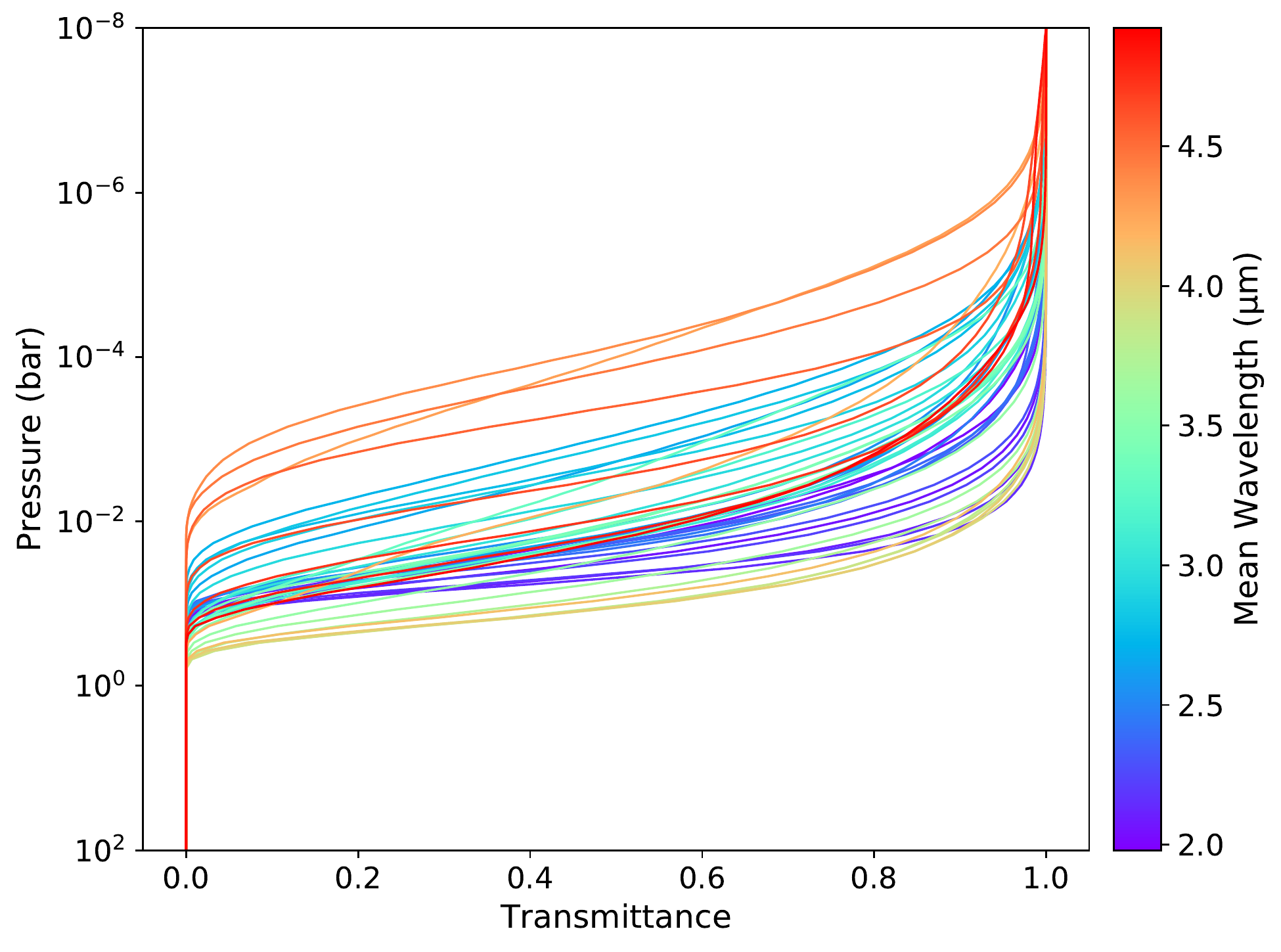}
\includegraphics[width=0.340\textwidth, clip=True]{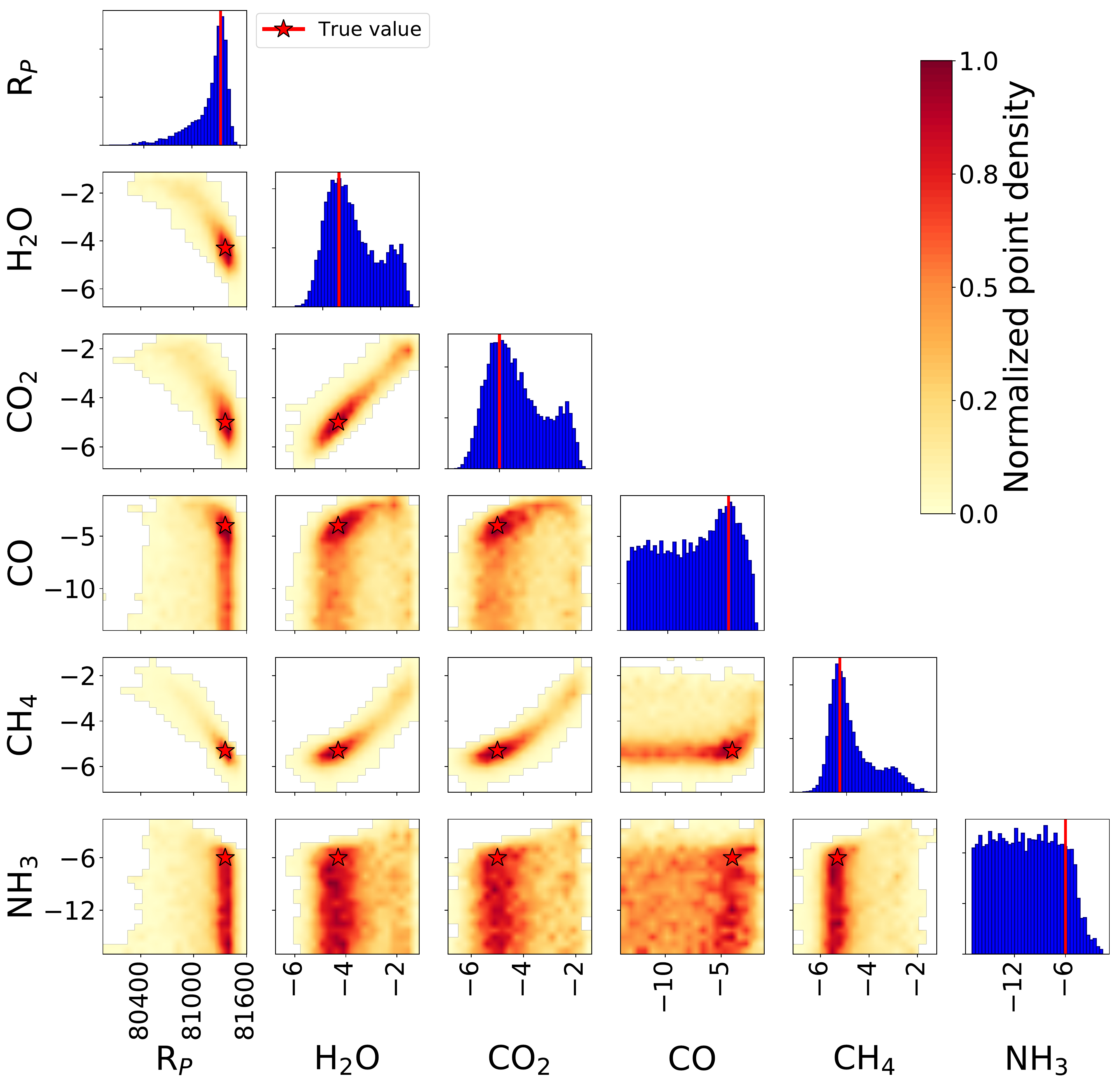}
\caption{Same as Figure \ref{fig:synth-retrievals-ecl}, but for transmission cases \edit1{of the \texttt{s01hjcleariso}, \texttt{s02hjclearnoinv}, and \texttt{s03hjclearinv} tests}.  \edit1{(Recall that low-resolution transmission spectra are relatively insensitive to temperature structure.)}
\label{fig:synth-retrievals-tra}}
\end{figure*}

\subsubsection{\citet{BarstowEtal2020mnrasRetrievalComparison} Synthetic Retrievals}
\label{sec:barstowrets}

\begin{table*}
\centering
\caption{{\BARTTest} Retrievals: Posterior Accuracy}
\label{tbl:retrievals-speis-ess}
\atabon\begin{tabular}{lcrrccc}
\hline\hline
Test  & Geometry & SPEIS  & ESS\sp{1} & \multicolumn{3}{c}{Credible Region Uncertainty} \\
                      &         &       &      & 68.27
                                               & 95.45
                                               & 99.73
\hline
\texttt{s01hjcleariso}   & Eclipse &  5131 &   97 & 4.65\% & 2.08\% & 0.52\% \\
                      & Transit &  9204 &  162 & 3.62\% & 1.62\% & 0.40\% \\
\texttt{s02hjclearnoinv} & Eclipse & 15506 &   96 & 4.68\% & 2.09\% & 0.52\% \\
                      & Transit &  8129 &  184 & 3.40\% & 1.52\% & 0.38\% \\
\texttt{s03hjclearinv}   & Eclipse & 14485 &  103 & 4.52\% & 2.02\% & 0.50\% \\
                      & Transit &  4553 &  329 & 2.55\% & 1.14\% & 0.28\% \\
\texttt{s04hjcleariso} {\NEMESIS} & Transit &  54 &  1851 & 1.08\% & 0.48\% & 0.12\% \\
\texttt{s04hjcleariso} {\CHIMERA} & Transit &  51 &   980 & 1.48\% & 0.66\% & 0.17\% \\
\texttt{s04hjcleariso} {\TauREx} & Transit &  65 &   769 & 1.68\% & 0.75\% & 0.19\% \\
\texttt{s05hjcloudiso} {\NEMESIS} & Transit &  511 & 978 & 1.49\% & 0.67\% & 0.17\% \\
\texttt{s05hjcloudiso} {\CHIMERA} & Transit &  412 & 970 & 1.49\% & 0.67\% & 0.17\% \\
\texttt{s05hjcloudiso} {\TauREx} & Transit &  812 & 862 & 1.58\% & 0.71\% & 0.18\% \\
\texttt{r01hd189733b}    & Eclipse &  2084 & 959 & 1.50\% & 0.67\% & 0.17\% \\
\hline
\multicolumn{7}{l}{\sp{1} Computed from the non-burned iterations for each case.} \\
\multicolumn{7}{l}{\sp{2} \parbox[t]{5.5 in}{Here and in the literature, these credible regions are labeled in analogy to the Gaussian, although they are not, generally, multiples of the posterior's standard deviation.}} \\
\end{tabular}\ataboff
\end{table*}

\edit1{
Figures \ref{fig:barstow-synth-retrievals-model0} and \ref{fig:barstow-synth-retrievals-model1} show the best-fit spectra and marginalized posteriors for the retrievals on the \citet{BarstowEtal2020mnrasRetrievalComparison} synthetic spectra produced by {\NEMESIS}, {\CHIMERA}, and {\TauREx} for Models 0 and 1, respectively, with an uncertainty of 60 ppm.
Tables \ref{tbl:retrievals-barstow-credreg-model0} and \ref{tbl:retrievals-barstow-credreg-model1} summarize the retrieved credible regions for each of the retrievals.  
The true parameters are contained within the 95.45\% credible regions, with most also being contained within the 68.27\% regions.}

\edit1{Comparing the reported 1\math{\sigma} credible regions for each retrieval shows minor differences.
For most cases, {\BART} finds narrower credible regions for temperature than the other codes.  
In the case of Model 0, the upper bounds of the 1\math{\sigma} temperature regions are consistently just below the known 1500 K temperature, while the true radius falls at the lower bound of the 2\math{\sigma} region of the {\BART} retrieval on the {\TauREx} forward model.
For the Model 0 cases, {\BART} favors high cloudtop pressures, consistent with the absence of clouds; slightly greater radii; and narrower credible regions for CO than the other codes.
The CO retrieval on the {\NEMESIS} spectrum falls in the 2\math{\sigma} region.
For H\sub{2}O in the Model 0 cases, {\BART} finds similar values, but with narrower credible regions compared to {\TauREx} and {\CHIMERA}.
For the Model 1 cases, {\BART} favors similar radii and lower cloudtop pressures. 
Compared to {\NEMESIS}, {\BART} finds narrower credible regions for CO and H\sub{2}O when retrieving on {\TauREx} and wider credible regions when retrieving on {\CHIMERA}.
Compared to {\TauREx}, {\BART} finds a similar amount of H\sub{2}O but with greater uncertainty; {\BART} also favors greater CO with similar or slightly smaller uncertainties.
Compared to {\CHIMERA}, {\BART} favors greater CO and H\sub{2}O when retrieving on {\NEMESIS}, similar CO and less H\sub{2}O when retrieving on {\TauREx}, and generally finds narrower credible regions.
For all cases except {\NEMESIS} on {\CHIMERA}, {\BART} agrees with the other codes at 1\math{\sigma} or less.  
The single exception agrees at just greater than 1\math{\sigma}.
}

\edit1{
Together with the tests in Section \ref{sec:synthrets}, this demonstrates {\BART}'s ability to retrieve parameters accurately from synthetic data produced by various RT codes.
}

\begin{figure*}[htp]
\includegraphics[width=0.550\textwidth, clip=True]{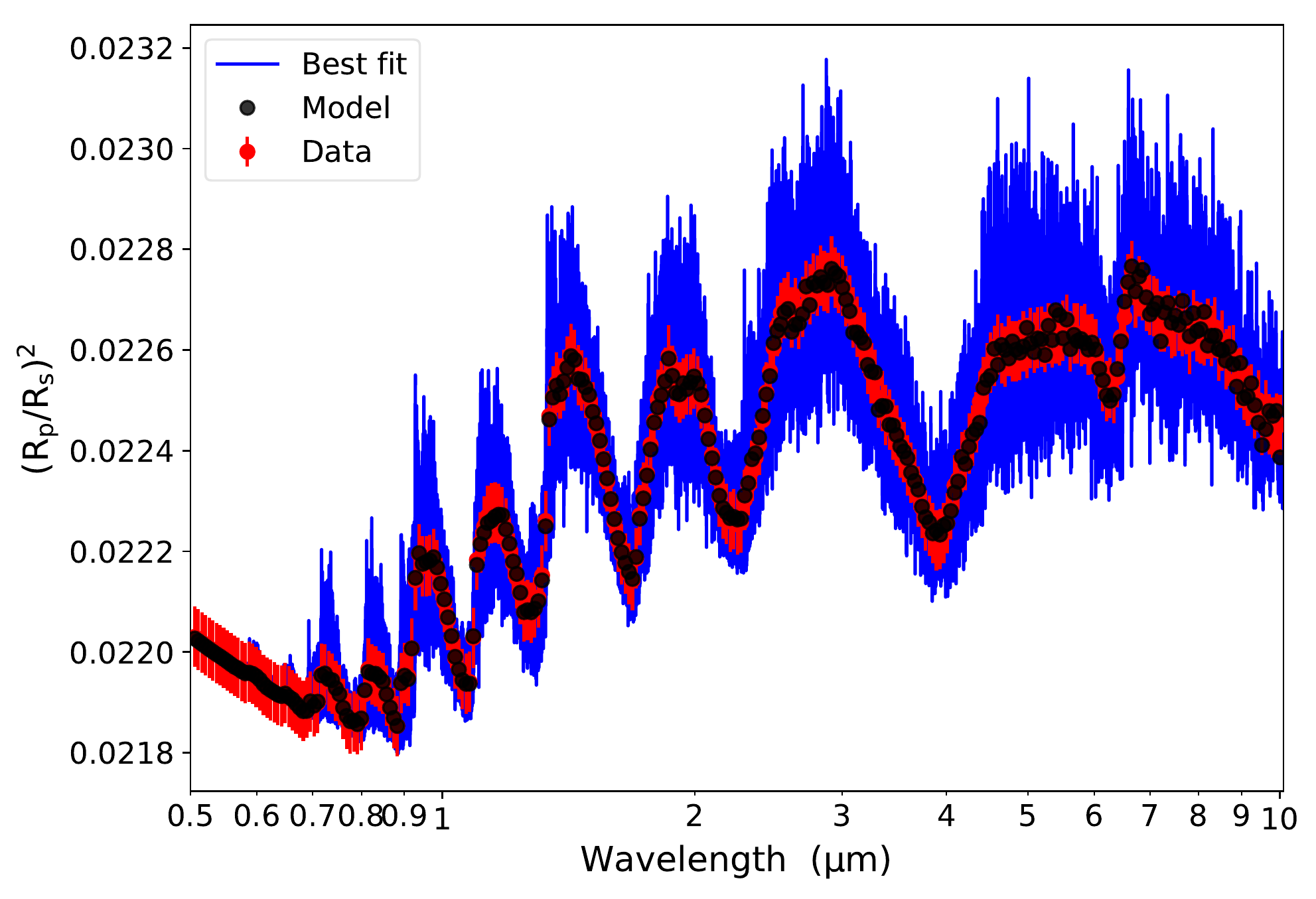}\hfill
\includegraphics[width=0.400\textwidth, clip=True]{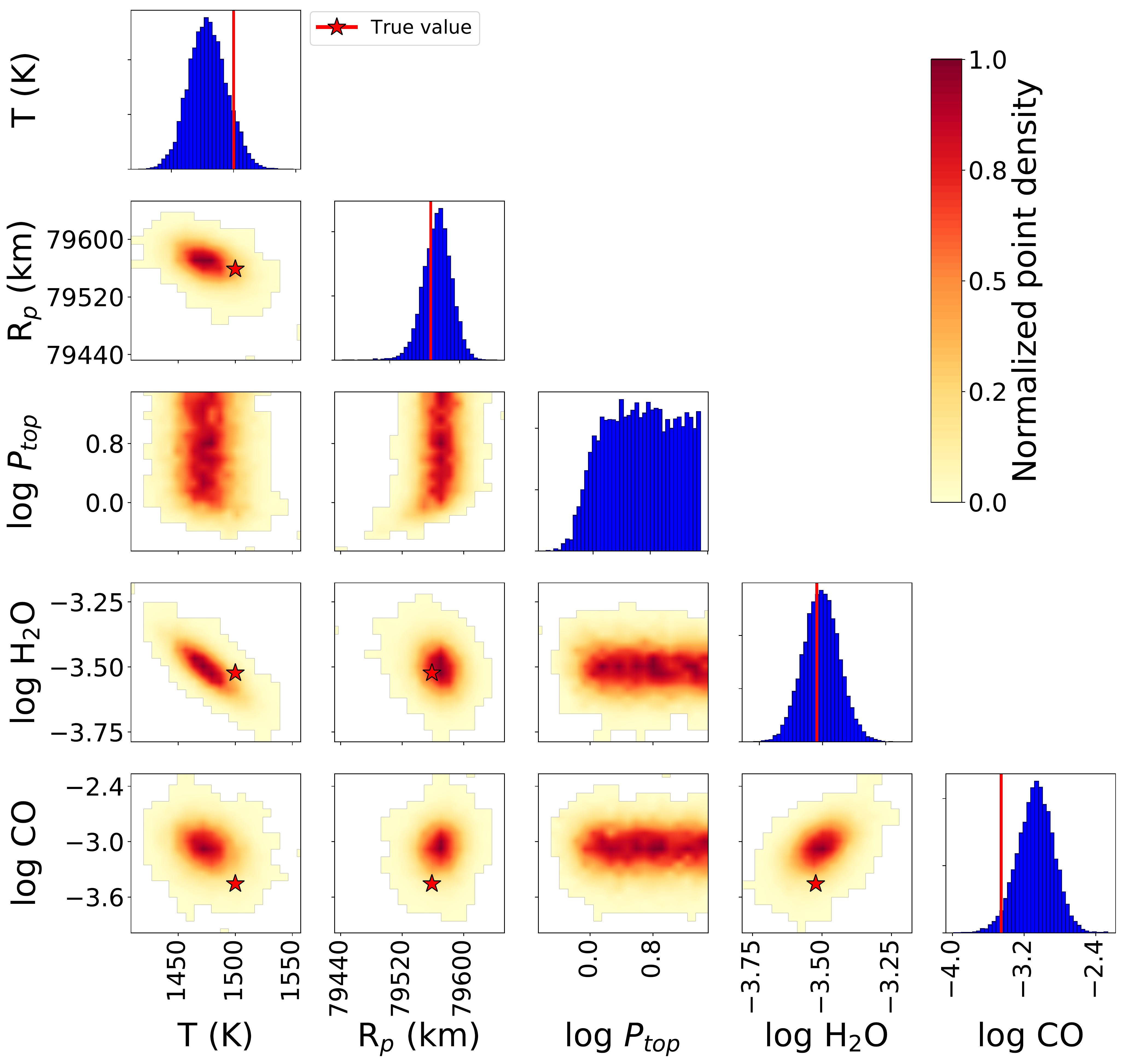}\\
\includegraphics[width=0.550\textwidth, clip=True]{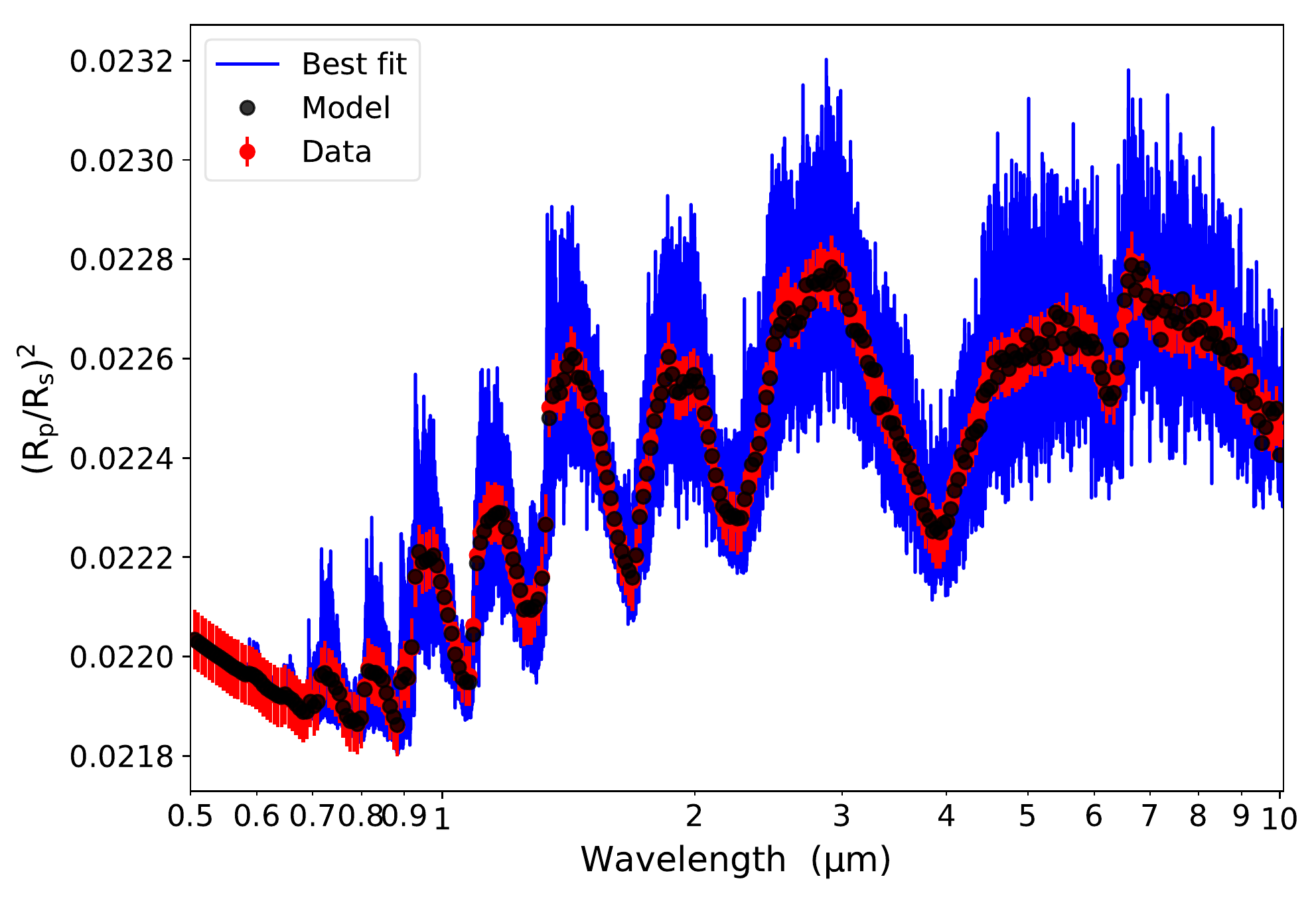}\hfill
\includegraphics[width=0.400\textwidth, clip=True]{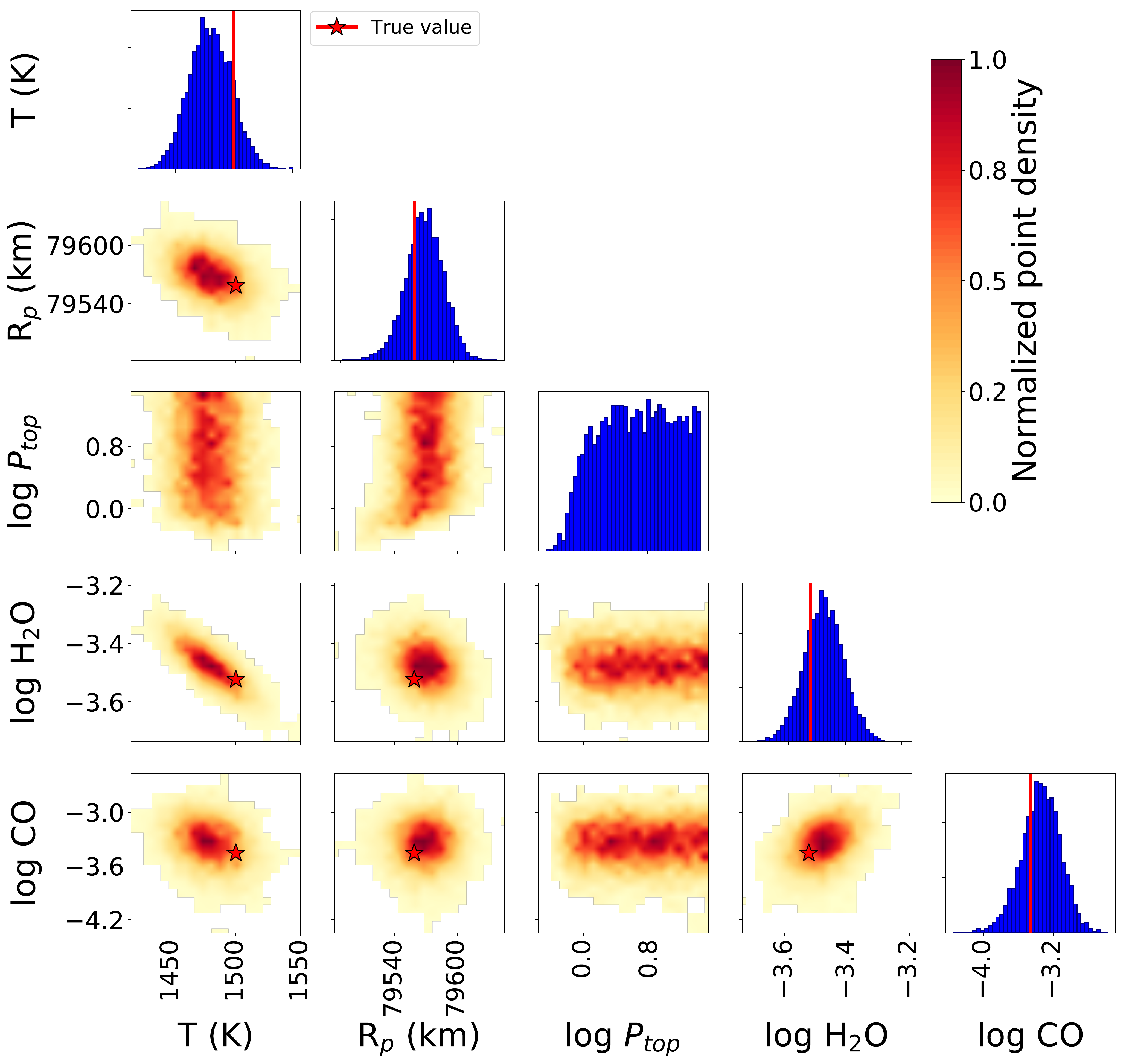}\\
\includegraphics[width=0.550\textwidth, clip=True]{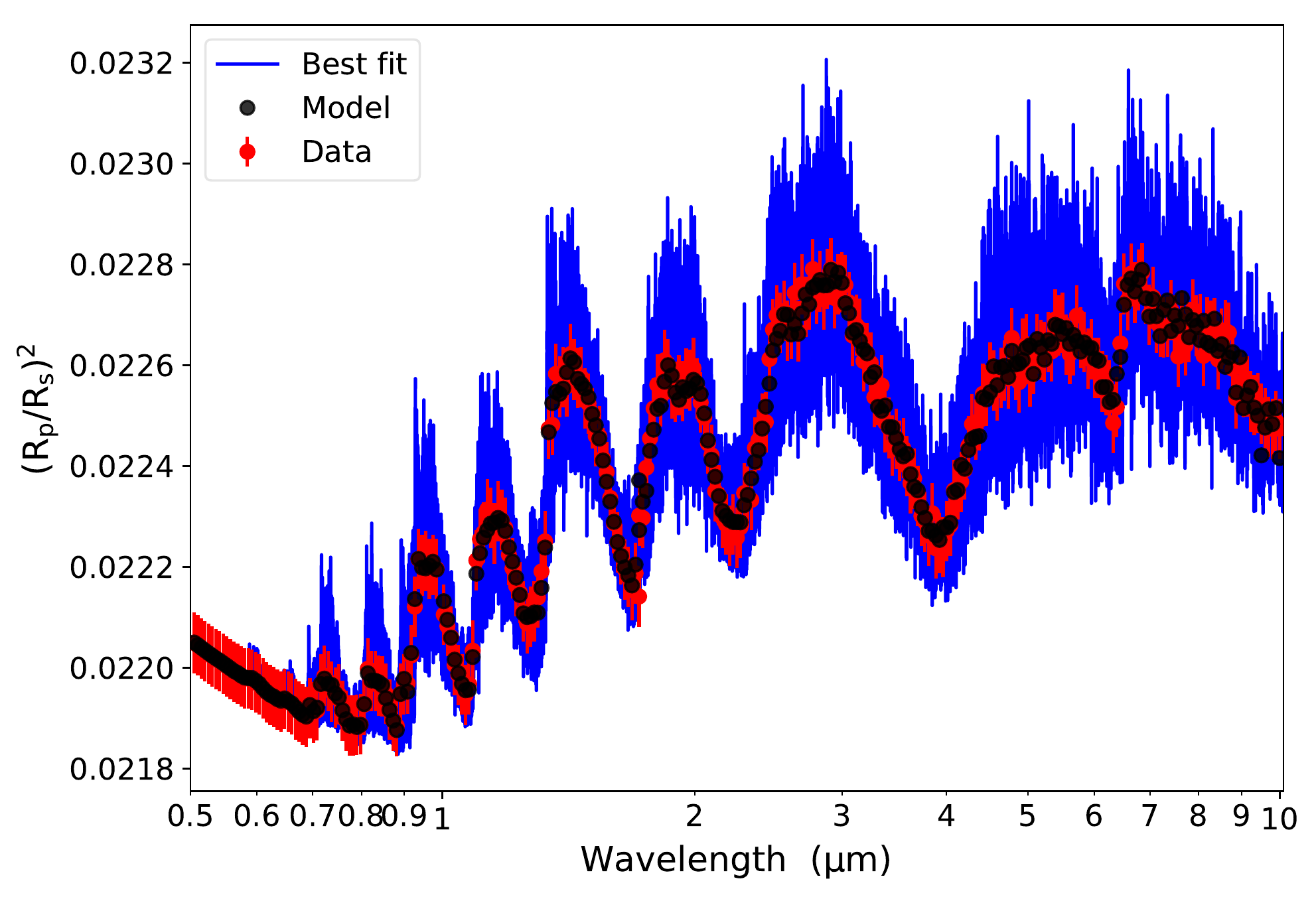}\hfill
\includegraphics[width=0.400\textwidth, clip=True]{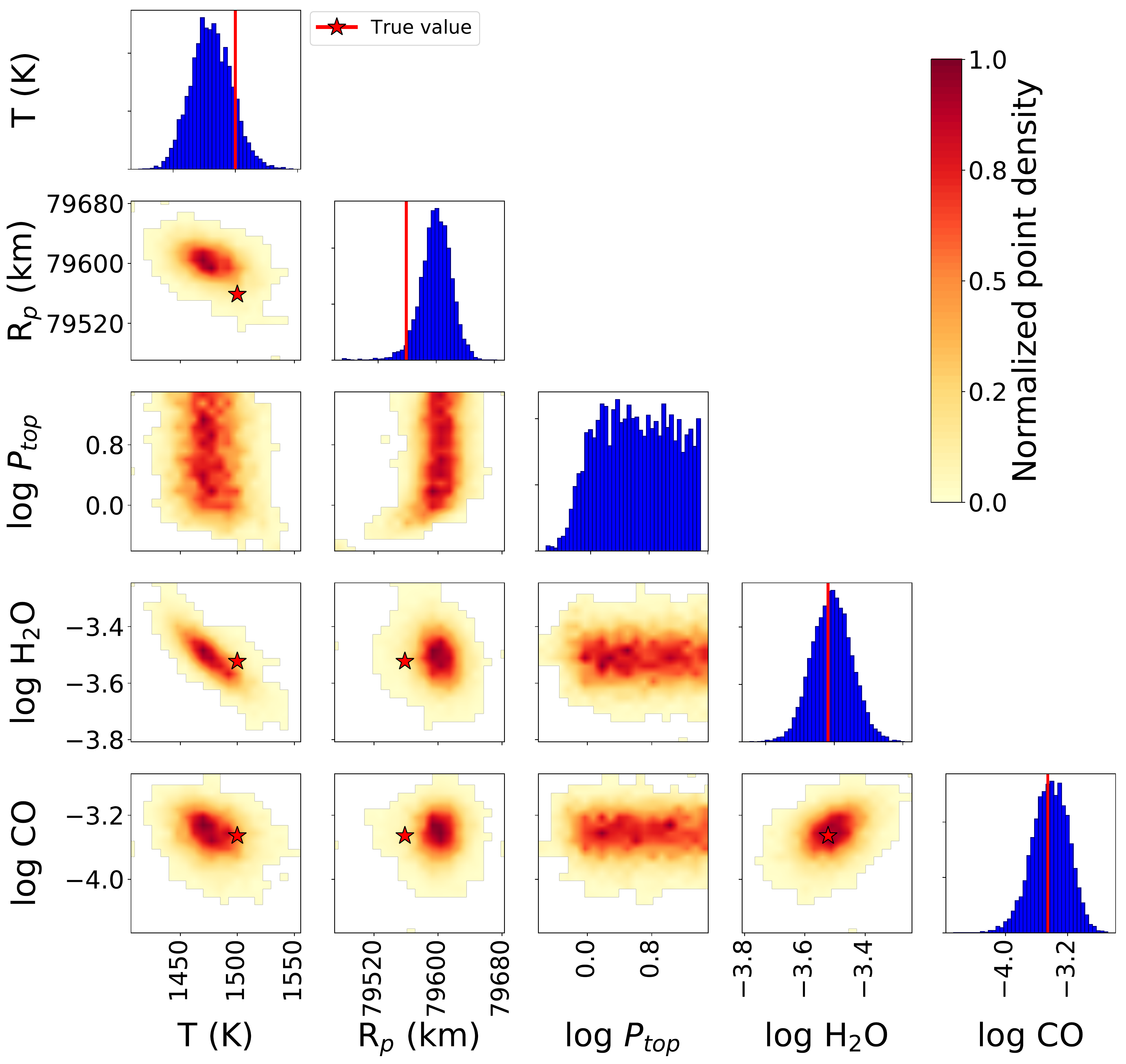}\\
\caption{Summary of retrieval results for the \edit1{\texttt{s04hjclearisoBarstowEtal} test, which retrieve on the spectra of} \citet{BarstowEtal2020mnrasRetrievalComparison} Model 0 cases with 60 ppm uncertainties.  Best-fit spectra (left column) and marginalized posteriors (right column) for the {\NEMESIS} (top row), {\CHIMERA} (middle row), and {\TauREx} (bottom row) data sets.
\label{fig:barstow-synth-retrievals-model0}}
\end{figure*}

\begin{figure*}[htp]
\includegraphics[width=0.550\textwidth, clip=True]{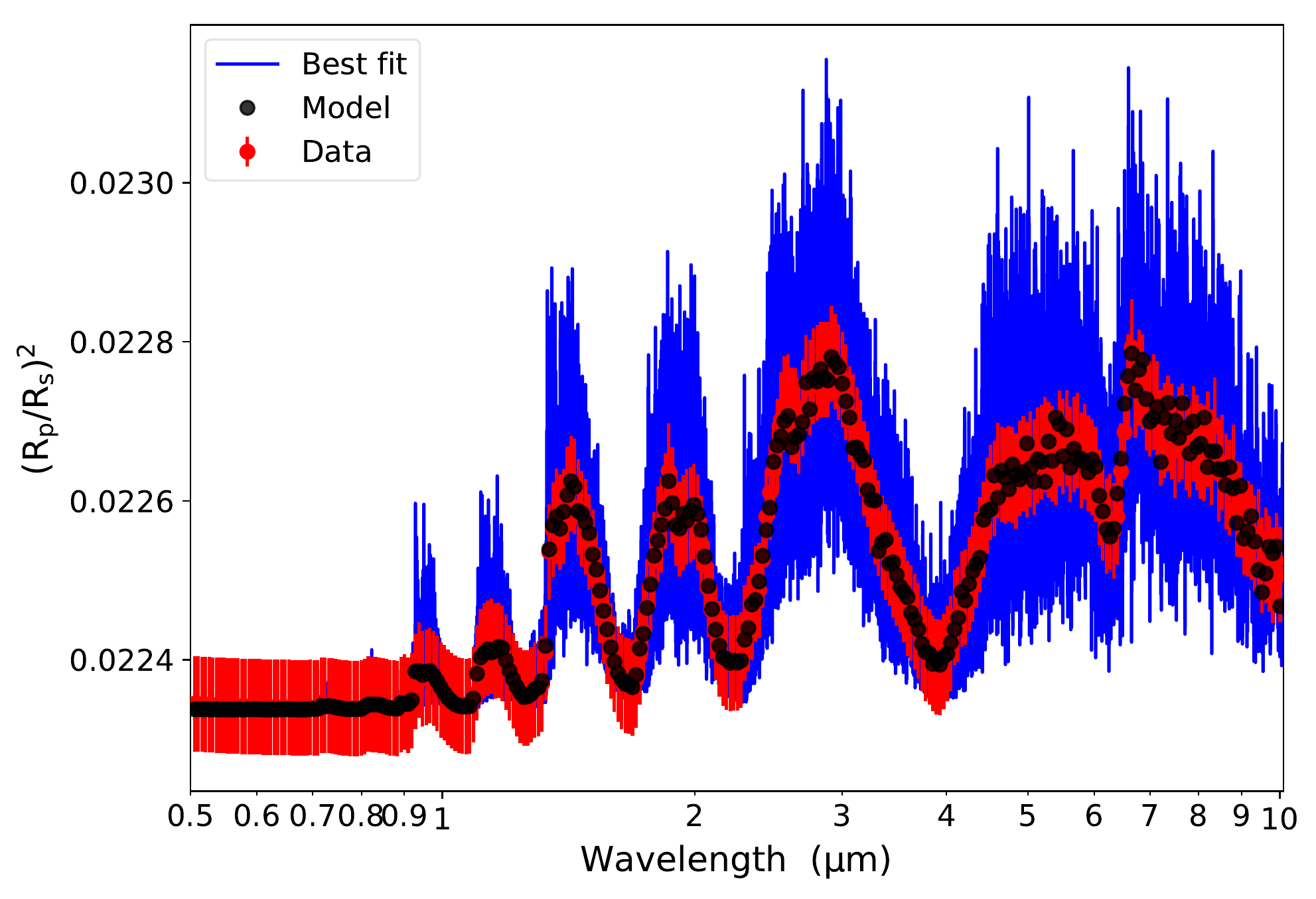}\hfill
\includegraphics[width=0.400\textwidth, clip=True]{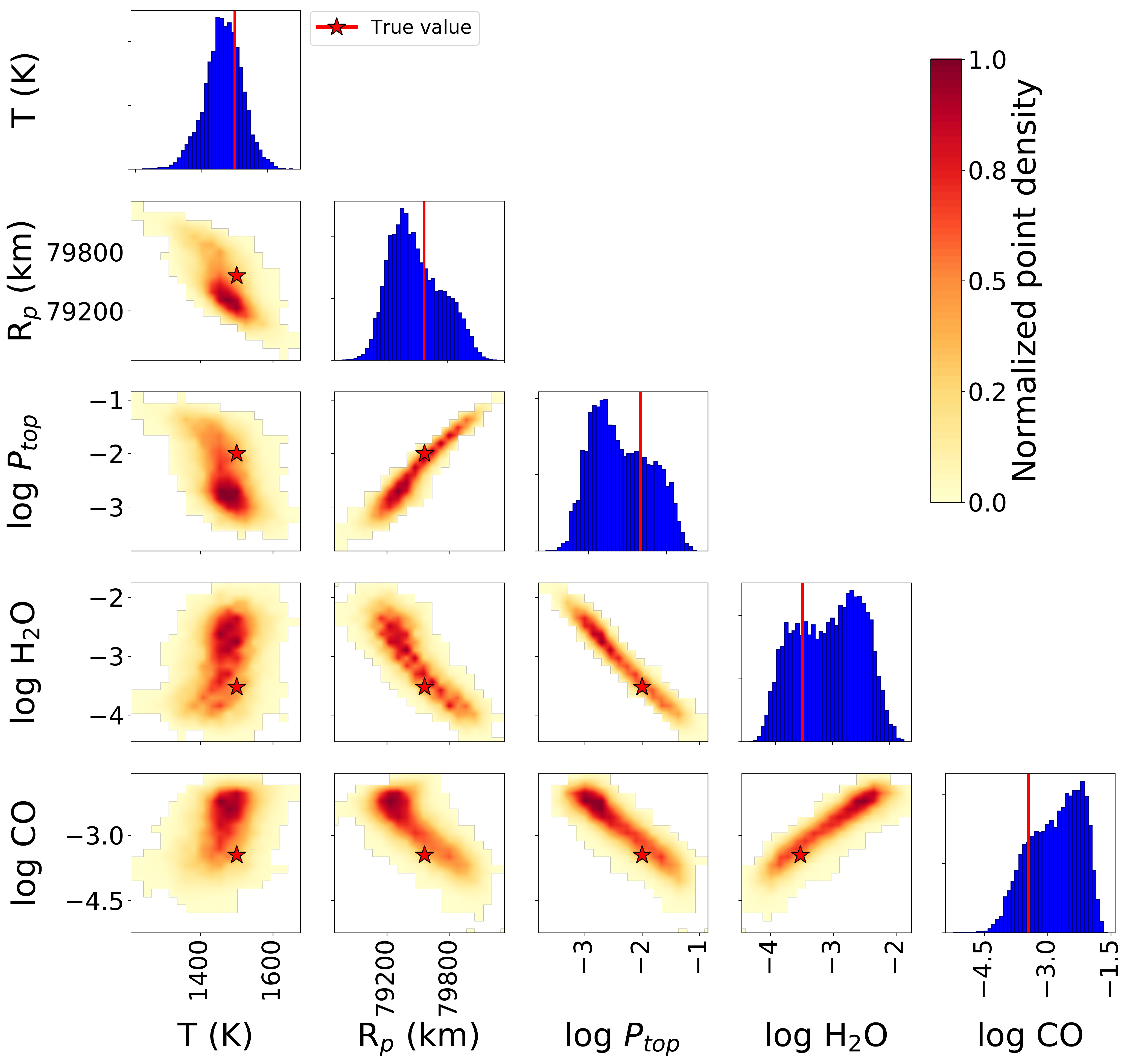}\\
\includegraphics[width=0.550\textwidth, clip=True]{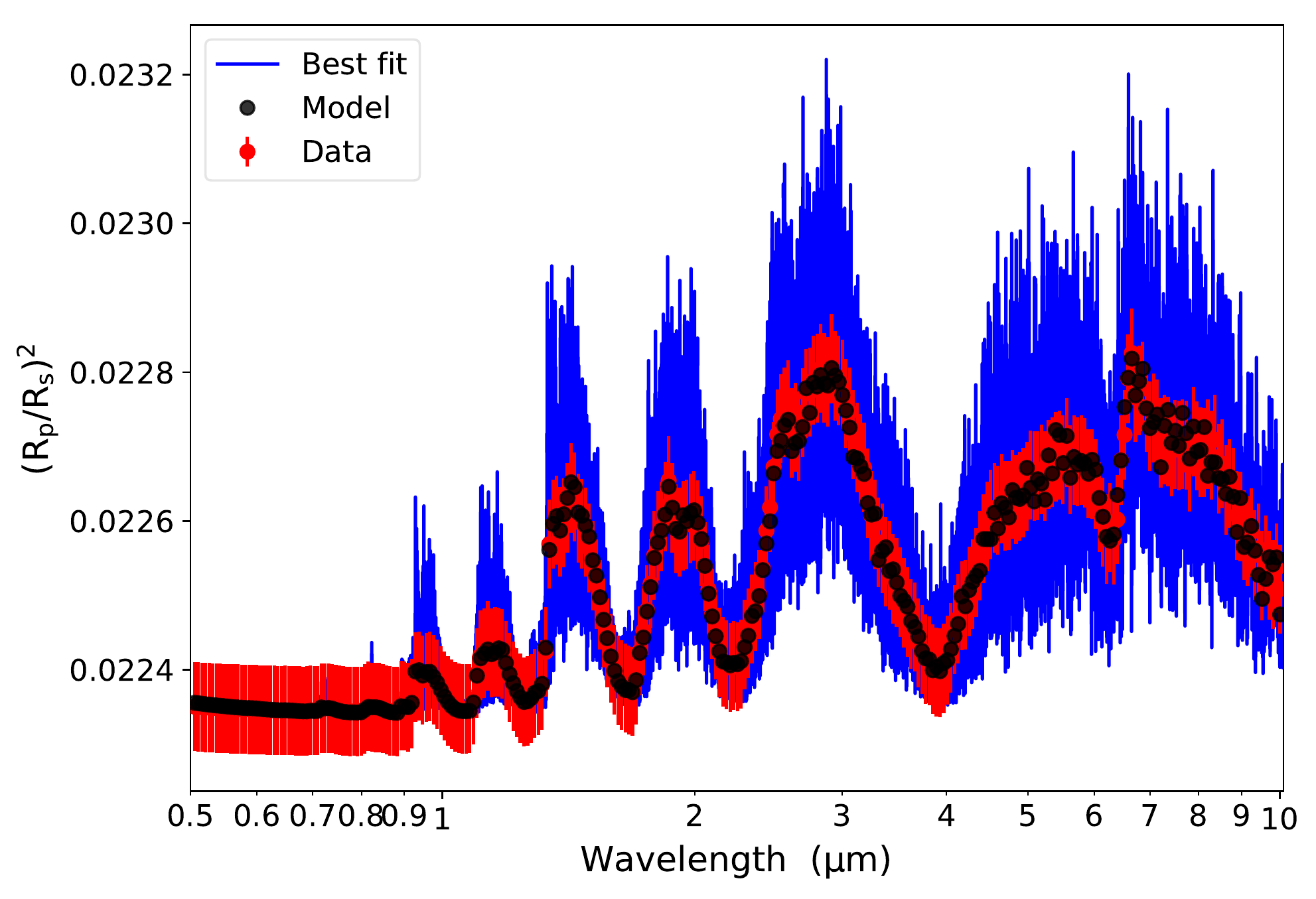}\hfill
\includegraphics[width=0.400\textwidth, clip=True]{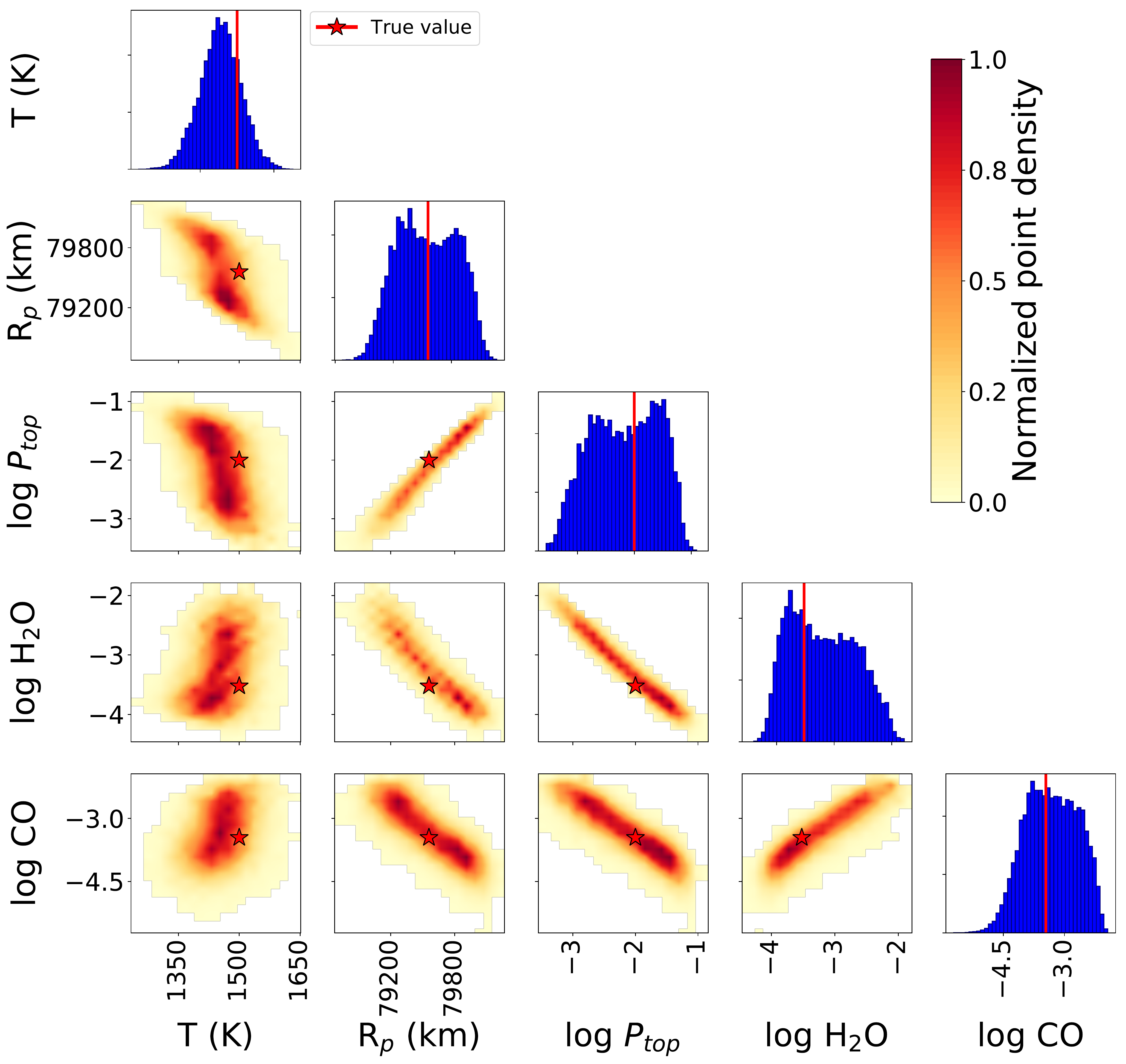}\\
\includegraphics[width=0.550\textwidth, clip=True]{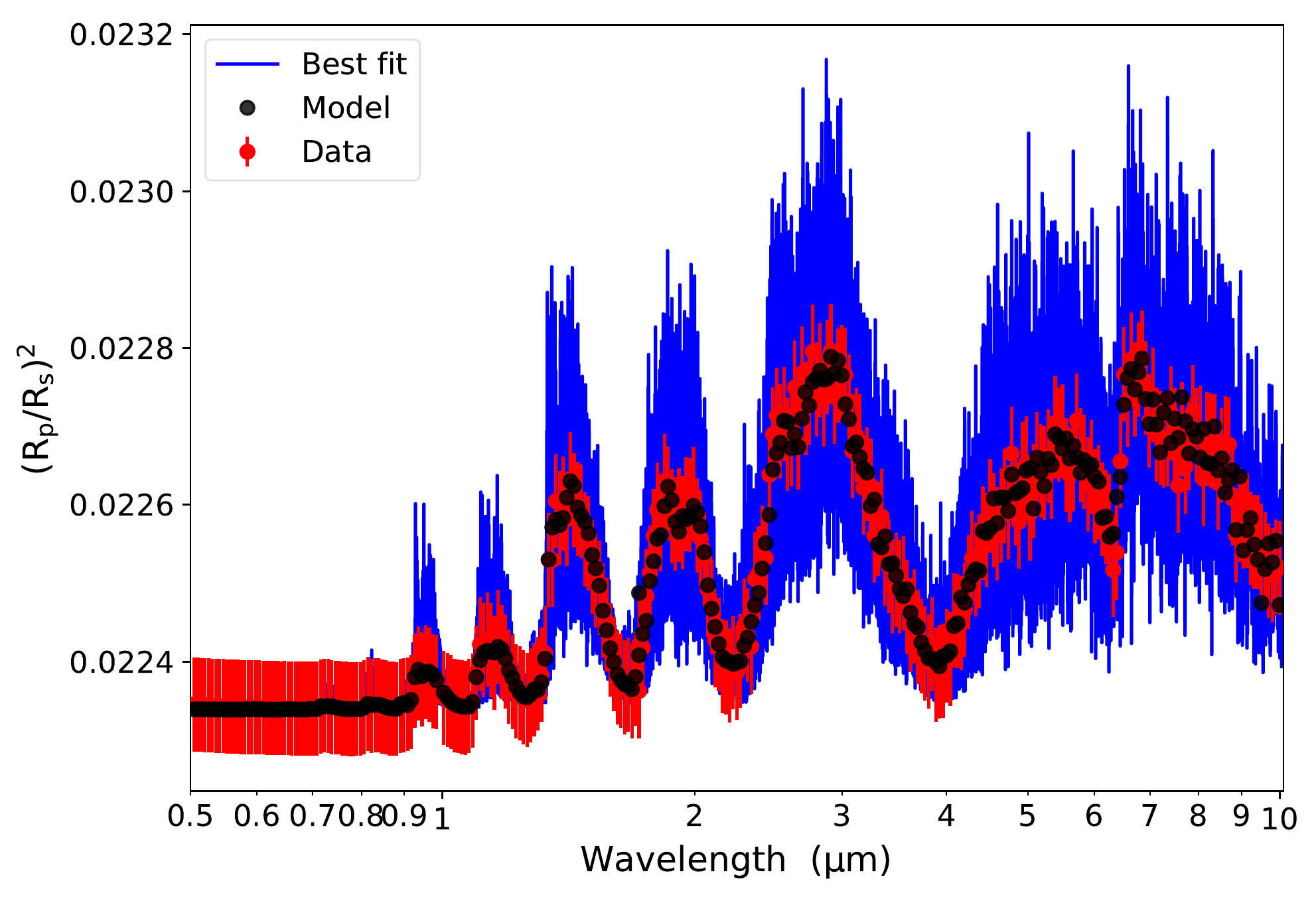}\hfill
\includegraphics[width=0.400\textwidth, clip=True]{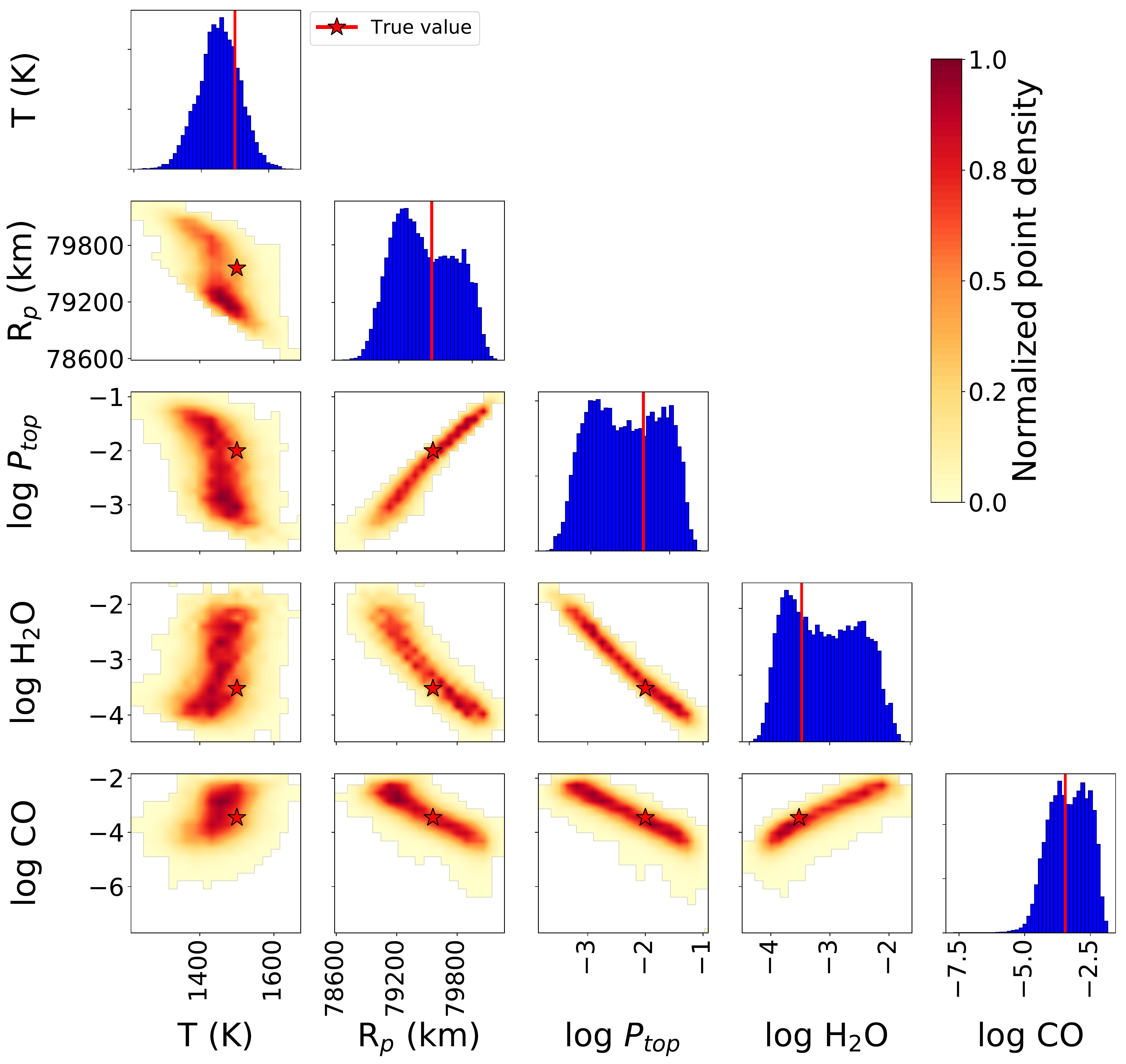}\\
\caption{Summary of retrieval results for the \edit1{\texttt{s05hjcloudisoBarstowEtal} test, which retrieve on the spectra of} \citet{BarstowEtal2020mnrasRetrievalComparison} Model 1 cases with 60 ppm uncertainties.  Best-fit spectra (left column) and marginalized posteriors (right column) for the {\NEMESIS} (top row), {\CHIMERA} (middle row), and {\TauREx} (bottom row) data sets.
\label{fig:barstow-synth-retrievals-model1}}
\end{figure*}

\begin{table*}
\centering
\caption{{\BARTTest} Retrievals, \citet{BarstowEtal2020mnrasRetrievalComparison} Model 0 Cases: Credible Regions}
\label{tbl:retrievals-barstow-credreg-model0}
\atabon\begin{tabular}{cccccc}
\hline\hline
Forward Model  & T (K)        & R\sub{p} (km) & H\sub{2}O (ppmv) & CO (ppmv)     & log \math{P\sb{cloud}} \\
\hline
True           & 1500         & 79558.718       & 300            & 350          & n/a\\
NEMESIS        & [1455, 1492] & [79549, 79588]  & [265, 369]     & [548, 1372]  & [ 0.17, 1.38]\\
               & [1440, 1513] & [79527, 79608]  & [223, 436]     & [325, 2126]  & [-0.12, 1.50]\\
               & [1421, 1531] & [79477, 79626]  & [190, 522]     & [178, 3476]  & [-0.40, 1.50]\\
CHIMERA        & [1462, 1499] & [79551, 79590]  & [286, 395]     & [282,  783]  & [ 0.24, 1.35]\\
               & [1443, 1517] & [79529, 79610]  & [242, 469]     & [153, 1289]  & [-0.15, 1.49]\\
               & [1424, 1540] & [79502, 79628]  & [204, 542]     & [ 76, 2094]  & [-0.38, 1.50]\\
{\TauREx}        & [1457, 1497] & [79581, 79621]  & [260, 365]     & [228,  670]  & [ 0.04, 1.17]\\
               & [1437, 1516] & [79557, 79643]  & [221, 432]     & [124, 1084]  & [-0.18, 1.49]\\
               & [1421, 1539] & [79471, 79660]  & [180, 535]     & [ 62, 1777]  & [-0.46, 1.50]\\
\hline
\multicolumn{6}{l}{For each data set, we report the 68.27\%, 95.45\%, and 99.73\% credible regions, \edit1{from top to bottom per model}.} \\
\end{tabular}\ataboff
\end{table*}

\begin{table*}
\caption{{\BARTTest} Retrievals, \citet{BarstowEtal2020mnrasRetrievalComparison} Model 1 Cases: Credible Regions}
\label{tbl:retrievals-barstow-credreg-model1}
\atabon\begin{tabular}{cccccc}
\hline\hline
Forward Model  & T (K)        & R\sub{p} (km) & H\sub{2}O (ppmv) & CO (ppmv) & log \math{P}\sub{cloud} \\
\hline
True           & 1500         & 79558.718       & 300            & 350          & -2.0 \\
NEMESIS        & [1416, 1524] & [79122, 79666]  & [214,  5198]   & [584, 10750] & [-3.13, -1.87] \\
               & [1344, 1578] & [79019, 80024]  & [ 82,  7905]   & [110, 15241] & [-3.32, -1.30] \\
               & [1270, 1630] & [78869, 80204]  & [ 52, 14246]   & [ 38, 22065] & [-3.55, -1.06] \\
CHIMERA        & [1402, 1512] & [79221, 79947]  & [104,  2418]   & [ 96,  2678] & [-2.82, -1.36] \\
               & [1342, 1564] & [79024, 80086]  & [ 75,  7096]   & [ 36,  6925] & [-3.22, -1.17] \\
               & [1291, 1618] & [78865, 80195]  & [ 52, 13029]   & [ 11,  9897] & [-3.50, -1.03] \\
{\TauREx}        & [1399, 1520] & [79062, 79946]  & [98,  568]     & [100,  391]  & [-3.21, -1.38] \\
               & [1330, 1573] & [78929, 80088]  & [69, 1068]     & [ 24,  891]  & [-3.43, -1.19] \\
               & [1273, 1633] & [78774, 80206]  & [47, 2006]     & [  3, 1448]  & [-3.70, -1.02] \\
\hline
\multicolumn{6}{l}{For each data set, we report the 68.27\%, 95.45\%, and 99.73\% credible regions, respectively.}
\end{tabular}\ataboff
\end{table*}

\subsection{Real-Data Retrieval Test}
\label{sec:realdataretrievaltest}

In a synthetic test, we may or may not know the answer as we work (e.g., in blind testing), but we know in principle what could have gone into the test.
A test on real data is a full analysis, including concerns for unknown systematic errors, time variability of the star and planet, the 3D structure of the atmosphere, physics and chemistry not included in the model, errors and incompleteness in line lists, and even the unknown existence of background sources within the point-spread function of the target star.
Given the state of exoplanet data today, we chose HD 189733 b, a system with a high planetary S/N, a relatively large number of observations in the literature, and little controversy over their interpretation.
Those implementing this test for comparison to our result must configure their codes as we have, in one dimension, without clouds, using the same line lists, and using exactly the same observations.
\edit1{The value of any comparison test lies in the number of comparisons, so we reiterate our invitation to those willing to contribute results of their own application of this test.}

In designing this test, we must mimic one of the two Bayesian models (discussed more in Section \ref{sec:hd189}), by \citet{LineEtal2014apjRetrievalCO} and \citet{WaldmannEtal2015apjTauRex2}.
Further, we wish the test to be accessible on a modest computer.
We chose to emulate the analysis of \citet{LineEtal2014apjRetrievalCO} with {\CHIMERA}, for two reasons.
First, {\CHIMERA} has been applied more broadly than \citet{WaldmannEtal2015apjTauRex2}'s {\TauREx}.
Second, the {\TauREx} analysis uses the \citet{YurchenkoTennyson2014mnrasExoMolCH4} CH\sub{4} line list, which is so complete that it exceeds the storage capacity of many modest computers.
\edit1{Although {\transit} can use this line list either directly or digested into a continuum opacity table and a separate list of the strongest lines \citep{Cubillos2017apjRepack}, not all codes can, and the {\TauREx} test did not.  
Additionally, \citet{HargreavesEtal2020apjsMethaneHITEMP} showed that ExoMol's CH\sub{4} list does not match laboratory studies.}

\edit1{This first test on real data is thus deliberately simple, as appropriate for a test suite, as it ignores clouds and the largest line lists.  
More-complete comparisons, with clouds, the ExoMol lists, numerous variations, and comparison to multiple other works, appear in BART2 and BART3.
As t}his test evaluates whether algorithms consistently fit data not produced by, and unlikely to be fit perfectly by, their respective underlying RT codes\edit1{, t}he metric of success is not \math{\chi\sp{2}} to the data, but similarity \edit1{among} retrievals.

Of course, as future observations and models improve, it may be desirable to \edit1{add more complete tests, with} more recent observations, line lists, physics, and code configurations.
Given the length and complexity of the analysis, we present our test, including \edit1{discussion of} the literature for this planet, in its own section.

\section{APPLICATION TO HD 189733 \lowercase{b}}
\label{sec:hd189}

Due to both its proximity to Earth and its high S/N, the atmosphere of HD 189733 b has been extensively studied since its discovery in 2005 \citep{BouchyEtal2005aapHD189733b}.
Being one of the most analyzed hot Jupiters to date, HD 189733 b is a prime candidate for a real-data retrieval test using published secondary-eclipse data.
Here we discuss previous retrieval analyses of HD 189733 b \citep{MadhusudhanSeager2009apjAtmRetrMeth, LeeEtal2012MNRASHD189733b, LineEtal2012ApJHD189733b, MosesEtl2013ApJC/ORatios, LineEtal2014apjRetrievalCO, WaldmannEtal2015apjTauRex2} and {\BART}'s retrieved atmospheric profiles and dayside emission spectra in the context of these prior analyses.

At an orbital semimajor axis of 0.0312\pm 0.00037 AU and with an eccentricity of 0.0041\pm 0.0025, it takes 2.21857312\pm 0.00000076 days for HD 189733 b to orbit its 5052\pm 16 K, K-type host star \citep{TriaudEtal2009A&AHD189params, StassunEtal2018AJHD189temp}.
In Jovian units, its mass is 1.130\pm 0.025 and its radius is 1.178\pm 0.023 \citep{TriaudEtal2009A&AHD189params}, making its bulk density just over three-quarters of Jupiter's \citep{Southworth2010mnrasPLanetParamsIII}.

Previous studies on HD 189733 b use data from NICMOS \citep{SwainEtal2009apjHD189733b}, IRAC \citep{CharbonneauEtal2008apjHD189733b, KnutsonEtal2009apjHD189733b, AgolEtal2010apjHD189IRAC, KnutsonEtal2012apjHD189IRAC}, IRS \citep{DemingEtal2006apjHD189733b, GrillmairEtal2007apjHD189IRS}, and MIPS \citep{CharbonneauEtal2008apjHD189733b, KnutsonEtal2009apjHD189733b}.

When \citet{MadhusudhanSeager2009apjAtmRetrMeth} published the first exoplanet retrieval via a parametric grid search, they derived that the atmospheric composition of HD 189733 b was high in CO and CO\sub{2}, had a moderate abundance of H\sub{2}O, and had minimal CH\sub{4}.
Further studies by \citet{LeeEtal2012MNRASHD189733b} and \citet{LineEtal2012ApJHD189733b} confirm these abundances, though \citet{MosesEtl2013ApJC/ORatios} indicate that the comparatively large abundance of CO\sub{2} is somewhat of an anomaly.
\edit1{\citet{BarstowEtal2014apjCloudsHD189} included clouds.}  
These studies used optimal estimation rather than a Bayesian sampler.
The upper limit on the CH\sub{4} abundances presented by \citet{LineEtal2012ApJHD189733b} is higher than previous results, such as \citet{MadhusudhanSeager2009apjAtmRetrMeth} and \citet{SwainEtal2009apjHD189733b}.
\citet{LineEtal2014apjRetrievalCO} attribute the discrepancy to non-Gaussian posterior distributions under Gaussian-assuming optimal estimation.

We adopt the following data set: NICMOS data from \citet{SwainEtal2009apjHD189733b}, with the 4 shortest-wavelength channels omitted; IRS data \edit1{from \citet{GrillmairEtal2008natHD189}}; IRAC 5.8 {\micron}, IRS 16 {\micron} photometric, and MIPS 24 {\micron} data of \citet{CharbonneauEtal2008apjHD189733b}; and IRAC \edit1{8.0} {\micron} data of \citet{AgolEtal2010apjHD189IRAC}.  \edit1{IRAC 3.6 and 4.5 data are adopted as 0.1533 \pm 0.0029\% and 0.1886 \pm 0.0071\%, consistent with \citet{LineEtal2014apjRetrievalCO}.}

\edit1{
While newer data (e.g., the secondary-eclipse measurements of \citealp{KnutsonEtal2012apjHD189IRAC}) and analyses (e.g., the IRS re-analysis by \citealp{TodorovEtal2014apjHD189IRS}) are available, this setup is meant as a real-data retrieval test to benchmark {\BART} and future retrieval codes.
This data set exactly matches that of \citet{LineEtal2014apjRetrievalCO}, allowing a direct comparison of results.
}
We note that \citet{LineEtal2014apjRetrievalCO} cite the IRS data as coming from \citet{GrillmairEtal2007apjHD189IRS} but use data from \citet{GrillmairEtal2008natHD189}.
The IRAC 5.8 {\micron} data are cited as coming from \citet{AgolEtal2010apjHD189IRAC}, who did not publish 5.8 {\micron} data.
Rather, they \edit1{use} the 5.8 {\micron} data of \citet{CharbonneauEtal2008apjHD189733b}.
\edit1{Similarly, the IRAC 3.6 and 4.5 {\micron} data are cited as \citet{KnutsonEtal2012apjHD189IRAC} but the data used do not appear to match any published data (M.\ Line, priv.\ comm.).
}
\comment{Michael Himes says: Double checked all of this, and it appears Mike's comments in his code were incorrect, the IRAC1 and 2 data do not appear to match any published data.}

The retrieval model has nine free parameters: five for the \math{T(p)} profile \edit1{\citep{LineEtal2013apjRetrieval1}} and four scaling factors for the vertically constant log abundances of CO, CO\sub{2}, CH\sub{4}, and H\sub{2}O.
\edit1{All priors are uniform, with the log parameters therefore having a log-uniform prior.}
The model atmosphere has 100 layers spanning 10\sp{\math{-8}} -- 100 bar, evenly spaced in log pressure.
We include HITEMP opacities for CO, CO\sub{2}, and H\sub{2}O \citep[valid at the temperature of HD 189733 b]{RothmanEtal2010jqsrtHITEMP}, and HITRAN opacities for CH\sub{4} \citep[measured at 296 K with some hot bands]{Rothman2013jqsrtHITRAN2012}.
We also include H\sub{2}-H\sub{2} and H\sub{2}-He CIAs \citep{RichardEtal2012jqsrtHITRANcia}.
For compatibility with the initial comparison studies, we do not include opacities for minor species whose abundances are not sought, although this is a good practice.

\edit1{We also consider two additional models that are identical in setup except for line lists, to explore their effect on the retrieval results.  
To match the setup of \citet{LineEtal2014apjRetrievalCO}, one model only differs for CH\sub{4}, using the theoretically derived Spherical Top Database System \citep[STDS,][]{WengerChampion1998jqsrtSTDS} at wavelengths greater than 1.7 {\microns} and HITRAN 2008 \citep{RothmanEtal2009jqsrtHITRAN2008} at shorter wavelengths.
The other model uses the latest HITEMP CH\sub{4} \citep{HargreavesEtal2020apjsMethaneHITEMP} and CO lists as well as ExoMol lists for H\sub{2}O  \citep{PolyanskyEtal2018mnrasExoMolH2O} and CO\sub{2} \citep{YurchenkoEtal2020mnrasExoMolCO2} processed via {\repack} \citep{Cubillos2017apjRepack}.
While both these fits are better, the setups are more complex, involving the relatively obscure STDS list and the large ExoMol database.
Since none of these comes close to a perfect spectrum fit, the test in {\BARTTest} is the simplest of the three, using just HITRAN 2012 and HITEMP 2010.
Those interested in reproducing the STDS and ExoMol runs can find the relevant lists, setup details, and plots in this article's electronic compendium (see below).}

Figure \ref{fig:HD189-retrieval-plots} shows {\BART}'s retrieved results for HD 189733 b.
Except for \edit1{\citet{LineEtal2014apjRetrievalCO}}, each of the studies \edit1{from the literature} differs from {\BART} in several fundamental ways, including data, line lists, and modeling approach.
Figure \ref{fig:hd189-abun-comp} and Table \ref{tbl:HD189bestfitvals} show the effect of these differences among the studies.

\begin{figure*}[ht]
\centering
\includegraphics[width=0.46\textwidth, trim=0cm 0cm 0cm 0cm, clip=True]{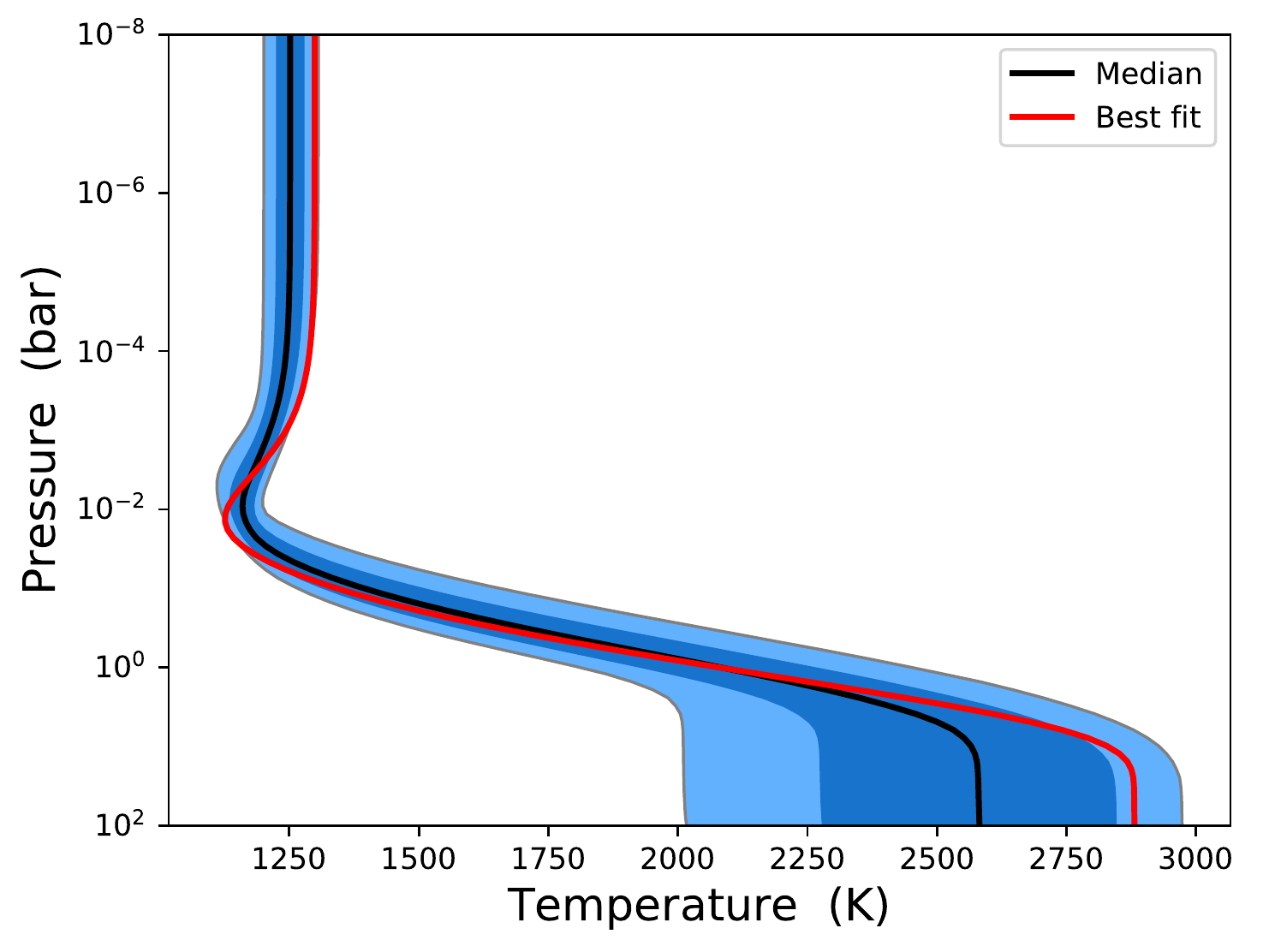}\hfill
\includegraphics[width=0.468\textwidth, clip=True]{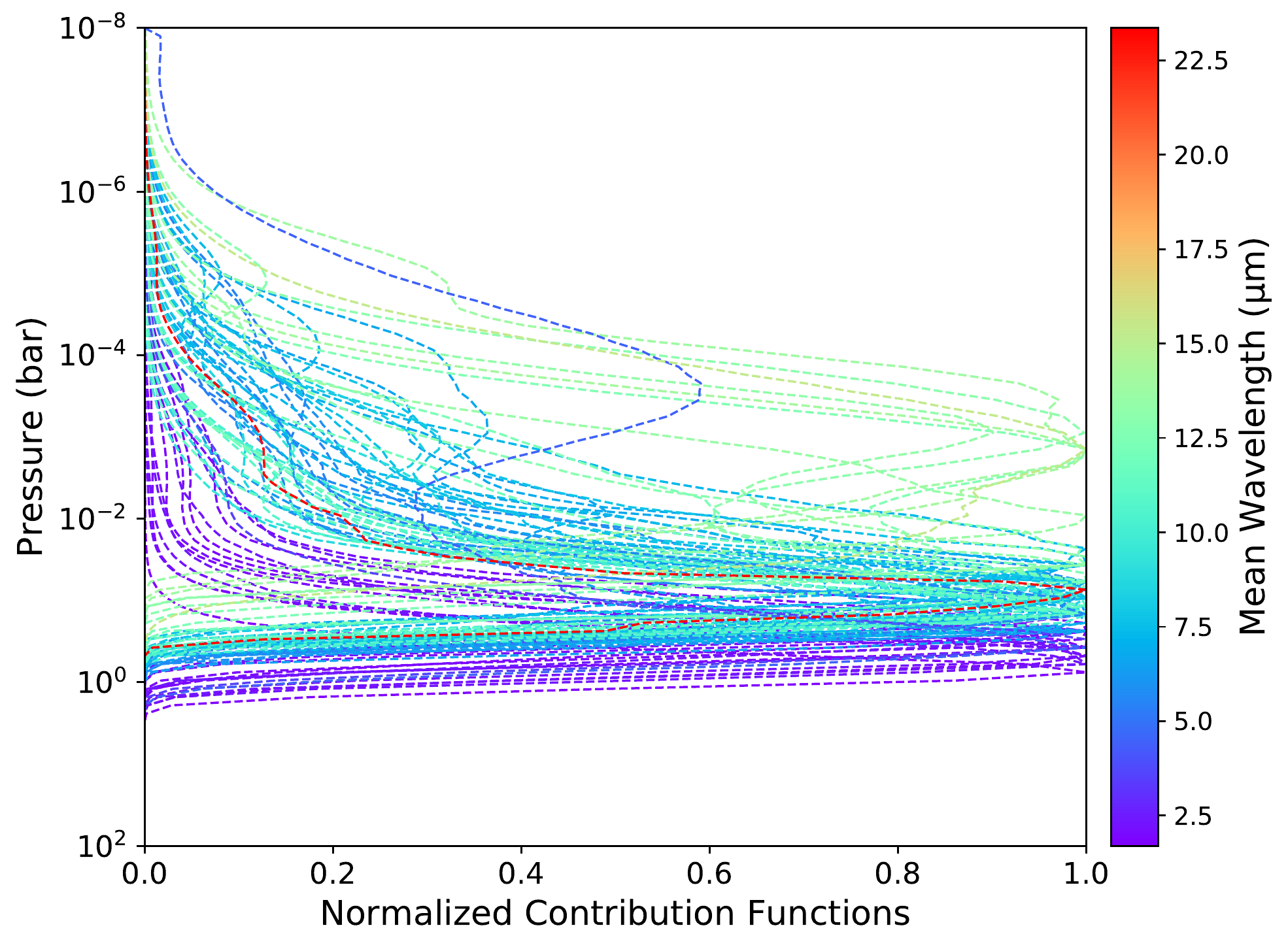}
\includegraphics[width=0.49\textwidth, trim=0cm 0cm 0cm 0cm, clip=True]{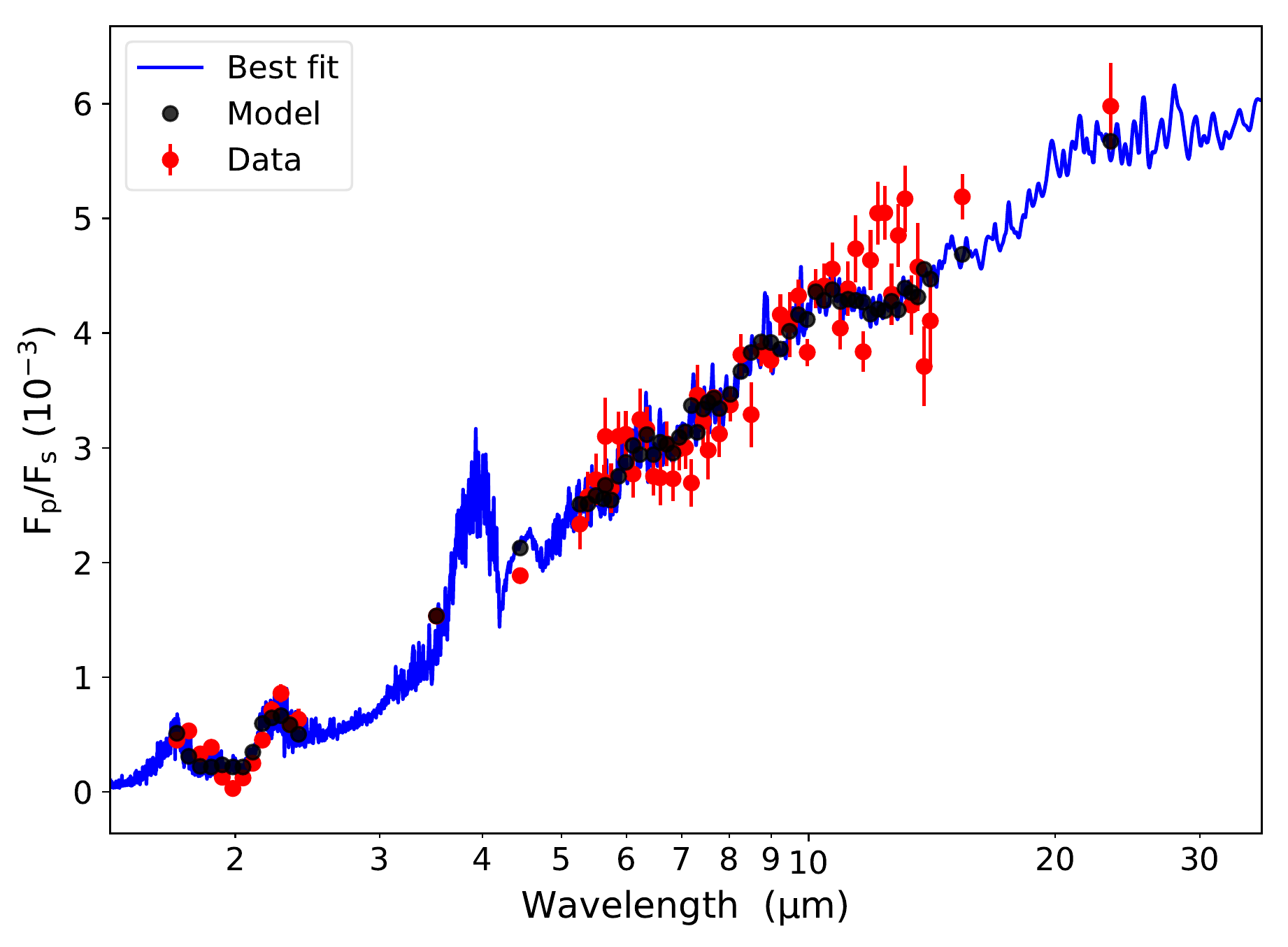}\hfill
\includegraphics[width=0.45\textwidth, clip=True]{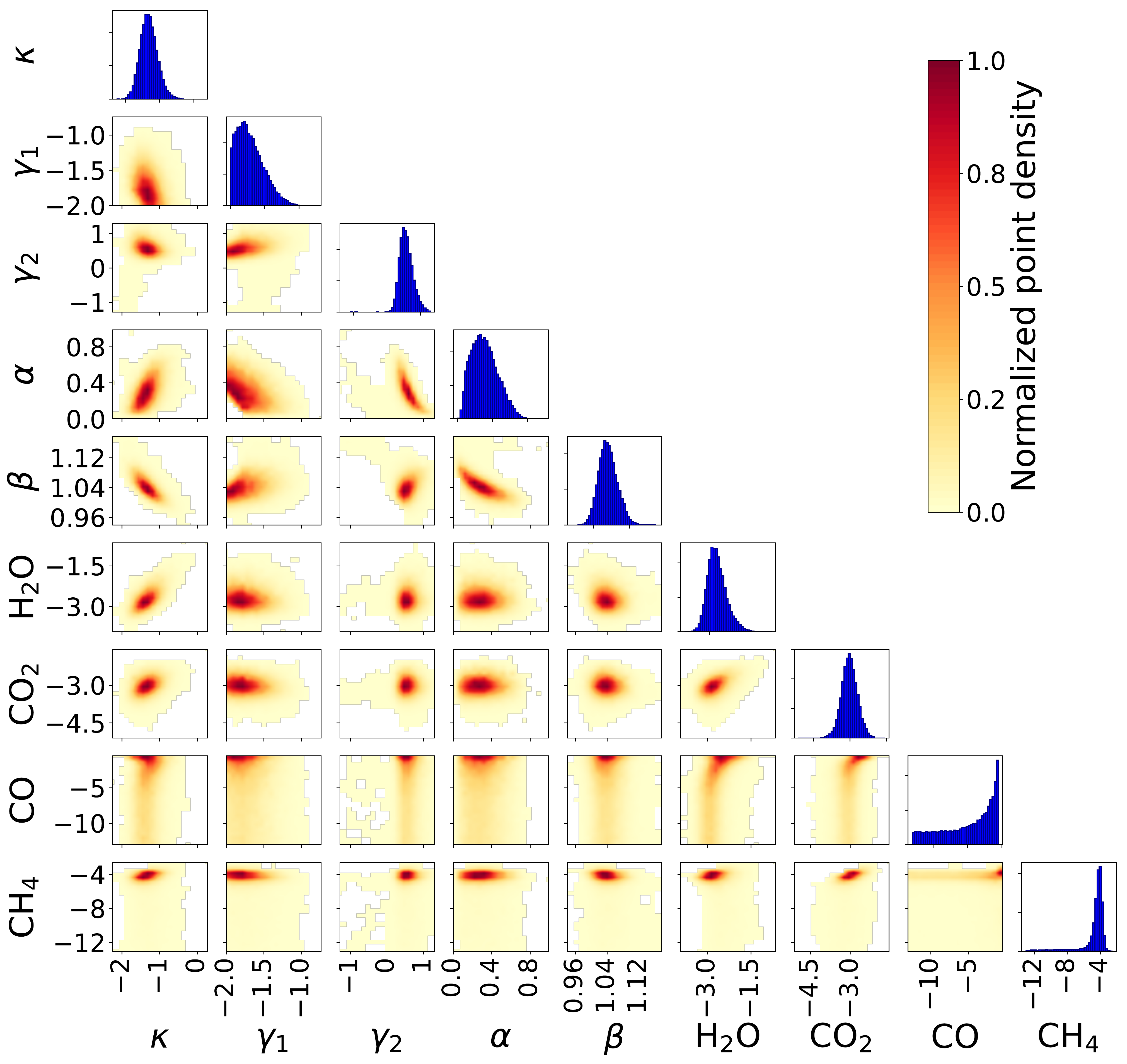}
\includegraphics[width=0.7\textwidth, trim=0cm 0cm 0cm 0cm, clip=True]{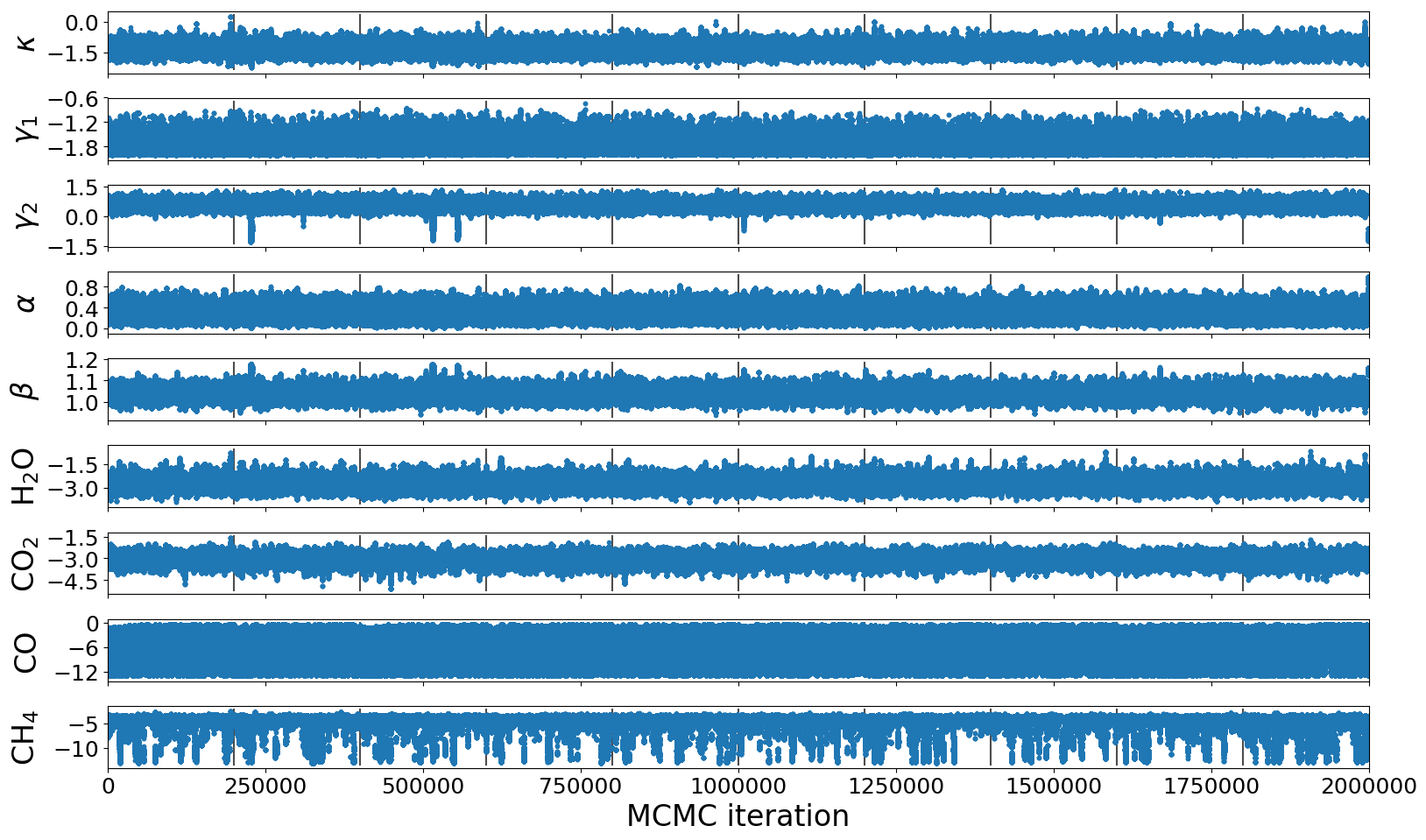}
\caption{Retrieval results for HD189733 b \edit1{(\texttt{r01hd189733b} test)}.
\textbf{Top left}: \math{T(p)} profiles explored by the MCMC (see Figure \ref{fig:synth-retrievals-ecl}).
\textbf{Top right}: Normalized contribution functions of filters (see Figure \ref{fig:synth-retrievals-ecl}).
\textbf{Middle left}: Best-fit spectrum.
\textbf{Middle right}: Pairwise correlations (marginalized over posterior) for all model parameters. Values are in base-10 logarithm, except for \math{\alpha} and \math{\beta}. 
\textbf{Bottom}: Trace plots of explored model parameters.  The 10 chains are concatenated.
\label{fig:HD189-retrieval-plots}}
\end{figure*}

\edit1{In the {\BARTTest} run, except for the IRAC channel 2 data point, the} best-fit spectrum \edit1{qualitatively} agrees with \edit1{that of \citet{LineEtal2014apjRetrievalCO}, and the retrieved abundances agree at 1\math{\sigma}, as shown in Table \ref{tbl:HD189bestfitvals} and Figure \ref{fig:hd189-abun-comp}.  
While \citet{LineEtal2014apjRetrievalCO} find no evidence of a thermal inversion, {\BART} favors a slight inversion in all three models, with retrieved \math{T(p)} parameters closely agreeing.
Except for CO\sub{2}, the retrieved molecular-abundance credible regions for the three models agree at 1\math{\sigma}. 
For CO\sub{2}, the ExoMol model agrees with the {\BARTTest} and STDS models at 1\math{\sigma}, while the {\BARTTest} and STDS models differ by less than 2\math{\sigma}.
When restricting {\BART} to non-inverted thermal profiles, {\chisq} increases significantly, favoring the inverted model by a maximum likelihood ratio of 30 -- 90, depending on the forward-model gridding (0.1 \textit{vs.}\ 1 cm\sp{-1}, respectively).
}

\edit1{We demonstrated 1\math{\sigma} agreement between {\BART} and the modern {\CHIMERA} in Section \ref{sec:barstowrets}, using synthetic cases.
In those tests, the forward model used in the retrieval, or one very similar to it, generated the test data, so a (near-)perfect match exists in the retrieval phase space.
Yet, despite similar spectra and mostly consistent posteriors, {\chisq} values in Table \ref{tbl:HD189bestfitvals} differ substantially.
The reduced {\chisq} values are greater than two, indicating model misspecification.
Real planets will always have physics that are not in any model, in this case including additional opacity sources, more sophisticated and varied clouds, some reflected stellar spectrum, and 3D temperature and compositional variation.
There are also well known, uncorrected systematics in the NICMOS data \citep[e.g.,][]{GibsonEtal2011mnrasNICMOSystematics, CrouzetEtal2012apjXO2bNICMOSystematics}.
The error from such misspecification must distribute somehow among the parameters, but model differences could distribute it differently.}

\edit1{There are some modest model differences.
BART's Bayesian sampler was DEMCzs \textit{vs.}\ DEMC for {\CHIMERA}.
All Bayesian methods should converge to the same posterior, within the noise of random sampling (see Appendix \ref{ap:MCMCsteps}). 
For example, we find similar results to {\CHIMERA}'s PyMultiNest with our DEMCzs in Section \ref{sec:barstowrets}, and these algorithms are less similar than DEMCzs and the DEMC of \citet{LineEtal2014apjRetrievalCO}.
So, we do not blame the samplers for the discrepancy among models.
BART uses Kurucz stellar models, while \citet{LineEtal2014apjRetrievalCO} use PHOENIX, but \citet{MartinsCoelho2007mnrasStellarLibraryComparisons} found that Kurucz and PHOENIX models are comparable in the near infrared for stars with effective temperatures greater than 4250 K, like HD 189733.
The pre-computed opacity grid of \citealp{LineEtal2014apjRetrievalCO} used 20 temperatures ranging 500--3000 K and 20 pressures ranging 20--\ttt{-6} bar, while {\BART} used 25 temperatures spanning 600--3000 K and 100 pressures ranging 100-\ttt{-8} bar.
This may explain {\BART}'s slight improvement in reduced {\chisq} for the STDS case, but not likely the inversion, as the contribution plots indicate little sensitivity in the extended regions and the pressure gridding should be sufficient for interpolation in both cases.
Priors on thermal-profile parameters differed (uniform \textit{vs.}\ Gaussian), which could have kept \citet{LineEtal2014apjRetrievalCO} from finding a {\chisq} minimum with an inverted profile.}

\edit1{The model differences above lead to a {\chisq} improvement of 10.29 from \citet{LineEtal2014apjRetrievalCO} to our STDS case, which uses the same line lists, a maximum likelihood ratio of 172.
One might expect even more improvement using the much-more-complete HITEMP and ExoMol line lists.
Instead, {\chisq} deteriorates to just 2.28 better than \citet[maximum likelihood ratio of 3.1]{LineEtal2014apjRetrievalCO}.
Evidently, the need to distribute the misspecification error dominates the improvement in the modern line lists.}

The next-closest study, and the only other Bayesian approach, \edit1{used {\TauREx} 2's Stage-1 MCMC} \citep{WaldmannEtal2015apjTauRex2}.
\edit1{With the release of version 3 \citep{AlRefaieEtal2019apjTauREx3}, many new results from {\TauREx} 2's Stage-1 MCMC are not anticipated, so we have not emulated it directly, but we can discuss it.}

They omit the IRAC 5.8 {\micron} point, which is quite constraining, due to its smaller uncertainties than the IRS spectrum at those wavelengths.
At 10 bar, their uncertainties on \math{T(p)} are small (their Figure 15), despite the data not probing to that depth.
This may be due to conditioning \math{T(p)}.
Our Figure \ref{fig:HD189-retrieval-plots} and similar plots elsewhere show the lack of contribution from that level and the resultant broadening of the \math{T(p)} credible region there.

They used the CH\sub{4} line list of \citet{YurchenkoTennyson2014mnrasExoMolCH4}.
Its higher limiting temperature yields many times the lines of the test's HITRAN list, and thus greater overall opacity.
This tends to reduce the abundance of CH\sub{4}, and could also reduce other species' abundances, if CH\sub{4} opacity appeared where there had been none previously.
This may explain their usually-lower retrieved abundances (Table \ref{tbl:HD189bestfitvals} and Figure \ref{fig:hd189-abun-comp}).
{\BART}'s retrieved \math{T(p)} profile \edit1{(all three cases) differs substantially, with a lower temperature in the upper atmosphere and a lower tropopause pressure (\ttt{-2} \textit{vs.}\ \ttt{-1} bar) than} {\TauREx}'s.
\edit1{Like \citet{LineEtal2014apjRetrievalCO}, they also find no inversion.}

In Table \ref{tbl:HD189bestfitvals} and Figure \ref{fig:hd189-abun-comp}, we also provide the fitted ranges for pre-Bayesian-retrieval abundances reported by \citet{MadhusudhanSeager2009apjAtmRetrMeth}, \citet{SwainEtal2009apjHD189733b}, \citet{LeeEtal2012MNRASHD189733b}, and \citet{LineEtal2012ApJHD189733b}.
We find agreement within 3\math{\sigma} for most molecular abundances.
{\BART}'s results differ at 3\math{\sigma} for CO\sub{2} and H\sub{2}O reported by \citet{SwainEtal2009apjHD189733b} and CO\sub{2} reported by \citet{MadhusudhanSeager2009apjAtmRetrMeth}.
Like \citet{LeeEtal2012MNRASHD189733b}, we similarly find that CO is poorly constrained.
In the case of \math{T(p)} profiles, \citet{LeeEtal2012MNRASHD189733b} found the upper atmosphere to be isothermal (\sim1100 K) down to \sim0.1 bar.
\citet{LineEtal2012ApJHD189733b} find a variety of potential \math{T(p)} profiles, some of which are consistent with \citet{LeeEtal2012MNRASHD189733b}, and some of which are consistent with the results of {\BART} and \citet{LineEtal2014apjRetrievalCO}.
The \math{T(p)} profiles explored by \citet{MadhusudhanSeager2009apjAtmRetrMeth} generally agree with these other analyses, with the exception of the upper atmosphere, which is cooler.
\citet{SwainEtal2009apjHD189733b} did not publish a \math{T(p)} profile.
These investigations used either a grid search or optimal estimation rather than a Bayesian method, and they used a different data set from that used here.
Both of these differences could explain discrepancies in retrieved parameters.

\begin{table*}
\centering
\caption{Comparison of Fitted Log Abundances for HD 189733 b}
\label{tbl:HD189bestfitvals}
\atabon\begin{tabular}{llllllllll}
\hline\hline
Item  
& \multicolumn{1}{p{1.2cm}}{\centering This work\sp{a}, HH\sp{b}} 
& \multicolumn{1}{p{1.2cm}}{\centering This work, STDS\sp{c}} 
& \multicolumn{1}{p{1.2cm}}{\centering This work, EM\sp{d}}
& \multicolumn{1}{p{1.7cm}}{\centering \citealp{WaldmannEtal2015apjTauRex2} Stage-1}
& \multicolumn{1}{p{1.3cm}}{\centering \citealp{LineEtal2014apjRetrievalCO}\sp{e}}
& \multicolumn{1}{p{1.3cm}}{\centering \citealp{LeeEtal2012MNRASHD189733b}\sp{f}}
& \multicolumn{1}{p{1.3cm}}{\centering \citealp{LineEtal2012ApJHD189733b}}
& \multicolumn{1}{p{1.8cm}}{\centering \citealp{MadhusudhanSeager2009apjAtmRetrMeth}\sp{g}}
& \multicolumn{1}{p{1.3cm}}{\centering \citealp{SwainEtal2009apjHD189733b}\sp{g}}\\
\hline
H\sub{2}O & [\n-3.1, -2.4] & [\n-3.4, -2.6] & [\n-3.4, -2.7] & -3.9 \pm 0.2 & [-3.5, -2.9] & [-4.5, -2.0]         & [-4.3, -3.5] & [-5.0, -3.0] & [-5.0, -4.0]\\
          & [\n-3.4, -1.9] & [\n-3.8, -2.2] & [\n-3.7, -2.3]\\
          & [\n-3.6, -1.4] & [\n-4.2, -1.6] & [\n-3.9, -1.8]\\
CH\sub{4} & [\n-4.7, -3.7] & [\n-4.5, -3.9] & [\n-5.1, -4.5] & -6.7 \pm 0.7 & [-5.0, -4.6] & < -4.0               & < -2.0       & < -5.2       & < -5.0\\
          & [-10.5,  -3.1] & [\n-4.8, -3.5] & [\n-5.2, -4.2] \\
          & [-12.7,  -3.1] & [\n-5.1, -3.1] & [\n-5.5, -3.7] \\
CO        & [\n-6.6, -0.5] & [\n-1.8, -0.5] & [\n-4.8, -0.5] & -2.7 \pm 1.4 & [-4.6, -1.5] & n/a & [-2.4, -1.4] & [-4.0, -2.0] & [-4.0, -3.5]\\
          & [-12.5,  -0.5] & [-12.4,  -0.5] & [-12.5,  -0.5] &              &              & (best fit: \\
          & [-12.9,  -0.5] & [-12.7,  -0.5] & [-12.9,  -0.5] &              &              & -2.5) \\
CO\sub{2} & [\n-3.3, -2.7] & [\n-2.6, -1.8] & [\n-2.9, -2.3] & -3.7 \pm 0.5 & [-2.9, -2.4] & [-3.8, -1.5]         & [-2.8, -2.2] & \sim -1.2    & [-7.0, -6.0]\\
          & [\n-3.7, -2.3] & [\n-2.9, -1.4] & [\n-3.2, -2.0] \\
          & [\n-4.0, -2.0] & [\n-3.2, -0.9] & [\n-3.5, -1.6] \\
{\chisq}  &         169.60 & 139.53         & 147.54         &   ---\sp{h}  & 149.82\sp{i} & ---                  & ---          & ---          & --- \\
Red. {\chisq}\sp{j} & 2.98  &   2.45        &   2.59         &   ---        &   2.63       & ---                  & ---          & ---          & --- \\
\hline
\multicolumn{10}{l}{\sp{a} 68.27\%, 95.45\%, and 99.73\% credible regions, stacked vertically.} \\
\multicolumn{10}{l}{\sp{b} Main {\BARTTest} model, which uses HITRAN 2012 and HITEMP 2010 line lists.} \\
\multicolumn{10}{l}{\sp{c} Same as HH, but using STDS CH\sub{4} line list at wavelengths \textgreater 1.7 {\microns} and HITRAN 2008 at wavelengths \textless 1.7 {\microns}.} \\
\multicolumn{10}{l}{\sp{d} Model using most recent ExoMol line lists for H\sub{2}O and CO\sub{2} and HITEMP line lists for CO and CH\sub{4}.} \\
\multicolumn{10}{l}{\sp{e} Reported 68.27\% interval.} \\
\multicolumn{10}{l}{\sp{f} Possible fit range for \math{{\Delta}{\chisq}/N < 1.0} case, with n/a indicating no constraint.} \\
\multicolumn{10}{l}{\sp{g} Reported range of model abundances.} \\
\multicolumn{10}{p{\textwidth}}{\sp{h} Comparing {\chisq} between different models fitting different data does not tell which model is better, so we do not report the {\chisq} of other studies besides \citet{LineEtal2014apjRetrievalCO}.} \\
\multicolumn{10}{l}{\sp{i} Calculated from their reported statistic of {\chisq}/\math{N}\sub{data}.} \\
\multicolumn{10}{p{\textwidth}}{\sp{j} 57 degrees of freedom, though the independence of adjacent spectral channels is questionable if they sense the same molecular band.} \\
\end{tabular}\ataboff
\end{table*}

\begin{figure}[htb]
\centering
\includegraphics[width=\linewidth, clip=True]{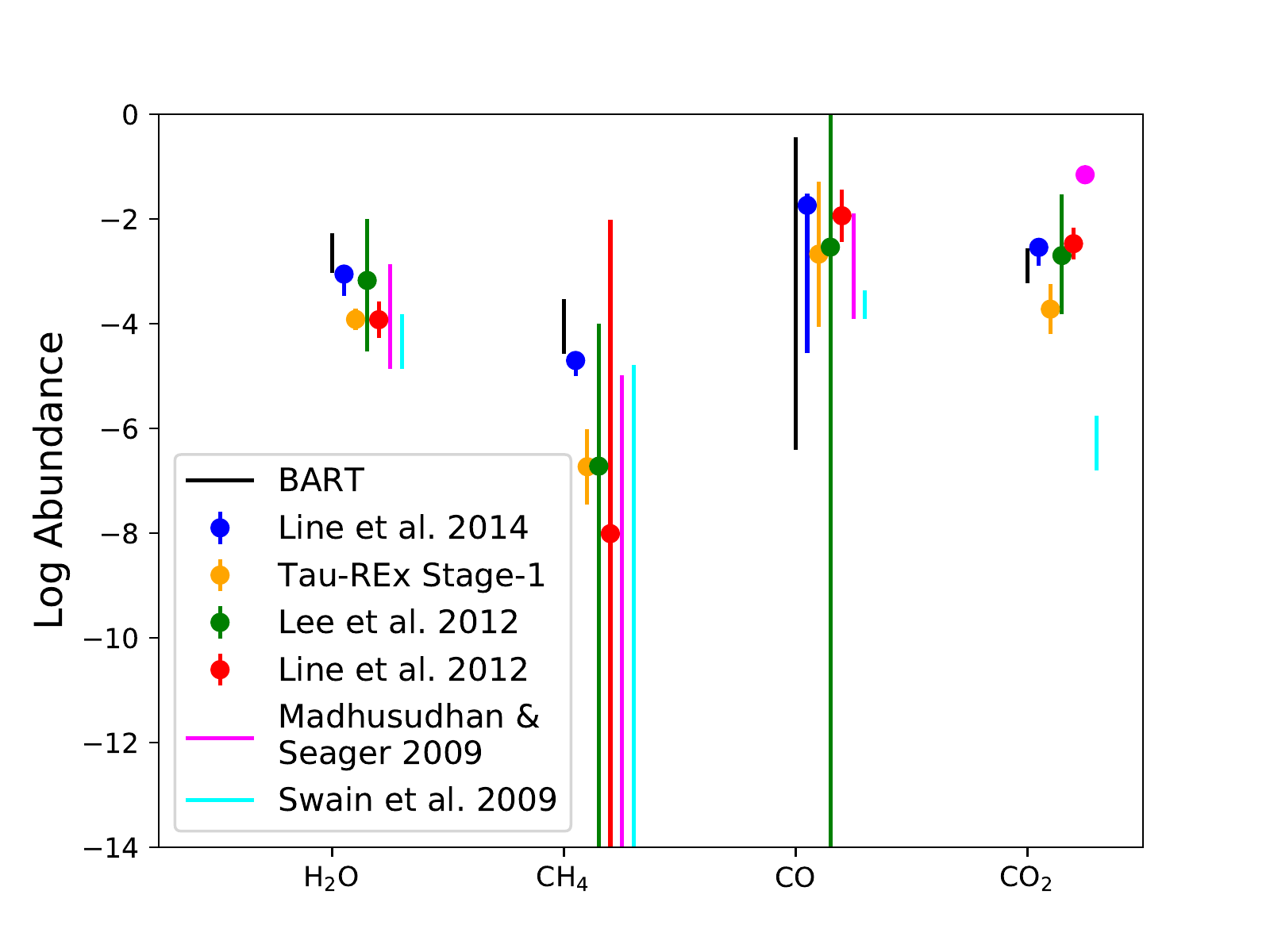}
\caption{Comparison of best-fit retrieved log abundances for HD 189733 b: {\BART}'s retrieved 68\% credible region, {\TauREx}'s Stage 1 result, {\CHIMERA}'s best fit and 68\% interval, the best fit and possible fit range reported by \citet{LeeEtal2012MNRASHD189733b}, and the range of values reported by \citet{LineEtal2012ApJHD189733b}, \citet{MadhusudhanSeager2009apjAtmRetrMeth}, and \citet{SwainEtal2009apjHD189733b}.
\label{fig:hd189-abun-comp}}
\end{figure}

\section{CONCLUSIONS}
\label{sec:concl}

In this paper, we present an open-source code, {\BART}, that makes atmospheric retrieval accessible to the entire community.
It has separate user and code documents and is intended for use and improvement by others.
Its modular, high-level code enables the examination of assumptions by allowing the user to swap out line lists, thermal profiles, and other aspects of the calculation.
The code passed the four categories of tests included in the new {\BARTTest} package, including analytic and comparison RT tests and synthetic and real-data retrieval tests.
The comparison and real-data tests depend on multiple codes implementing them and contributing the results to a central repository, which we invite the community to do.

In applying {\BART} to HD 189733 b, we attempt to validate it by comparing to six retrievals in the literature\edit1{, one of which we emulated in detail}.
In these as well as other comparisons we have conducted, we find that \edit1{physics modeled, line lists, and data can each have a significant effect on the results.}
\edit1{While synthetic tests may match very well} when utilizing the same line lists and data, \edit1{if there is model misspecification (and there always is, on a real planet), its impact on parameter estimates will differ for even slightly different codes.
Figure \ref{fig:hd189-abun-comp} offers some comfort, in that models with significantly different best-fit {\chisq}, and thus degrees of misspecification, can still give consistent parameter estimates.}

\edit1{The level of disagreement among studies using different modeling approaches, and sometimes different data, for the same planet indicates what we might expect going forward, when better instruments make similar-quality 1D measurements of potentially habitable planets.
Thus, when a discovery analysis enters the literature, we should ask whether the conclusion about habitability would change if \math{T(p)} or abundances changed by \math{3\sigma} or even more, due to the impact of the inevitable model misspecifications on parameter estimates and even on the interpretations of the parameters.}

Given the high temperatures of the most easily observed exoplanets, it is critical to have lists that include lines excited by those high temperatures.
By extension, given the high opacities of radicals and ions, it is likely important to include those species as well, when the temperatures warrant it.
Conversely, it is important {\textit{not}} to include species above the temperatures at which they \edit1{dissociate}, although we have seen this in the literature.
So, although we have supplied a free tool that allows anyone to perform a retrieval, \edit1{care is necessary to do so correctly.}

\edit1{Much of what we have learned in creating, testing, and using {\BART} applies to retrieval analyses generally, and even well beyond that scope, giving rise to the four appendices that follow.}

\edit1{We have not found a satisfactory stopping condition (sufficient number of iterations) for Monte Carlo Bayesian analysis in the literature.
The decision of when to stop has varied among practitioners of exoplanet retrieval and light-curve fitting, and has never been justified statistically, to our knowledge.
Many practitioners incorrectly stop at Gelman-Rubin convergence (this is when the statistical posterior may mimic the true posterior sufficiently that we can \textit{start} using samples).
Credible regions and other quantities derived from a sampled posterior become more accurate with more accepted steps.
Appendix \ref{ap:MCMCsteps} derives a new expression relating credible-region accuracy to the number of steps, allowing practitioners to run until they have achieved a desired accuracy.}

\edit1{While developing synthetic-data tests, we obtained poor reproducibility until we discovered that such spectra must be created at extremely high resolution, much higher than needed in the retrieval spectra.
As others may run afoul of this requirement, we demonstrate the effect and experimentally derive the needed resolution in Appendix \ref{ap:synthreterrors}.}

In reproducing the work of others to validate {\BART}, we have found inconsistencies between things as basic as the data cited and those actually used.
Frequently, published descriptions are inadequate to reproduce retrieval work without extensive direct conversations with and even data requests of their investigators.
As time passes, such conversations become much more difficult: old research records are lost and people leave the field.
By providing a checklist for authors and reviewers in \edit1{Appendix \ref{ap:reports}}, we hope that future retrieval reports \edit1{from all codes} will be more complete and reproducible.

By releasing {\BART} under a license that requires full disclosure and archiving of a retrieval calculation, including any code modifications, we ensure that {\BART}'s body of work, at least, will be reproducible in the long term.
\edit1{See Appendix \ref{ap:rr} to learn why this accelerates science and how to manage it without undue additional work.}
We hope others will embrace reproducible research in their own work, and we look forward to a deeper conversation on openness and reproducibility in astrophysics and planetary science.

The description of {\BART}'s algorithms, initialization, and post-processing routines continues in Papers 2 and 3.

A compendium of this paper's software and analyses is available at \url{https://physics.ucf.edu/~jh/BART-compendia/BART1/}.
(It will be uploaded to Zenodo.org upon acceptance\edit1{, and the sentence above will be updated appropriately.  Reviewers wishing anonymous access can do so via the free VPN service at protonvpn.com.}).
\comment{Old compendia:
https://github.com/exosports/BART/HarringtonEtal2019_BART1
}

\begin{acknowledgments}
We thank J.\ Fortney for co-advising the dissertations of JB and PC,
and for discussions.
\edit1{We also thank M.\ Line for helpful clarification on the HD 189733 b data used in \citet{LineEtal2014apjRetrievalCO}.}
We appreciate references to historical sources provided by J.\ Pasachoff.
We appreciate feedback and feature requests provided by members of the {\BART} mailing lists and others who tried out early versions.
We thank contributors to NumPy, SciPy, Matplotlib, AstroPy, the Python Programming Language, GitHub.io, the NASA Astrophysics Data System, and the free and open-source \edit1{software} communit\edit1{ies} for software and services.
Part of this work is based on observations made with the {\em Spitzer Space Telescope}, which is operated by the Jet Propulsion Laboratory, California Institute of Technology under a contract with NASA.
This work was supported by NASA Planetary Atmospheres grant NNX12AI69G, NASA Astrophysics Data Analysis Program grant NNX13AF38G, and NASA Exoplanets Research Program grant NNX17AB62G, held by JH.
\edit1{MDH held NASA Fellowship Activity fellowship 80NSSC20K0682.}
JB held NASA Earth and Space Science Fellowship NNX12AL83H.
IDD and JB held NASA Exoplanets Research Program grant NNX17AC03G.
P\edit1{E}C was supported by the Fulbright Program for Foreign Students.
P\edit1{M}R acknowledges support from CONICYT project Basal AFB-170002.
\end{acknowledgments}

\software{
{\BART}\footnote{\href{https://github.com/exosports/BART}
                      {https://github.com/exosports/BART}}
(This work,
\edit1{BART2}\comment{\citealp{CubillosEtal2021psjBART2}},
\edit1{BART3}\comment{\citealp{BlecicEtal2021psjBART3}}),
MC3\footnote{\href{https://github.com/pcubillos/mc3}
                      {https://github.com/pcubillos/mc3}}
\citep{CubillosEtal2017apjRednoise},
{\TEA}\footnote{\href{https://github.com/dzesmin/TEA}
                     {https://github.com/dzesmin/TEA}}
\citep{BlecicEtal2016apjsTEA},
NumPy \citep{Oliphant2006bookNumpy, vanderWaltEtal2011numpy},
SciPy \citep{JonesEtal2001scipy, VirtanenEtal2020NatMethSciPy1},
Sympy \citep{MeurerEtal2017pjcsSYMPY},
Matplotlib \citep{Hunter2007ieeeMatplotlib},
and
AASTeX6.\edit1{3.1} \citep{AASteamHendrickson2018aastex62}.
}

\facilities{HST(NICMOS),
Spitzer(IRAC),
Spitzer(IRS),
Spitzer(MIPS)}

\object{HD 189733 b}

\newpage
\appendix
\counterwithin{figure}{section}
\counterwithin{table}{section}

\section{METHODS FOR REPRODUCIBLE RESEARCH}
\label{ap:rr}

\edit1{
In this Appendix, we discuss research methods, applied herein, that, if widely adopted, could (and in other fields, do) accelerate the progress of science, by making it easier to find out exactly what was done in a numerical study and by making efforts at reproducing work more straightforward.
}

\edit1{
Reproducibility distinguishes science from all other knowledge systems.
The first question we should ask of a new discovery is whether it could just be a fluke.
Before the advent of computers, research reports carried all the information needed to reproduce an experiment or observation.
They detailed experimental setups and presented all the raw data.
Observatories kept extensive archives of sketches, photographic plates, and astronomers' original notes.
Visits to access original data were common.
}

\edit1{
With the advent of computers, studies became much more complex, and the barriers to reproducing them rose dramatically.
It is now challenging, and often impossible, to learn the exact methods applied in a numerical study, because the number of settings, implementation choices, and input data is too large to describe in any reasonable paper.
Even if one could, writing the code for a significant investigation from scratch can take person-years of effort.
This means that most studies are never independently checked, even to the extent of simply running the same code to confirm that it indeed produces the reported output.
}

\edit1{
This is particuarly worrisome at the cutting edge of discovery, where signal is low, systematics abound, the physical situation under observation is not well understood, and the quantity of independent data points is insufficient to constrain a complete physical model.
This describes the field of exoplanet characterization for the forseeable future.
As demonstrated in the main text, modest differences in methods or data can lead to significantly different results.
It is often entirely opaque why two different studies obtained different results on the same data until someone methodically reproduces both studies and other possible variants to determine which differences explain the disparity.
Without access to original codes and data, this rarely happens.
}

\edit1{
The Reproducible Research (RR, \citealp{FomelClarebout2009ciseRR, Stodden2009CSELegalRR, Donoho2010biostatInviteRR, Barba2018corrTermRR}) movement simply proposes that all reports based on computation include a compendium of the codes, data, settings, and output that support each scientific claim in a paper.
This allows full transparency without bloating the report.
Readers can answer a host of methodological questions just by looking at the code.
They can reprocess the output, plotting it differently or comparing it to the output of similar studies.
They can even rerun the code, adjusting parameters to determine how the results change under different assumptions.
}

\edit1{
Further, in any computational analysis, the code and inputs, not a text written about them, are the most complete, detailed, and accurate record of what was actually done.
While the written description is critical, its main function is summary, informing those using the result and providing a guide to any subsequent reproduction effort.
No-one would want to skip the written description, but only the code is authoritative.
}

\edit1{
RR also benefits science directly by improving practice.
}

\edit1{
Being human, and under significant pressure to publish quickly and frequently, investigators sometimes take shortcuts or even hide inconvenient details that could impede publication.
When others attempt replication from their insufficiently detailed reports, it may be impossible to identify why the results do not match.
This can lead to months of fruitless work on the part of the second investigator, often a junior scientist just trying to understand and build on prior work.
}

\edit1{
Also, by nature, computer codes evolve as investigators add features to handle new situations.
Thus, the code used for the first-run cases in a report may differ from that used later, even though there is only one description.
Worse, if an author receives a question years after publication, the code as published may no longer exist.
The investigators may even no longer be available.
}

\edit1{
At first glance, RR looks much more time-consuming than current practice.
It is, if compendium assembly is done retroactively, especially if corners have been cut.
We argue that good codes should build their own compendia, and careful workers thus spend little additional time.
BART, for example, makes a directory for each run, copies the configuration files to it, and puts all the output there.
It also records the software version.
As there are many files, {\BART} includes a guide to the file structure that helps readers find the most-sought items.
To build the compendium, the investigator simply collects these run directories and adds a top-level text file mapping the runs to the claims in the paper (e.g., ``Figure 7 comes from run 9'').
In conducting the work, the investigator logs the purpose of each run and keeps a list of the best runs for each claim, to locate the right directories later among what could be many trial runs.
Careful investigators already do this, so there is little extra burden.
}

\edit1{
Upon submission, the investigators upload the compendium, usually as a single, compressed file, to one of many RR archive services, provides the permanent identifier in the paper, and never touches it again.
}

\edit1{
Why not continue with the status quo?
The purest replication would not even look at the original code.
If code access is really needed, the authors can propose a collaboration with the original authors, or just ask for access.
}

\edit1{
Unfortunately, the time required to implement many computational analyses from scratch strongly deters reproduction, and often we just want to answer a question, rather than fully rerun an analysis.
If little is reproduced or even questioned, science strays from its founding strength.
Also, authors often consider their codes to be proprietary, a sort of moat against competition.
This attitude, time since the original work's publication, and simple busyness can reduce the level of cooperation below that needed for a successful confirmation.
}

\edit1{
On the other hand, with access to the full investigation, readers can ask, ``what did they really do?'', ``how did they do that?'', and even, ``what happens if we do this instead?''
Most of these questions are too small to merit bothering original authors with, but they could easily be answered with a look at the compendium.
This quickly builds trust in the work (and citations) or identifies errors before they mislead the field.
RR makes hiding dishonesty and cutting corners much harder, too.
}

\edit1{
RR is different from open-source software (OSS).
A proprietary code could appear in an RR compendium with a license that allows reading and even running the code, but prohibits redistribution, modification, or publishing new calculations without permission from and even co-authorship with the code owners.
We promote the OSS approach, as it allows others to stand on our shoulders, as we have stood on the shoudlers of those who came before us.
}

\edit1{
We have released {\BART} under the same Reproducible Research Software License (RRSL) used for {\TEA} \citep{BlecicEtal2016apjsTEA}.
This license ensures that all work produced by {\BART} and its derivatives is subject to confirmation.
It also addresses concerns often raised about openness.
When publishing, code users are required to cite the papers describing the code.
Investigators can develop the code in the open without fear of others publishing with it, or portions of it, before they finish it and publish, as this is forbidden.
Those improving the code must share their improvements if they publish results from them.
This prevents ``castling'', where investigators privatize a public code they did not write by improving it, but not sharing.
Authors may still write less-restrictive licenses to specific parties, possibly in exchange for compensation.
For commercial or other non-scientific purposes, the RRSL acts as a permissive OSS license.
}

\edit1{
The RR approach produces a higher quality of work in original papers without unduly greater effort, and makes evident exactly what was done without longer papers.
There are also abundant political, social, financial, and access benefits of RR, which we leave to white papers and national studies \citep[references above and][]{NASEM2018NAPNASAOpenSource}.
The RRSL eliminates barriers to openness that do not exist in the commercial world of conventional software licenses, but softer solutions might be better received.
I}deal\edit1{ly}, fund\edit1{ers}, journals, and peers would provide i\edit1{ncentives, and r}esearchers could choose whether and when to participate.
Our main hope is thus to \edit1{raise} a conversation and \edit1{propose norms} on how we \edit1{conduct and report} computational work.

\section{WHAT TO INCLUDE IN PUBLISHED REPORTS}
\label{ap:reports}

Above, and elsewhere, we have attempted to demonstrate {\BART}'s correctness and flexibility by duplicating several studies in the literature.
There are many inputs and settings in a {\BART} analysis (\edit1{observations} and \edit1{their} uncertainties, filter functions, line lists, atmospheric profiles).
Only in \edit1{a} few cases (and none in this paper) have we been able to set up runs to reproduce another's work based on published reports alone.
While we are grateful to the authors of the original papers, communicating with an author should never be required to reproduce a study.
We thus offer the following {\em minimal}\/ checklist of items that authors and reviewers should ensure are in all retrieval papers and/or their compendia, ideally in machine-readable form, \edit1{whether from {\BART} or any other code, and whether following RR or not}:

\begin{enumerate}
    \item Input data, including uncertainties and filter transmission functions.
    \item Atmospheric pressure/altitude limits and gridding.
    \item Analytic form(s) (possibly by reference) and initial values of all retrieved vertical profiles.
    \item Line lists (including CIAs), including statement of their temperature limits.
    \item Statement that the atmosphere was thick enough that no rays reached the abyssal layer, or the properties of a surface or thick cloud with realistic reflectance/scattering.
    \item \edit1{Discussion of molecular stability under the retrieved \math{T(p)} conditions, if relevant.}
    \item List of retrieval parameters and any conditioning (e.g., smoothing), allowed ranges (can be shown on histograms), and initial values.
    If there is conditioning, any reduction in the number of effective free parameters must be stated.
    \item How initial values were chosen (e.g., thermochemical equilibrium at a given pressure and temperature).
    \item List of non-retrieval values that affected the spectrum (e.g., a constant abundance of C\sub{2}H\sub{4}).
    \item Publication reference for retrieval code.
    \item Any modifications to the code since the publication.
    \item Statement of convergence and how it was tested.
    \item Number of iterations, SPEIS, and ESS; SPEIS calculation method; and precision of confidence intervals.
    \item Values and uncertainties and/or confidence intervals of all retrieved parameters, including any nuisance parameters.
    \item Plots (at least in an electronic supplement or compendium) of posterior histograms, pairwise parameter correlations, and parameter traces.
    \item {\chisq} \edit1{and reduced {\chisq}} value\edit1{s}.
    \item Retrieved vertical profile and filter contribution functions on same or shared-axis, 2-panel plot \edit1{(uniform profiles excepted)}.
\end{enumerate}

This is a long list.
It combines what is needed to reproduce and to validate a run.
It also includes the answers to most questions that readers will have.
\edit1{Users of proprietary codes that cannot be included in a Reproducible Research Compendium (RRC), and those unwilling to share one, can still provide these details in their texts and electronic supplements, greatly improving reproducibility.}

Per {\BART}'s license (above), users must publish all of this information in reviewed reports of {\BART} runs in a\edit1{n RRC}.
\edit1{To make this easier,} {\BART} saves nearly all of this information in its output directory.
In addition to the points above, a full RRC includes \edit1{the code or access to it,} full configuration information (including command line) of each reported run\edit1{,} and the scripts that produced all plots and tables.

If the user takes proper scientific care to document runs as they execute, including recording their command lines and the general purpose and features of each run, then making the RRC is a matter of collecting the {\BART} output directories for the runs in the paper into one directory and writing a README file.
That file should give the command line for each run and identify the specific runs that support the claims in the paper (e.g., table entries, plots, and statements in the text).
\edit1{{\BART}'s user manual is useful for locating specific information in the compendium, as it documents the inputs and outputs in detail.}
We warn that compendium assembly can be \textit{quite time consuming}\/ if the user is sloppy and does not later recall which figure or table entry came from which run.

\section{Required MCMC Effective Sample Size}
\label{ap:MCMCsteps}

MCMC-based posterior sampling uses a finite set of dependent random samples from a posterior distribution to approximate the values of integrals that summarize the posterior.
We address here the question of how large a set of posterior samples one should collect to compute common posterior summaries with sufficient precision.

For a problem with a parameter, \math{\theta}, and posterior distribution \math{p(\theta)}, common summaries we might want to compute include the posterior mean,
\begin{equation}
\mu \equiv \int \theta\, p(\theta)\; d\theta,
\label{mu-def}
\end{equation}
the posterior standardard deviation, \math{\sigma}, with
\begin{equation}
\sigma\sp{2} \equiv \int [\theta - \mu]\sp{2}\, p(\theta)\; d\theta,
\label{sig-def}
\end{equation}
and the probability content, \math{C}, of a credible region \math{\mathcal{R}},
\begin{equation}
C \equiv \int I\sub{\mathcal{R}}(\theta)\, p(\theta)\; d\theta,
\label{C-def}
\end{equation}
where \math{I\sub{\mathcal{R}}(\theta)} is the indicator function for the region, taking the value 1 inside the region and 0 outside it.

Note that the last two summaries are quantifications of uncertainty.
We are computing them with a Monte Carlo algorithm, so estimates of these uncertainty quantifications will themselves have uncertainties (due to use of a finite and random Monte Carlo sample).
This leads to some awkward but necessary linguistic constructions, e.g., ``the standard error of the estimate of the standard deviation''.
Note further that the reason we use MCMC algorithms is that \math{p(\theta)} is not a simple distribution like a Gaussian, which would be amenable to analytic computation.
As a result, there is no general relationship between \math{\sigma} and \math{C} (e.g., the \math{\mu\pm\sigma} interval will not in general have \math{C\approx 0.683}), so, even if we have a precise estimate of \math{\sigma}, we still must separately compute \math{C}.
Similarly, if we find a region with \math{C\approx 0.683}, it does not follow that expanding that region by a factor of two produces a region with \math{C\approx 0.954} (as it would for a one-dimensional Gaussian).
If we want to report credible regions of different sizes, each one needs its own computation and will have its own Monte Carlo uncertainty.

We here estimate how large a Monte Carlo sample we need to compute these summaries with usefully small uncertainty.
We present guidelines for the number of samples from an algorithm that produces \textit{independent, identically distributed}\/ (IID) samples from \math{p(\theta)}.
For MCMC algorithms, the samples are dependent; these guidelines should be interpreted in terms of the \textit{effective sample size}\/ (ESS) for the MCMC output (which may be estimated with various standard techniques that quantify the strength of dependence in the Markov chain output).

We use hats to denote a Monte Carlo estimate of a posterior summary that replaces the integral over \math{p(\theta)} with an average over IID posterior samples.
Using a Monte Carlo sample size of \math{N} values, \math{\{\theta\sub{i}\}}, the estimate of the posterior mean, \math{\mu}, is
\begin{equation}
\hat{\mu} = \frac{1}{N} \sum\sbp{i=1}{N} \theta\sub{i}.
\end{equation}
An analogous expression holds for \math{\hat{\sigma}}, based on Equation (\ref{sig-def}), and substituting \math{\hat{\mu}} for \math{\mu}:
\begin{equation}
\hat{\sigma}\sp{2} = \frac{1}{N} \sum\sub{i} [\theta\sub{i} - \hat{\mu}]\sp{2}
\label{sig-hat-def}.
\end{equation}
(An unbiased estimator for \math{\sigma\sp{2}} would divide by \math{(N-1)}, but we are interested in cases with large \math{N}, so we ignore the difference between \math{N} and \math{N-1}; in any case, the slightly biased estimator using \math{N} has slightly smaller standard error; also, we are interested in estimating \math{\sigma} rather than \math{\sigma\sp{2}}.) For \math{\hat{C}}, the sum is over values of the indicator function; this amounts to simply counting the number of samples in \math{\mathcal{R}}, which we denote \math{N\sub{\mathcal{R}}}, so
\begin{equation}
\hat{C} = \frac{N\sub{\mathcal{R}}}{N}.
\end{equation}

We will use the standard error (root-mean-square variability) as a measure of the uncertainty in one of these estimates.
For \math{\hat{\mu}}, the standard error is the standard deviation of the \math{p(\theta)} distribution, divided by \math{\sqrt{N}}.
Of course, that standard deviation is unknown, but we can estimate it using Equation (\ref{sig-hat-def}).
So, an estimate of the standard error for \math{\hat{\mu}} is
\begin{equation}
s\sub{\hat{\mu}} = \frac{\hat{\sigma}}{N\sp{1/2}}.
\end{equation}
The standard error for the \math{\hat{\sigma}} estimate is trickier to compute.
\citet{BoothSarkar1998AmstatMontecarlo} give a simple approximation, suitable for ballpark estimates of the required sample size.
Their calculation is in the context of bootstrap resampling; we adapt it to our purposes as follows.
Divide Equation (\ref{sig-hat-def}) by the (unknown) true value of \math{\sigma\sp{2}}, giving
\begin{equation}
\frac{\hat{\sigma}\sp{2}}{\sigma\sp{2}} = \frac{1}{N} \sum\sub{i} \frac{[\theta\sub{i} - \hat{\mu}]\sp{2}}{\sigma\sp{2}}
\label{sig-ratio}.
\end{equation}
Adopt a Gaussian approximation for the distribution of \math{(\theta\sub{i} - \hat{\mu})}.
Then, the distribution for the sum will be approximately \math{\chi\sp{2}} with \math{N-1} degrees of freedom (since \math{\mu} is estimated by \math{\hat{\mu}}).
Recall that a \math{\chi\sp{2}} random variable with \math{N} degrees of freedom has expectation value \math{N}, and standard deviation \math{\sqrt{2N}}.
Thus (again ignoring the difference between \math{N} and \math{N-1}), we expect
\begin{equation}
\frac{\hat{\sigma}\sp{2}}{\sigma\sp{2}} \approx 1 \pm \sqrt{\frac{2}{N}}.
\label{sig-chisqr}
\end{equation}
The relative error of the \math{\hat{\sigma}\sp{2}} (variance) estimate is
\begin{equation}
\delta \equiv \left|\frac{\hat{\sigma}\sp{2} - \sigma\sp{2}}{\sigma\sp{2}}\right|
  = \left|\frac{\hat{\sigma}\sp{2}}{\sigma\sp{2}} - 1\right|.
\end{equation}
Using Equation (\ref{sig-chisqr}), the \textit{expected}\/ relative error in the variance is thus
\begin{equation}
\langle\delta\rangle \approx \sqrt{\frac{2}{N}}.
\label{delta-est}
\end{equation}
Let \math{s\sub{\hat{\sigma}}} denote the standard error in \math{\hat{\sigma}}.
By propagation of errors (the delta method), the expected relative error in the standard deviation is half of \math{\langle\delta\rangle}, giving
\begin{equation}
\frac{s\sub{\sigma}}{\sigma} \approx \sqrt{\frac{1}{2N}}.
\end{equation}
So, if we are aiming for 5

Turning now to the error in \math{\hat{C}}, note that estimating \math{C} corresponds to estimating the probability parameter for a binomial distribution, based on \math{N\sub{\mathcal{R}}} successes in \math{N} trials.
We can treat this as a basic Bayesian inference problem.
Adopting a flat prior on \math{C} and a binomial sampling distribution, the posterior distribution for \math{C} is a beta distribution, \math{\mathop{\mathrm{Beta}}(N\sub{\mathcal{R}}+1, N+1)}.
The posterior mode is \math{\hat{C}}, and the posterior standard deviation is
\begin{equation}
s\sub{\hat{C}} = \sqrt{\frac{\hat{C}(1-\hat{C})}{N+3}}.
\end{equation}
If we specify target values for \math{\hat{C}} and \math{s\sub{\hat{C}}}, the necessary sample size (ignoring the insignificant \math{+3}) is
\begin{equation}
N \approx \frac{\hat{C}(1-\hat{C})}{s\sbp{\hat{C}}{2}}.
\end{equation}
Some useful examples:
\begin{itemize}
\item Targeting a \math{68.27\%} credible region (``\math{1\sigma}'' for a Gaussian) with \math{2\%} (absolute) error requires \math{N\approx 500}.

\item Targeting a \math{95.45\%} credible region (``\math{2\sigma}'') with \math{0.5\%} error requires \math{N\approx 1700}.

\item Targeting a \math{99.73\%} credible region (``\math{3\sigma}'') with \math{0.1\%} error requires \math{N\approx 2700}.
\end{itemize}

The latter two sample sizes are fairly large because accurately estimating the probability content of larger regions requires adequate sampling of the tail of the posterior distribution.
A ``\math{4\sigma}'' region extends far into the tail, with \math{C = 0.99994}; to capture its first non-nine digit accurately would require \math{N\approx 250,000}.

Given the prevalence and usefulness of ``\math{2\sigma}'' and ``\math{3\sigma}'' intervals, it is a good practice to generate an ESS of 3000, when computationally feasible.
\citet{FlegalEtal2008StatsciMarkov} advocate a more detailed approach; they use a batch means algorithm to estimate the standard errors for every quantity of interest as an MCMC calculation proceeds, stopping the run when target values are achieved.

\edit1{During the preparation of this manuscript, \citet{VehtariEtal2019arxivRhatESS} introduced an improvement on the convergence criteria of \citet{GelmanRubin1992} by considering both the within-chain and between-chain variances to determine when all chains have converged to the same stationary distribution.  
They also present a related method to estimate SPEIS (and from it, ESS) via combining the autocorrelation estimates of each chain by similarly considering the within- and between-chain variances, which they describe as a conservative estimate.
As of this writing, their paper is still under review.
}

\edit1{
We have elected to use the convergence metric of \citet{GelmanRubin1992} and the SPEIS method described in this manuscript rather than their methods for a few reasons.  
First, our SPEIS method is more conservative.
It yields a greater SPEIS and correspondingly lower ESS, ensuring that our credible region uncertainties are not smaller than they would be if estimated using the \citeauthor{VehtariEtal2019arxivRhatESS} criterion.
Second, for our runs, we find similar values for both convergence criteria.
Additionally, their recommendation regarding posterior draws (``only using the sample if \math{\hat{R} < 1.01}'') would require significantly more MCMC iterations for a negligible change in the results shown in this paper.  
We highlight that their recommended cutoff of 1.01 is consistent with some papers in the literature (including here), although many others use greater values \citep[see Table 1 of][]{VatsKnudson2018arxivGelmanRubin}.
}

\edit1{
Recognizing the likely emergence of a new criterion from recognized experts, we recommend, for the time being, evaluating both criteria and using the more conservative estimate.
A true convergence determination is theoretically impossible.
Other studies, especially with the legacy MCMC explorers still included in {\BART}, may find that either criterion may be more conservative.
}

\section{Retrieval Errors due to Wavenumber Sampling Grid Mismatch}
\label{ap:synthreterrors}

\begin{figure*}
\centering
\includegraphics[width=0.49\linewidth, clip=True]{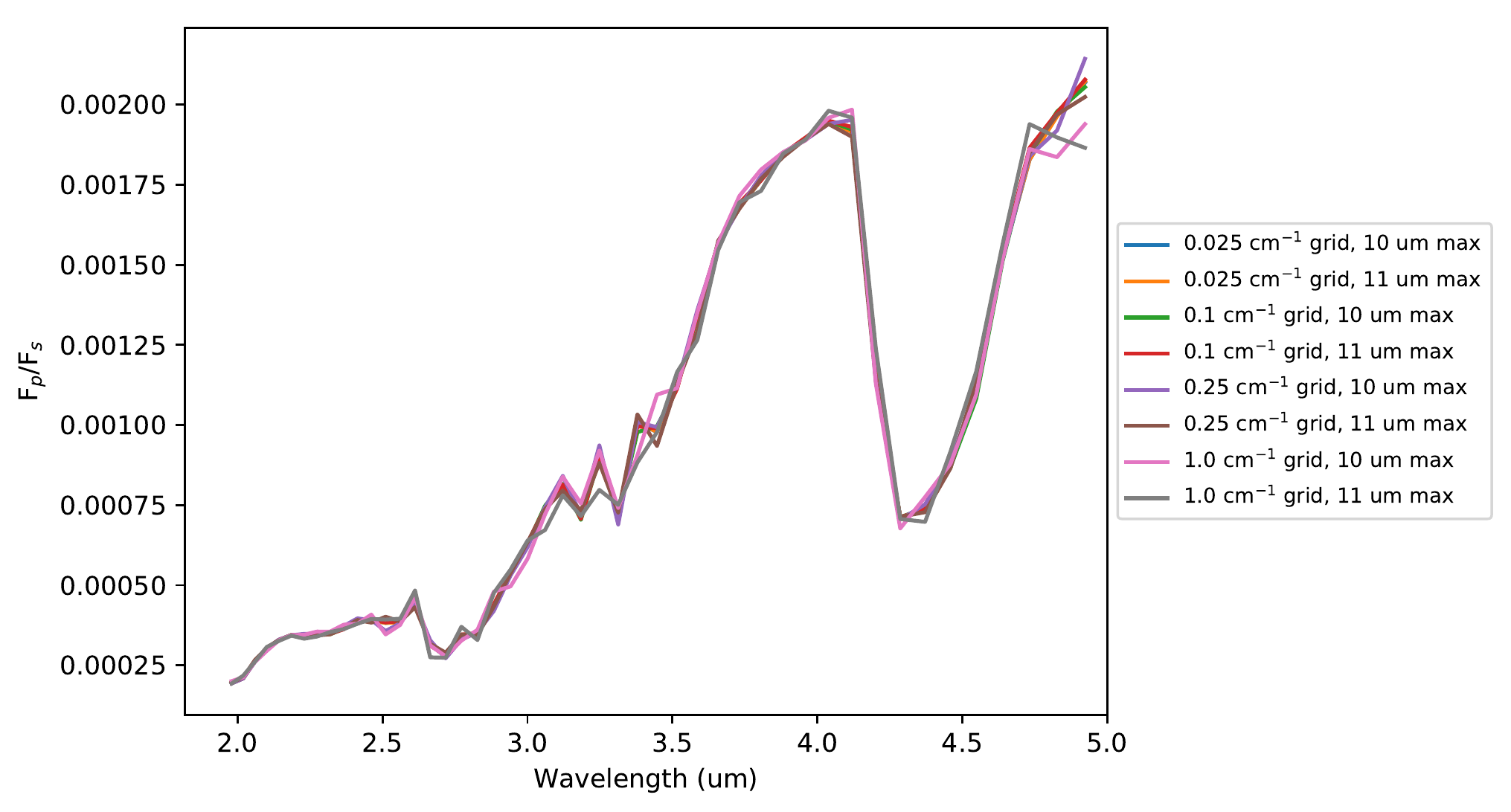}\hfill
\includegraphics[width=0.49\linewidth, clip=True]{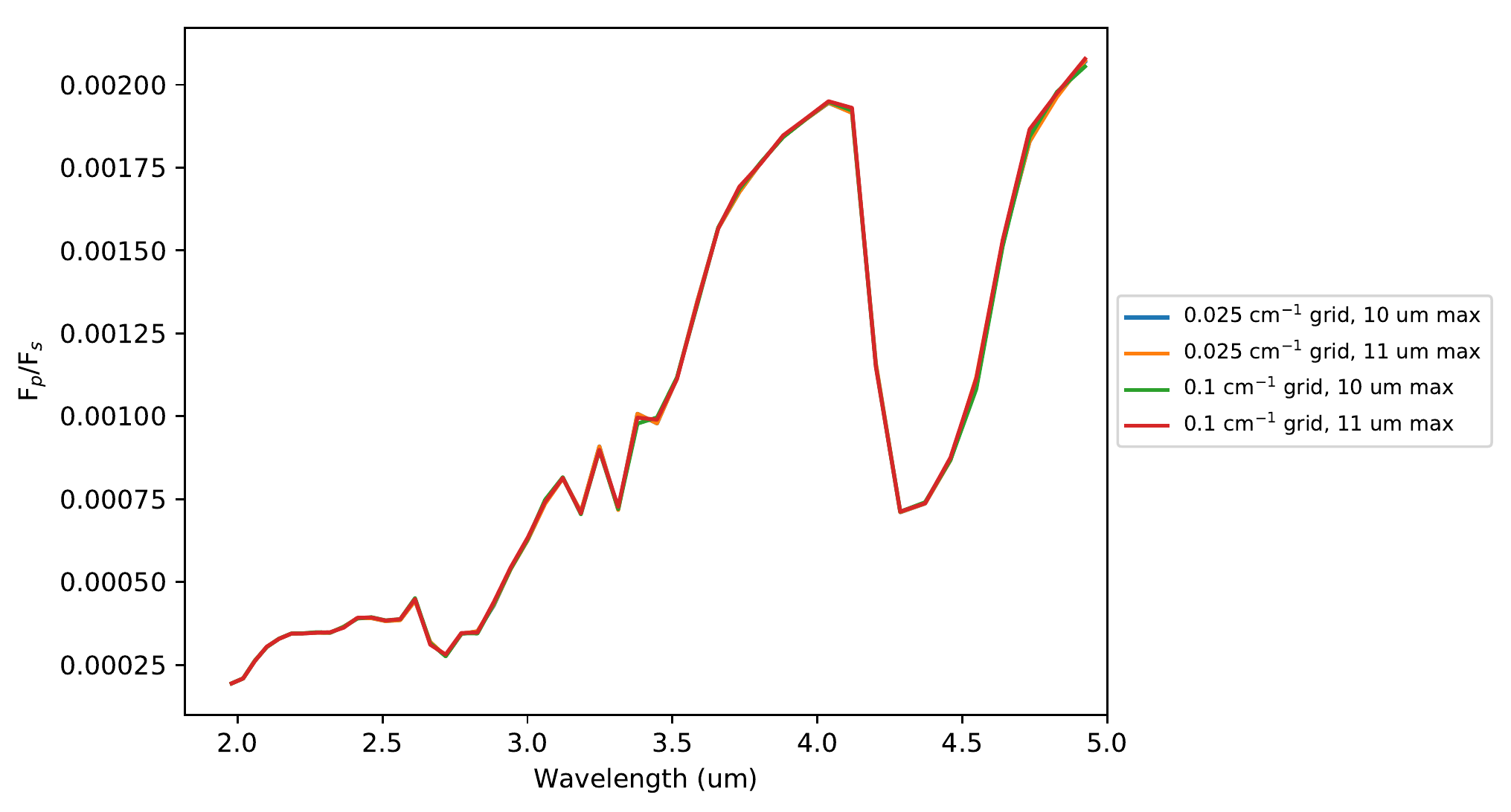}
\caption{Example of the effect of wavenumber griddings when band integrating spectra.
\textbf{Left:} 0.025, 0.1, 0.25, and 1.0 cm\sp{-1} griddings.
\textbf{Right:} 0.025 and 0.1 cm\sp{-1} griddings from the left plot.
All spectra are computed using identical setups except for the wavenumber gridding.
The gridding of the spectra computed with a maximum wavelength of 11 {\micron} is offset by \math{\sim}0.09 cm\sp{-1} compared to the spectra with a maximum wavelength of 10 {\micron}.
Note the model spread in the 2.5 -- 3.5 and \textgreater4.5 {\micron} regions of the left plot.
By comparison, the right plot shows little spread in the band-integrated spectra.
\label{fig:grid-effect-forward}}
\end{figure*}

\edit1{
When performing the synthetic retrievals described in Section \ref{sec:synthrets}, we observed a numerical effect that can bias the Bayesian sampler towards slightly incorrect answers when the forward and retrieval models use differing wavenumber grids.  
Here we describe this effect and how to minimize it.
We note that this error most strongly manifests when the RT code that produced the synthetic spectrum matches the RT code used when retrieving on the synthetic spectrum, and the error appears to be negligible when retrieving on real data at current resolutions.
}

\edit1{
Line-by-line calculations necessitate a discrete sampling of the resulting spectrum.  
Yet, spectra are continuous, and this discrete sampling will therefore introduce error when, e.g., band integrating a spectrum during a retrieval.
At commonly-used grid samplings (e.g., 1.0 cm\sp{-1}) for current- and next-generation telescope resolutions, these errors can drive the Bayesian sampler to an incorrect part of the phase space.
}

\edit1{
We consider forward models with 4 different grid spacings: 0.025, 0.1, 0.25, and 1.0 cm\sp{-1}.  
For each grid spacing, we also consider a horizontal shift of all values by \math{\sim}0.09 cm\sp{-1}.  
Figure \ref{fig:grid-effect-forward} shows that, for the 0.25 and 1.0 cm\sp{-1} griddings, there is disagreement in the 2.5 -- 3.5 and \textgreater4.5 {\micron} regions.  
By comparison, when only considering the 0.025 and 0.1 cm\sp{-1} griddings, the differences tend to be comparable to the width of the plotted lines.
}

\edit1{
We further investigate this effect by simulating a spectrum at 0.1 cm\sp{-1}, band integrating according to some filters, and retrieving with wavenumber grids that only differ in their spacing.  The retrieval models use grid spacings of 0.1, 0.25, 0.5, and 1.0 cm\sp{-1}.  We ensure that all wavenumbers in the retrieval model grids exactly overlap with wavenumbers from the forward model grid.  Figure \ref{fig:grid-effect-retrieval} shows that choosing coarser griddings than that of the forward model leads to a reduction in accuracy: the 0.1 cm\sp{-1} gridding recovers all of the parameters within 1\math{\sigma}, while the 0.25 and 0.5 cm\sp{-1} griddings recovering most parameters at \textgreater1\math{\sigma}.  The 1.0 cm\sp{-1} gridding entirely misses 4 out of the 5 parameters.
}

\begin{figure*}
\centering
{\centering \textbf{\large{0.1 cm\sp{-1} grid\hspace{6cm}0.25 cm\sp{-1} grid}}\par\medskip}
\includegraphics[width=0.47\linewidth, clip=True]{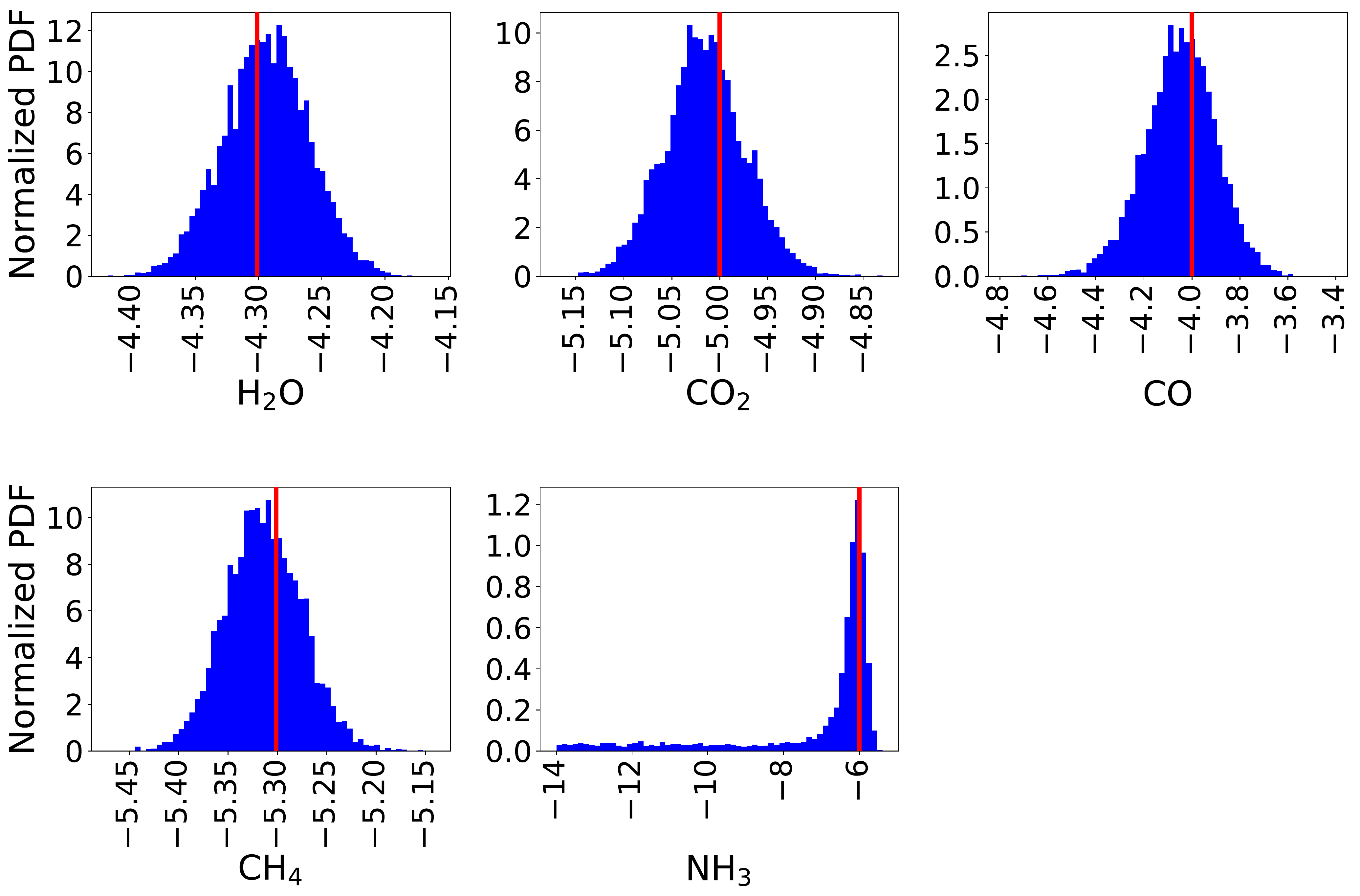}\hfill
\includegraphics[width=0.47\linewidth, clip=True]{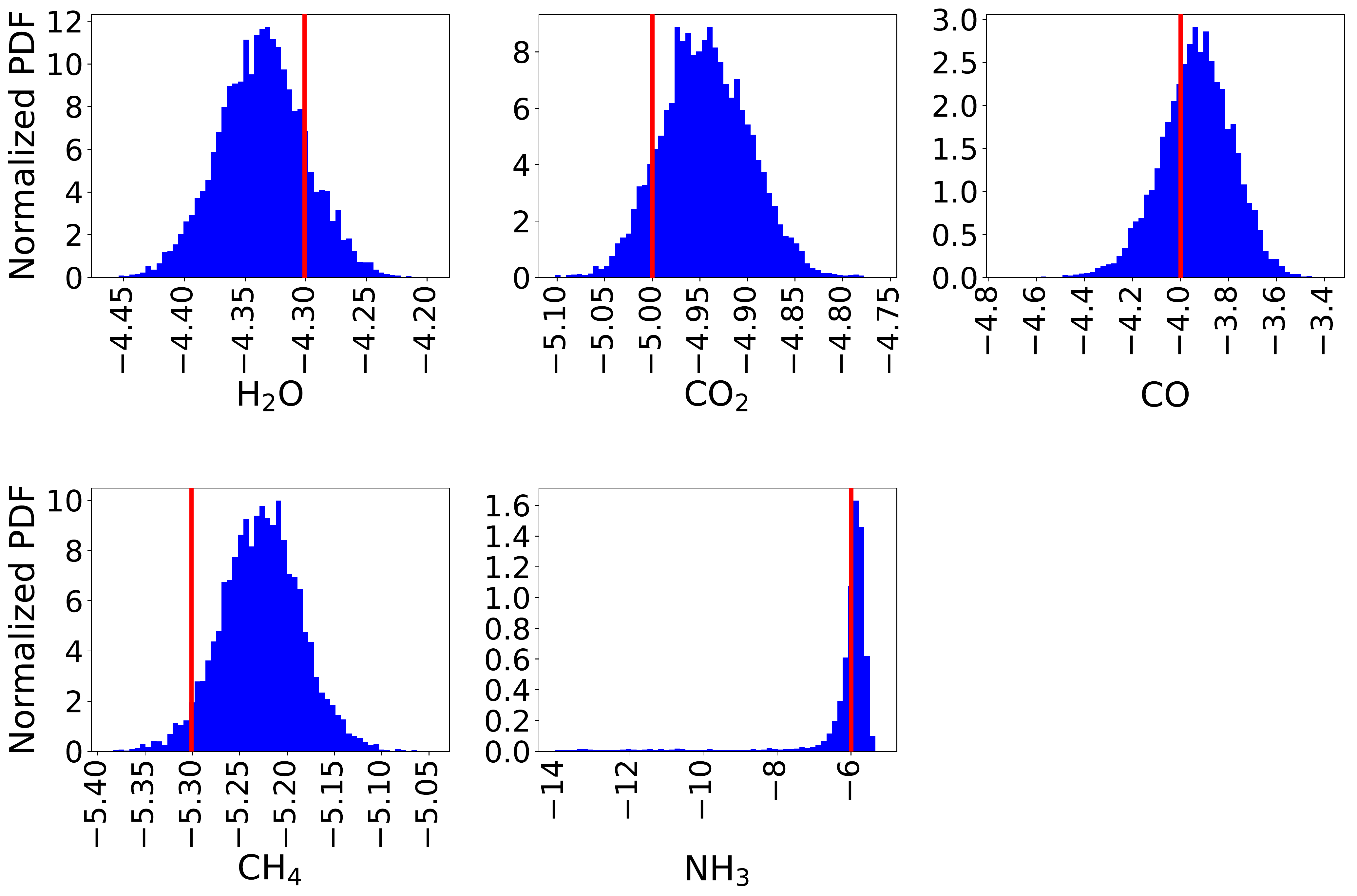}\\
\vspace{0.8cm}
{\centering \textbf{\large{0.5 cm\sp{-1} grid\hspace{6cm}1.0 cm\sp{-1} grid}}\par\medskip}

\includegraphics[width=0.47\linewidth, clip=True]{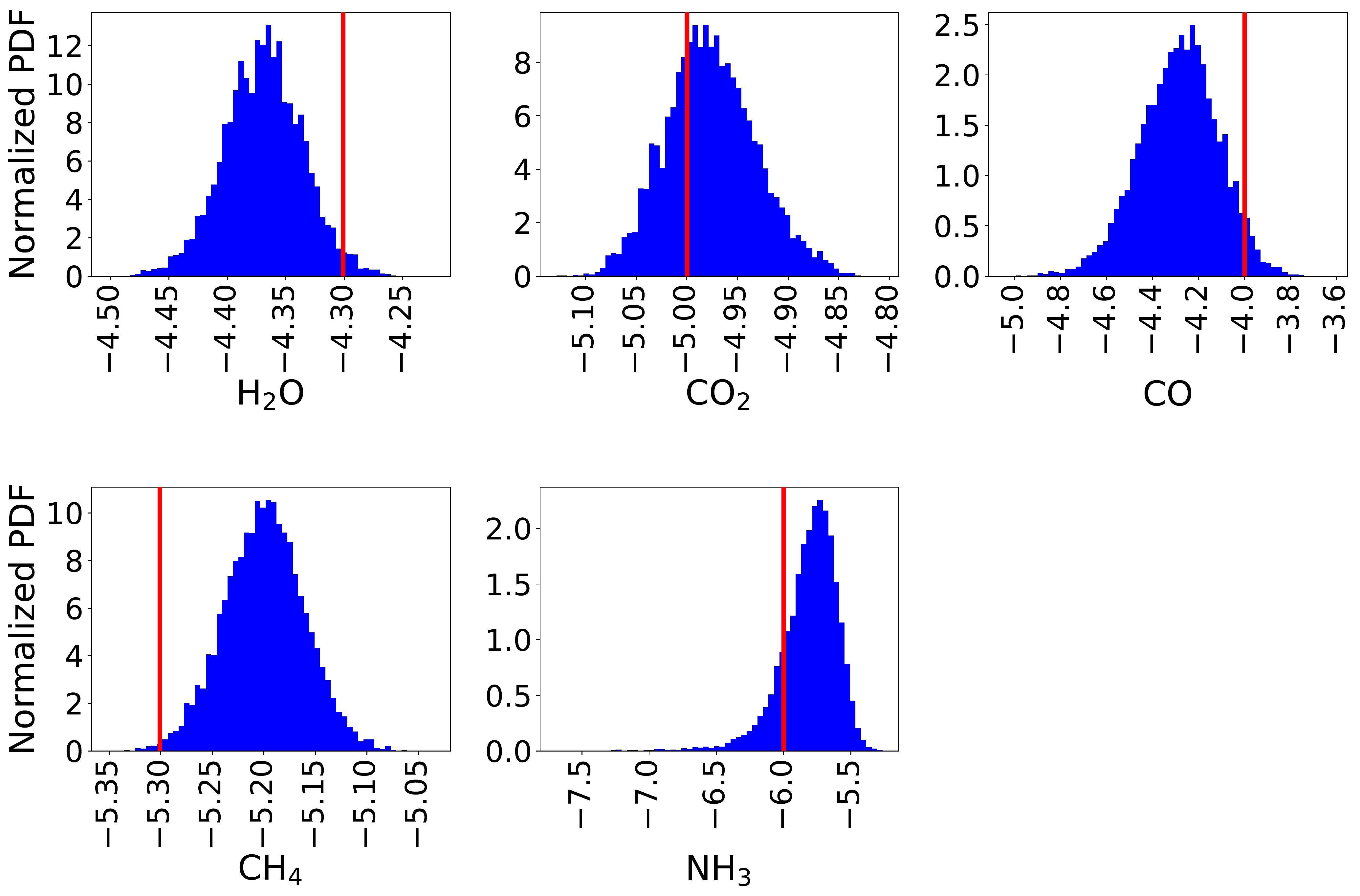}\hfill
\includegraphics[width=0.47\linewidth, clip=True]{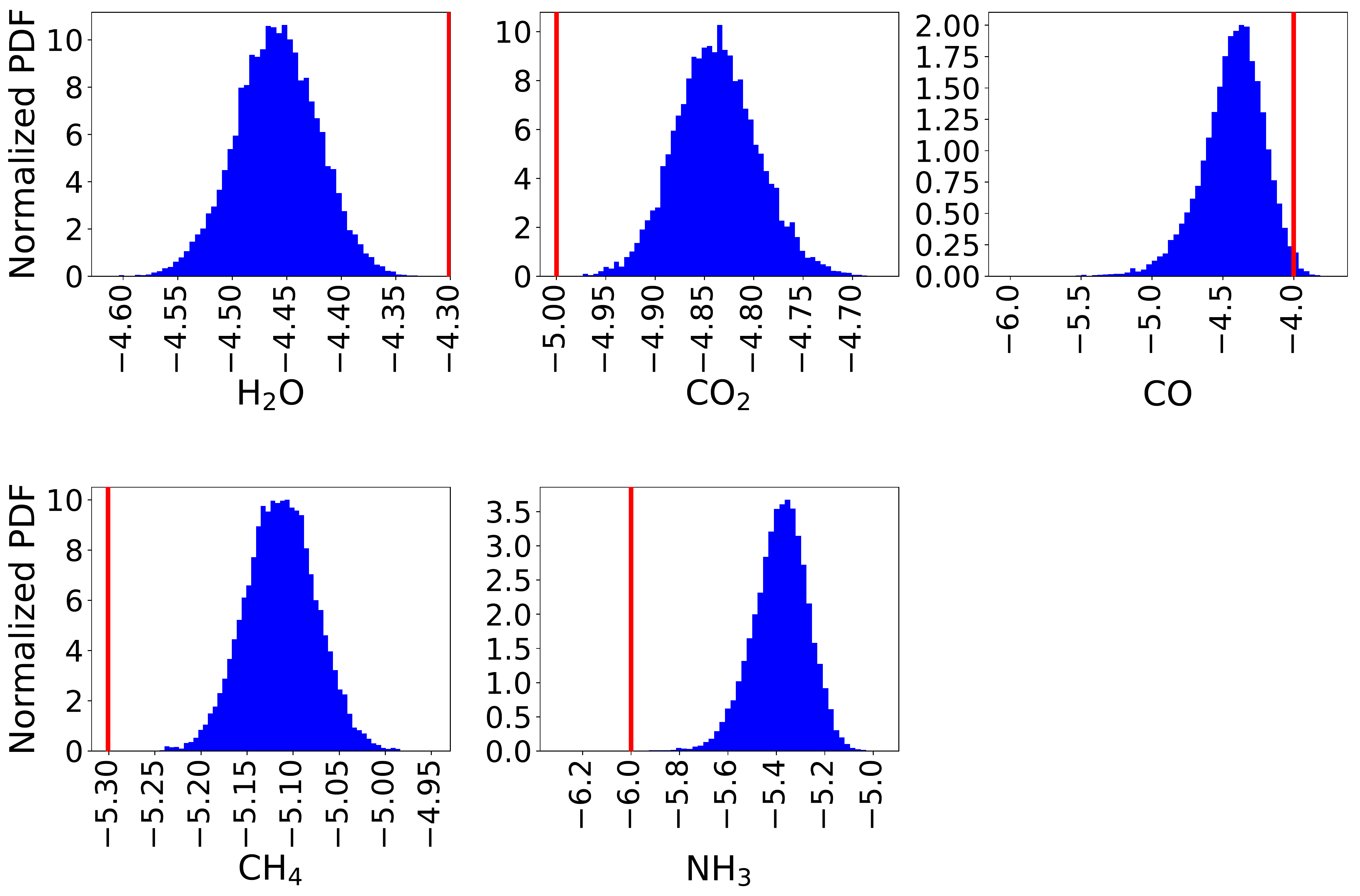}
\caption{Retrieved 1D marginalized posterior distributions for each molecule considered in the synthetic retrieval.  The forward model was computed using a 0.1 cm\sp{-1} grid, while the retrieval models were computed using 0.1 (\textbf{top left}), 0.25 (\textbf{top right}), 0.5 (\textbf{bottom left}), and 1.0 (\textbf{bottom right}) cm\sp{-1} grids.  Note that only the 0.1 cm\sp{-1} grid correctly retrieves the underlying parameters within 1\math{\sigma}; as the gridding becomes coarser, the accuracy decreases, with the 1.0 cm\sp{-1} gridding entirely missing 4 out of 5 of the parameters.
\label{fig:grid-effect-retrieval}}
\end{figure*}

\edit1{
This effect would be expected to be most pronounced in the case where the same RT code is used for the forward and retrieval models, since the retrieval RT model can, in theory, exactly match the forward RT model.  To test whether this effect arises when the forward and retrieval RT codes differ, we ran the \citet{BarstowEtal2020mnrasRetrievalComparison} cases at two different resolutions (0.1 and 1.0 cm\sp{-1}, Figure \ref{fig:grid-effect-diff-rt}).  The inferred temperatures and radii are generally insensitive to the grid selection, though the retrieved molecular abundances only slightly overlap, with the lower/upper credible region boundaries varying by around an order of magnitude.  On the other hand, when retrieving using the real data of HD 189733 b at the aforementioned grid resolutions, we find only minor differences (Figure \ref{fig:grid-effect-real-spec}).  The posterior of thermal profiles (calculated via the \math{\kappa}, \math{\gamma\sb{1}}, and \math{\beta} parameters) are nearly identical, and H\math{\sb{2}}O and CO\math{\sb{2}} are essentially unaffected.  CO and CH\math{\sb{4}} are minorly affected, with the coarser grid preferring slightly higher CO and lower CH\math{\sb{4}} abundances.  These minor differences are not significant enough to change the interpretation of the results in the case of current-resolution data for HD 189733 b.  However, the numerical effect described in this section may manifest in real-data retrievals as resolution improves.
}

\begin{figure*}
\centering
{\centering \textbf{\large{NEMESIS\hspace{6cm}CHIMERA}}\par\medskip}
\includegraphics[width=0.49\linewidth, trim={0 0 11.5cm 0}, clip=True]{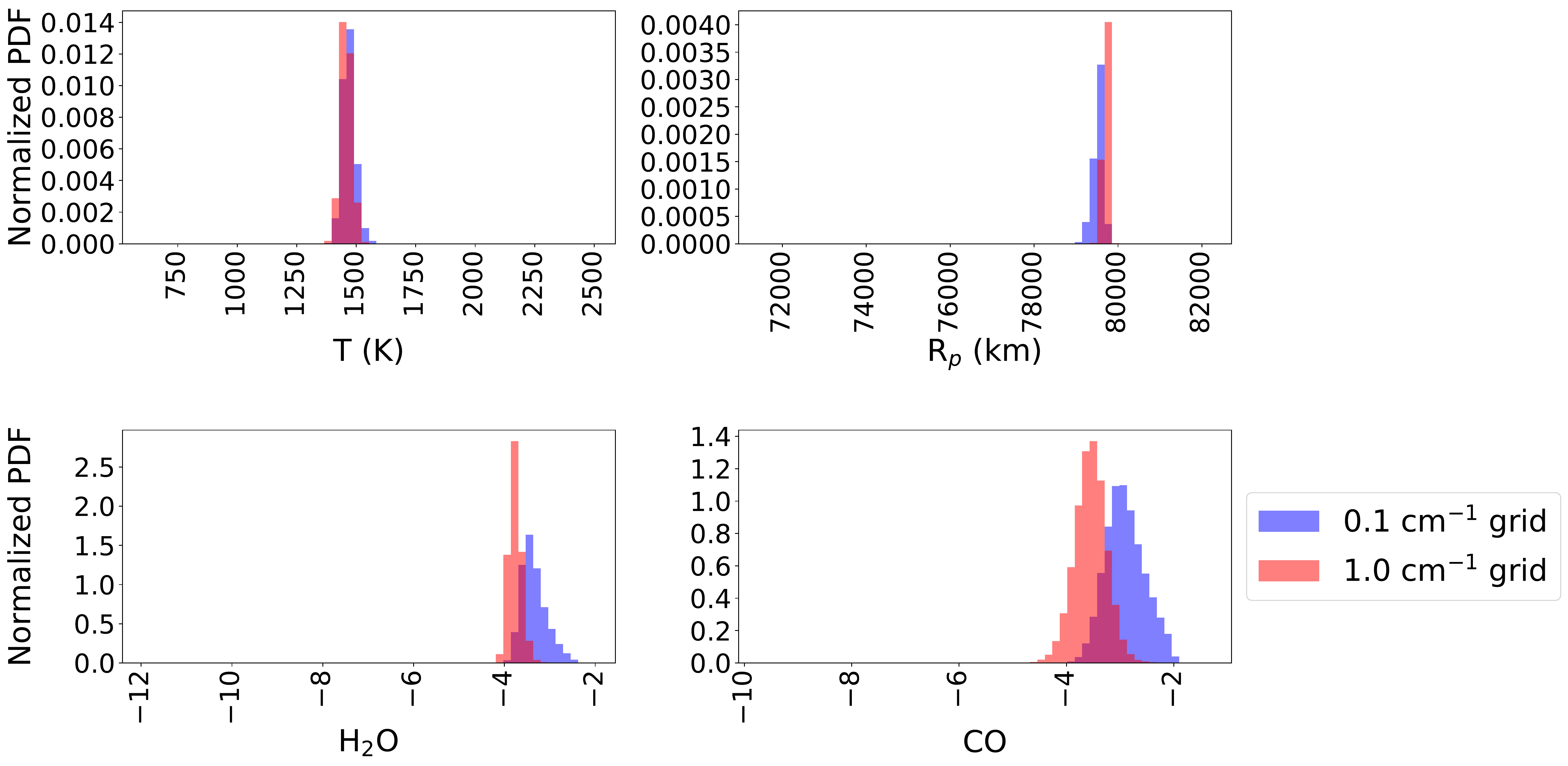}\hfill
\includegraphics[width=0.49\linewidth, trim={0 0 11.5cm 0}, clip=True]{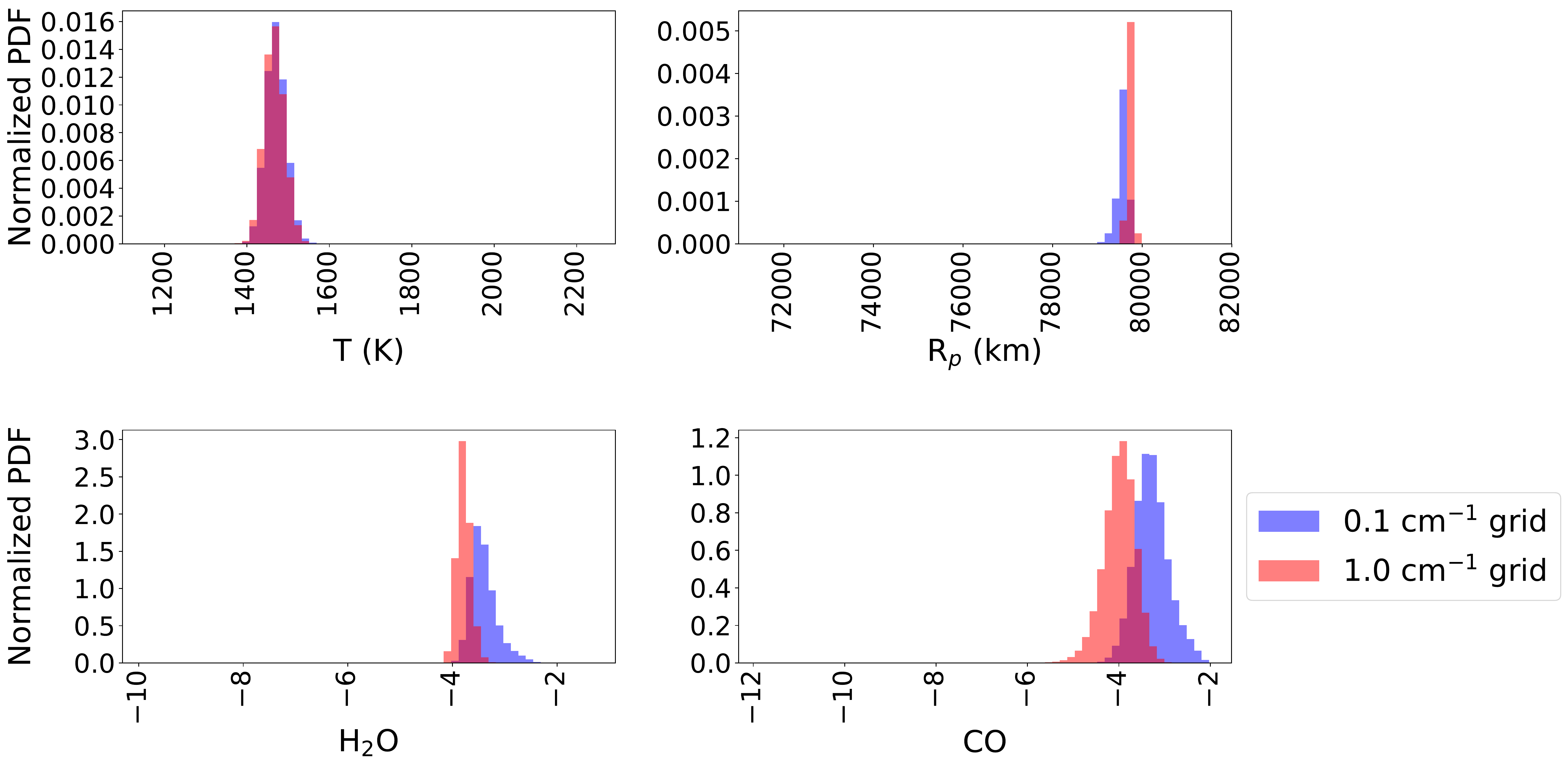}\\
{\centering \textbf{\large{{\TauREx}}}\par\medskip}
\includegraphics[width=0.62\linewidth, clip=True]{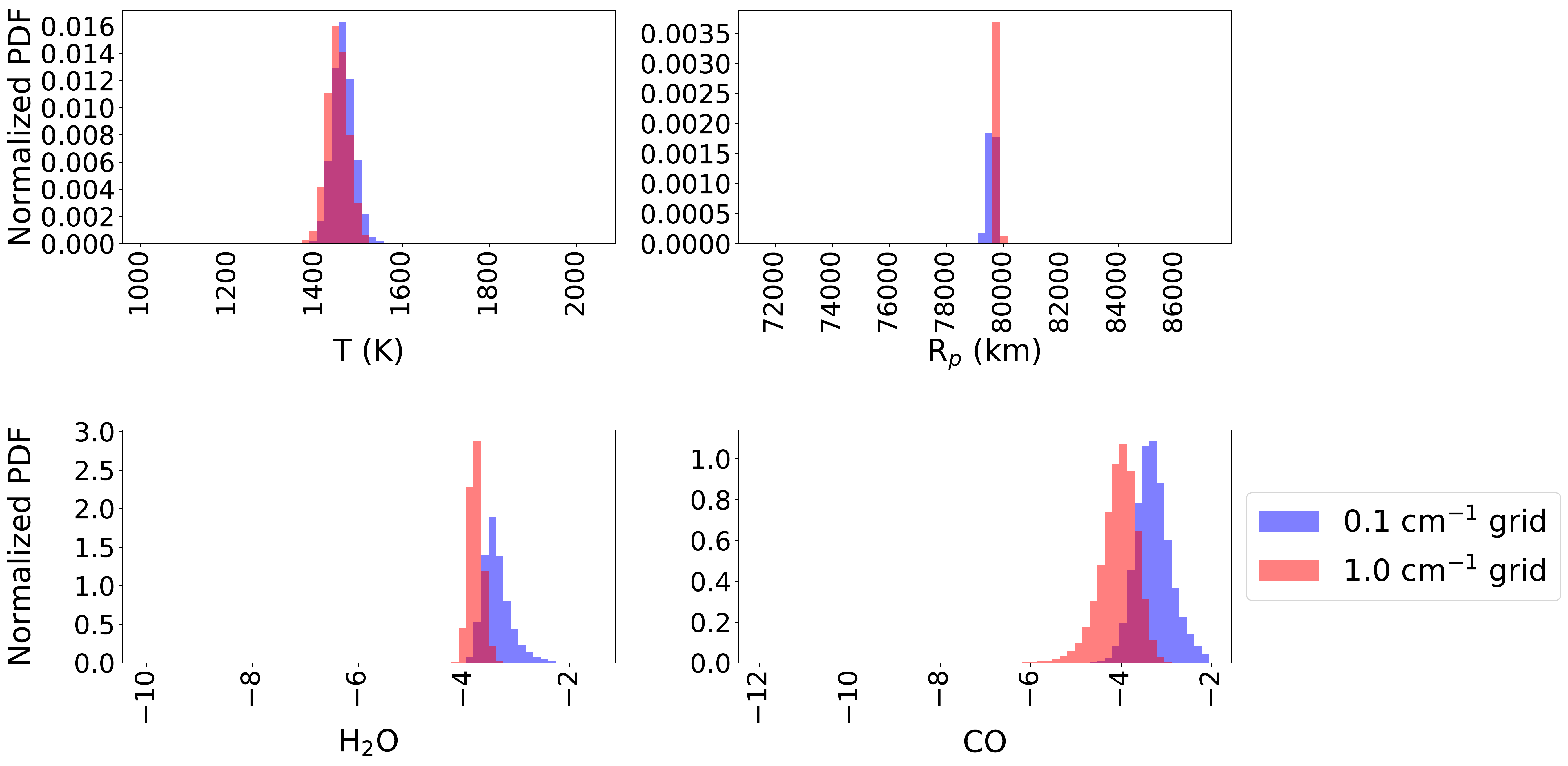}
\caption{Comparison of retrieved posteriors for the three indicated \citet{BarstowEtal2020mnrasRetrievalComparison} model 0 spectra at two different wavenumber grid resolutions.  While the retrieved temperatures and radii are generally unaffected, the molecular abundances are more sensitive to the grid resolution.  
\label{fig:grid-effect-diff-rt}}
\end{figure*}

\begin{figure}
\centering
\includegraphics[width=0.49\linewidth, clip=True]{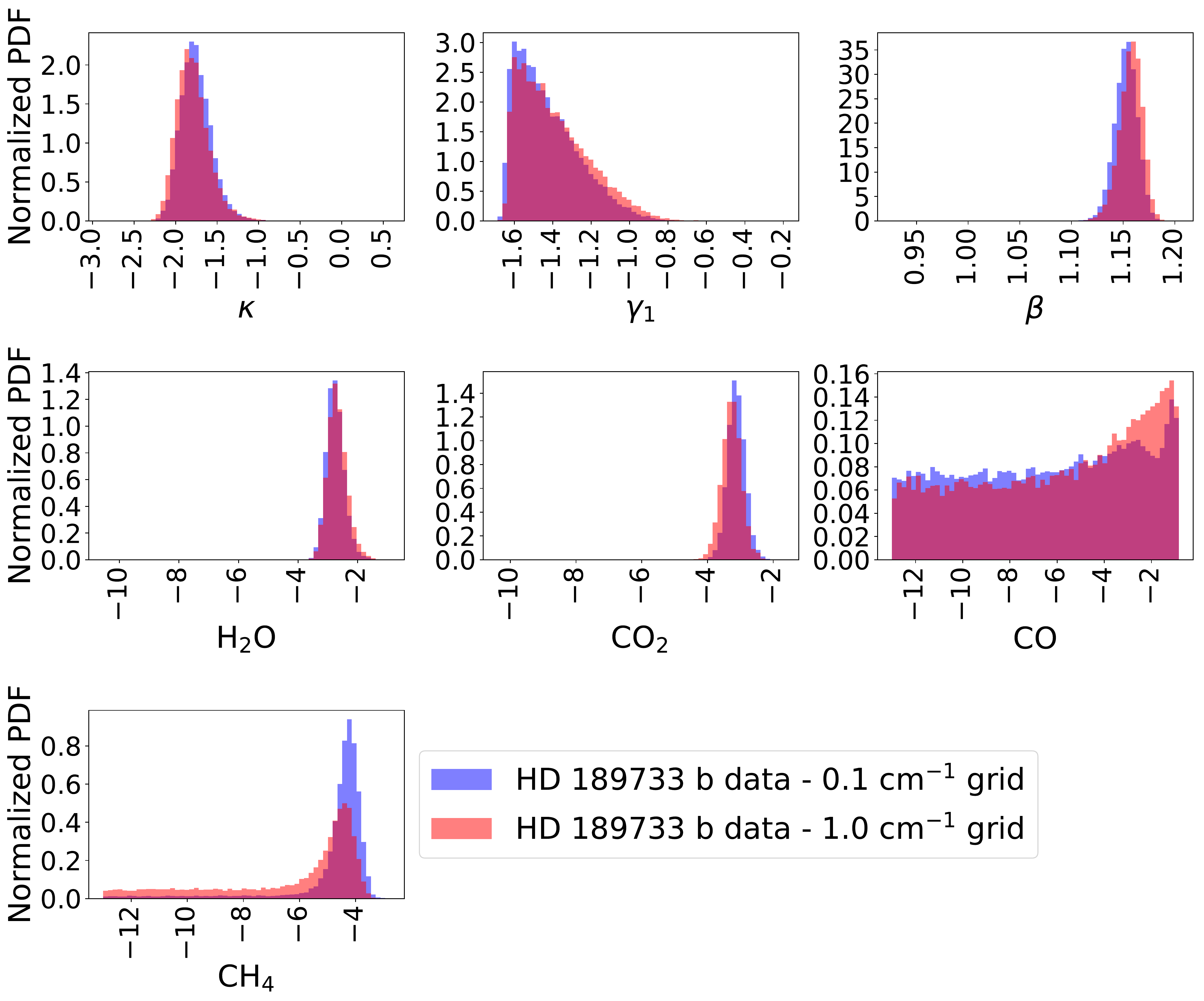}
\caption{Comparison of retrieved posteriors for HD 189733 b at two different wavenumber grid resolutions.  In general, the posteriors match; only CO and CH\math{\sb{4}} show minor differences, though they are not significant enough to affect conclusions drawn from the posterior.
\label{fig:grid-effect-real-spec}}
\end{figure}

\nojoe{\bibliography{BART1}}

\begin{thebibliography}{}
\expandafter\ifx\csname natexlab\endcsname\relax\def\natexlab#1{#1}\fi
\providecommand{\url}[1]{\href{#1}{#1}}
\providecommand{\dodoi}[1]{doi:~\href{http://doi.org/#1}{\nolinkurl{#1}}}
\providecommand{\doeprint}[1]{\href{http://ascl.net/#1}{\nolinkurl{http://ascl.net/#1}}}
\providecommand{\doarXiv}[1]{\href{https://arxiv.org/abs/#1}{\nolinkurl{https://arxiv.org/abs/#1}}}

\bibitem[{{AAS Journals Team} \&
  {Hendrickson}(2018)}]{AASteamHendrickson2018aastex62}
{AAS Journals Team}, \& {Hendrickson}, A. 2018, {AASJournals/AASTeX60}: Version
  6.2 Official Release, v6.2,  Zenodo, \dodoi{10.5281/zenodo.1209290}

\bibitem[{{Agol} {et~al.}(2010){Agol}, {Cowan}, {Knutson}, {Deming}, {Steffen},
  {Henry}, \& {Charbonneau}}]{AgolEtal2010apjHD189IRAC}
{Agol}, E., {Cowan}, N.~B., {Knutson}, H.~A., {et~al.} 2010, \apj, 721, 1861,
  \dodoi{10.1088/0004-637X/721/2/1861}

\bibitem[{{Al-Refaie} {et~al.}(2019){Al-Refaie}, {Changeat}, {Waldmann}, \&
  {Tinetti}}]{AlRefaieEtal2019apjTauREx3}
{Al-Refaie}, A.~F., {Changeat}, Q., {Waldmann}, I.~P., \& {Tinetti}, G. 2019,
  \apj, submitted.
\newblock \doarXiv{1912.07759}

\bibitem[{{Anderson} {et~al.}(1994){Anderson}, {Wang}, {Hoke}, {Kneizys},
  {Chetwynd}, {Rothman}, {Kimball}, {McClatchey}, {Shettle}, {Clough},
  {Gallery}, {Abreu}, \& {Selby}}]{AndersonEtal1994spieHistHITRAN}
{Anderson}, G.~P., {Wang}, J., {Hoke}, M.~L., {et~al.} 1994, in \procspie, Vol.
  2309, Passive Infrared Remote Sensing of Clouds and the Atmosphere II, ed.
  D.~K. {Lynch}, 170--183, \dodoi{10.1117/12.196674}

\bibitem[{Barba(2018)}]{Barba2018corrTermRR}
Barba, L.~A. 2018, CoRR, abs/1802.03311.
\newblock \doarXiv{1802.03311}

\bibitem[{{Barstow} {et~al.}(2014){Barstow}, {Aigrain}, {Irwin}, {Hackler},
  {Fletcher}, {Lee}, \& {Gibson}}]{BarstowEtal2014apjCloudsHD189}
{Barstow}, J.~K., {Aigrain}, S., {Irwin}, P.~G.~J., {et~al.} 2014, \apj, 786,
  154, \dodoi{10.1088/0004-637X/786/2/154}

\bibitem[{{Barstow} {et~al.}(2020){Barstow}, {Changeat}, {Garland}, {Line},
  {Rocchetto}, \& {Waldmann}}]{BarstowEtal2020mnrasRetrievalComparison}
{Barstow}, J.~K., {Changeat}, Q., {Garland}, R., {et~al.} 2020, \mnras, 493,
  4884, \dodoi{10.1093/mnras/staa548}

\bibitem[{{Becker}(2017)}]{Becker2017bookUnravelStarlight}
{Becker}, B.~J. 2017, {Unravelling Starlight} (Cambridge University Press)

\bibitem[{{Benneke} \& {Seager}(2012)}]{BennekeSeager2012apjRetrieval}
{Benneke}, B., \& {Seager}, S. 2012, \apj, 753, 100,
  \dodoi{10.1088/0004-637X/753/2/100}

\bibitem[{{B{\'e}tr{\'e}mieux} \&
  {Swain}(2018)}]{BetremieuxSwain2018SmallPlanetLimbRef}
{B{\'e}tr{\'e}mieux}, Y., \& {Swain}, M.~R. 2018, \apj, 865, 12,
  \dodoi{10.3847/1538-4357/aad80f}

\bibitem[{{\jhstud{Blecic}} {et~al.}(2016){\jhstud{Blecic}},
  {\jhauth{Harrington}}, \& {\jhstud{Bowman}}}]{BlecicEtal2016apjsTEA}
{\jhstud{Blecic}}, {\jhstud{J}}., {\jhauth{Harrington}}, {\jhauth{J}}., \&
  {\jhstud{Bowman}}, {\jhstud{M.~O}}. 2016, \apjs, 225, 4,
  \dodoi{10.3847/0067-0049/225/1/4}

\bibitem[{{\jhstud{Blecic}} {et~al.}(2021){\jhstud{Blecic}},
  {\jhauth{Harrington}}, {\jhstud{Cubillos}}, {Rojo}, {\jhstud{Lust}},
  {\jhstud{Challener}}, {\jhstud{Bowman}}, {\jhstud{Foster}}, {\jhstud{Stemm}},
  \& {Loredo}}]{BlecicEtal2021psjBART3}
{\jhstud{Blecic}}, {\jhstud{J}}., {\jhauth{Harrington}}, {\jhauth{J}}.,
  {\jhstud{Cubillos}}, {\jhstud{P}}., {et~al.} 2021, \psj, submitted

\bibitem[{Booth \& Sarkar(1998)}]{BoothSarkar1998AmstatMontecarlo}
Booth, J.~G., \& Sarkar, S. 1998, The American Statistician, 52, 354,
  \dodoi{10.1080/00031305.1998.10480596}

\bibitem[{{Bouchy} {et~al.}(2005){Bouchy}, {Udry}, {Mayor}, {Moutou}, {Pont},
  {Iribarne}, {da Silva}, {Ilovaisky}, {Queloz}, {Santos}, {S{\'e}gransan}, \&
  {Zucker}}]{BouchyEtal2005aapHD189733b}
{Bouchy}, F., {Udry}, S., {Mayor}, M., {et~al.} 2005, \aap, 444, L15,
  \dodoi{10.1051/0004-6361:200500201}

\bibitem[{Buchner(2016)}]{Buchner2016statcomputNestedSamplingTests}
Buchner, J. 2016, Statistics and Computing, 26, 383,
  \dodoi{10.1007/s11222-014-9512-y}

\bibitem[{{Castelli} \&
  {Kurucz}(2003)}]{CastelliKurucz2003iausATLAS9GridModelAtmospheres}
{Castelli}, F., \& {Kurucz}, R.~L. 2003, in IAU Symposium, Vol. 210, Modelling
  of Stellar Atmospheres, ed. N.~{Piskunov}, W.~W. {Weiss}, \& D.~F. {Gray},
  A20.
\newblock \doarXiv{astro-ph/0405087}

\bibitem[{{Charbonneau} {et~al.}(2008){Charbonneau}, {Knutson}, {Barman},
  {Allen}, {Mayor}, {Megeath}, {Queloz}, \&
  {Udry}}]{CharbonneauEtal2008apjHD189733b}
{Charbonneau}, D., {Knutson}, H.~A., {Barman}, T., {et~al.} 2008, \apj, 686,
  1341, \dodoi{10.1086/591635}

\bibitem[{{Charbonneau} {et~al.}(2005){Charbonneau}, {Allen}, {Megeath},
  {Torres}, {Alonso}, {Brown}, {Gilliland}, {Latham}, {Mandushev}, {O'Donovan},
  \& {Sozzetti}}]{CharbonneauEtal2005apjTrES-1}
{Charbonneau}, D., {Allen}, L.~E., {Megeath}, S.~T., {et~al.} 2005, \apj, 626,
  523, \dodoi{10.1086/429991}

\bibitem[{{Crouzet} {et~al.}(2012){Crouzet}, {McCullough}, {Burke}, \&
  {Long}}]{CrouzetEtal2012apjXO2bNICMOSystematics}
{Crouzet}, N., {McCullough}, P.~R., {Burke}, C., \& {Long}, D. 2012, \apj, 761,
  7, \dodoi{10.1088/0004-637X/761/1/7}

\bibitem[{{\jhstud{Cubillos}} {et~al.}(2021){\jhstud{Cubillos}},
  {\jhauth{Harrington}}, {\jhstud{Blecic}}, {Rojo}, {\jhstud{Lust}},
  {\jhstud{Challener}}, {\jhstud{Bowman}}, {\jhstud{Foster}}, {\jhstud{Stemm}},
  \& {Loredo}}]{CubillosEtal2021psjBART2}
{\jhstud{Cubillos}}, {\jhstud{P}}., {\jhauth{Harrington}}, {\jhauth{J}}.,
  {\jhstud{Blecic}}, {\jhstud{J}}., {et~al.} 2021, \psj, submitted

\bibitem[{{Cubillos}(2017)}]{Cubillos2017apjRepack}
{Cubillos}, P.~E. 2017, \apj, 850, 32, \dodoi{10.3847/1538-4357/aa9228}

\bibitem[{{\jhstud{Cubillos}} {et~al.}(2017){\jhstud{Cubillos}},
  {\jhauth{Harrington}}, {Loredo}, {\jhstud{Lust}}, {\jhstud{Blecic}}, \&
  {\jhstud{Stemm}}}]{CubillosEtal2017apjRednoise}
{\jhstud{Cubillos}}, {\jhstud{P.\ E}}., {\jhauth{Harrington}}, {\jhauth{J}}.,
  {Loredo}, T.~J., {et~al.} 2017, \aj, 153, 3,
  \dodoi{10.3847/1538-3881/153/1/3}

\bibitem[{{\jhstud{Cubillos}} {et~al.}(2014){\jhstud{Cubillos}},
  {\jhauth{Harrington}}, {Madhusudhan}, {\jhstud{Foster}}, {\jhstud{Lust}},
  {\jhstud{Hardy}}, \& {\jhstud{Bowman}}}]{CubillosEtal2014apjTrES1}
{\jhstud{Cubillos}}, {\jhstud{P.\ E}}., {\jhauth{Harrington}}, {\jhauth{J}}.,
  {Madhusudhan}, N., {et~al.} 2014, \apj, 797, 42,
  \dodoi{10.1088/0004-637X/797/1/42}

\bibitem[{{de Wit} {et~al.}(2012){de Wit}, {Gillon}, {Demory}, \&
  {Seager}}]{deWitEtal2012aapMappingHD189}
{de Wit}, J., {Gillon}, M., {Demory}, B.-O., \& {Seager}, S. 2012, \aap, 548,
  A128, \dodoi{10.1051/0004-6361/201219060}

\bibitem[{{Deming} {et~al.}(2006){Deming}, {Harrington}, {Seager}, \&
  {Richardson}}]{DemingEtal2006apjHD189733b}
{Deming}, D., {Harrington}, J., {Seager}, S., \& {Richardson}, L.~J. 2006,
  \apj, 644, 560, \dodoi{10.1086/503358}

\bibitem[{{Deming} {et~al.}(2005){Deming}, {Seager}, {Richardson}, \&
  {Harrington}}]{DemingEtal2005natHD209}
{Deming}, D., {Seager}, S., {Richardson}, L.~J., \& {Harrington}, J. 2005,
  \nat, 434, 740, \dodoi{10.1038/nature03507}

\bibitem[{{Deming} \& {Seager}(2017)}]{DemingSeager2017jgrpIllRealAtmExo}
{Deming}, L.~D., \& {Seager}, S. 2017, Journal of Geophysical Research
  (Planets), 122, 53, \dodoi{10.1002/2016JE005155}

\bibitem[{Donoho(2010)}]{Donoho2010biostatInviteRR}
Donoho, D.~L. 2010, Biostatistics, 11, 385,
  \dodoi{10.1093/biostatistics/kxq028}

\bibitem[{Flegal {et~al.}(2008)Flegal, Haran, \&
  Jones}]{FlegalEtal2008StatsciMarkov}
Flegal, J.~M., Haran, M., \& Jones, G.~L. 2008, Statistical Science, 23, 250,
  \dodoi{10.1214/08-STS257}

\bibitem[{{Fomel} \& {Claerbout}(2009)}]{FomelClarebout2009ciseRR}
{Fomel}, S., \& {Claerbout}, J.~F. 2009, Computing in Science \& Engineering,
  11, 5, \dodoi{10.1109/MCSE.2009.14}

\bibitem[{{Fraine} {et~al.}(2014){Fraine}, {Deming}, {Benneke}, {Knutson},
  {Jord{\'a}n}, {Espinoza}, {Madhusudhan}, {Wilkins}, \&
  {Todorov}}]{FraineEtal2014NatHATP11b}
{Fraine}, J., {Deming}, D., {Benneke}, B., {et~al.} 2014, \nat, 513, 526,
  \dodoi{10.1038/nature13785}

\bibitem[{{Gelman} \& {Rubin}(1992)}]{GelmanRubin1992}
{Gelman}, A., \& {Rubin}, D.~B. 1992, Statistical Science, 7, 457

\bibitem[{Gibson {et~al.}(2011)Gibson, Pont, \&
  Aigrain}]{GibsonEtal2011mnrasNICMOSystematics}
Gibson, N.~P., Pont, F., \& Aigrain, S. 2011, Monthly Notices of the Royal
  Astronomical Society, 411, 2199, \dodoi{10.1111/j.1365-2966.2010.17837.x}

\bibitem[{{Griffith}(2014)}]{Griffith2014rsptaExoDegenreateSolutions}
{Griffith}, C.~A. 2014, Philosophical Transactions of the Royal Society of
  London Series A, 372, 20130086, \dodoi{10.1098/rsta.2013.0086}

\bibitem[{{Grillmair} {et~al.}(2007){Grillmair}, {Charbonneau}, {Burrows},
  {Armus}, {Stauffer}, {Meadows}, {Van Cleve}, \&
  {Levine}}]{GrillmairEtal2007apjHD189IRS}
{Grillmair}, C.~J., {Charbonneau}, D., {Burrows}, A., {et~al.} 2007, \apjl,
  658, L115, \dodoi{10.1086/513741}

\bibitem[{{Grillmair} {et~al.}(2008){Grillmair}, {Burrows}, {Charbonneau},
  {Armus}, {Stauffer}, {Meadows}, {van Cleve}, {von Braun}, \&
  {Levine}}]{GrillmairEtal2008natHD189}
{Grillmair}, C.~J., {Burrows}, A., {Charbonneau}, D., {et~al.} 2008, \nat, 456,
  767, \dodoi{10.1038/nature07574}

\bibitem[{{Hargreaves} {et~al.}(2020){Hargreaves}, {Gordon}, {Rey}, {Nikitin},
  {Tyuterev}, {Kochanov}, \& {Rothman}}]{HargreavesEtal2020apjsMethaneHITEMP}
{Hargreaves}, R.~J., {Gordon}, I.~E., {Rey}, M., {et~al.} 2020, \apjs, 247, 55,
  \dodoi{10.3847/1538-4365/ab7a1a}

\bibitem[{{Heng} \&
  {Kitzmann}(2017)}]{HengKitzmann2017mnrasTransmissionSpectraTheory}
{Heng}, K., \& {Kitzmann}, D. 2017, \mnras, 470, 2972,
  \dodoi{10.1093/mnras/stx1453}

\bibitem[{{Hunter}(2007)}]{Hunter2007ieeeMatplotlib}
{Hunter}, J.~D. 2007, Computing in Science and Engineering, 9, 90,
  \dodoi{10.1109/MCSE.2007.55}

\bibitem[{{Irwin} {et~al.}(2008){Irwin}, {Teanby}, {de Kok}, {Fletcher},
  {Howett}, {Tsang}, {Wilson}, {Calcutt}, {Nixon}, \&
  {Parrish}}]{IrwinEtal2008jqsrtNEMESIS}
{Irwin}, P.~G.~J., {Teanby}, N.~A., {de Kok}, R., {et~al.} 2008, \jqsrt, 109,
  1136, \dodoi{10.1016/j.jqsrt.2007.11.006}

\bibitem[{Jones {et~al.}(2001)Jones, Oliphant, Peterson,
  {et~al.}}]{JonesEtal2001scipy}
Jones, E., Oliphant, T., Peterson, P., {et~al.} 2001, {SciPy}: {Open} Source
  Scientific Tools for {Python}.
\newblock \url{http://www.scipy.org/}

\bibitem[{Kirchhoff \& Bunsen(1860)}]{KirchhoffBunsen1860AdPchemanalspec}
Kirchhoff, G., \& Bunsen, R. 1860, Annalen der Physik, 186, 161,
  \dodoi{10.1002/andp.18601860602}

\bibitem[{{Knutson} {et~al.}(2009){Knutson}, {Charbonneau}, {Cowan}, {Fortney},
  {Showman}, {Agol}, {Henry}, {Everett}, \&
  {Allen}}]{KnutsonEtal2009apjHD189733b}
{Knutson}, H.~A., {Charbonneau}, D., {Cowan}, N.~B., {et~al.} 2009, \apj, 690,
  822, \dodoi{10.1088/0004-637X/690/1/822}

\bibitem[{{Knutson} {et~al.}(2012){Knutson}, {Lewis}, {Fortney}, {Burrows},
  {Showman}, {Cowan}, {Agol}, {Aigrain}, {Charbonneau}, {Deming}, {D{\'e}sert},
  {Henry}, {Langton}, \& {Laughlin}}]{KnutsonEtal2012apjHD189IRAC}
{Knutson}, H.~A., {Lewis}, N., {Fortney}, J.~J., {et~al.} 2012, \apj, 754, 22,
  \dodoi{10.1088/0004-637X/754/1/22}

\bibitem[{{Lee} {et~al.}(2012){Lee}, {Fletcher}, \&
  {Irwin}}]{LeeEtal2012MNRASHD189733b}
{Lee}, J.-M., {Fletcher}, L.~N., \& {Irwin}, P.~G.~J. 2012, \mnras, 420, 170,
  \dodoi{10.1111/j.1365-2966.2011.20013.x}

\bibitem[{{Line} {et~al.}(2014){Line}, {Knutson}, {Wolf}, \&
  {Yung}}]{LineEtal2014apjRetrievalCO}
{Line}, M.~R., {Knutson}, H., {Wolf}, A.~S., \& {Yung}, Y.~L. 2014, \apj, 783,
  70, \dodoi{10.1088/0004-637X/783/2/70}

\bibitem[{{Line} {et~al.}(2012){Line}, {Zhang}, {Vasisht}, {Natraj}, {Chen}, \&
  {Yung}}]{LineEtal2012ApJHD189733b}
{Line}, M.~R., {Zhang}, X., {Vasisht}, G., {et~al.} 2012, \apj, 749, 93,
  \dodoi{10.1088/0004-637X/749/1/93}

\bibitem[{{Line} {et~al.}(2013){Line}, {Wolf}, {Zhang}, {Knutson}, {Kammer},
  {Ellison}, {Deroo}, {Crisp}, \& {Yung}}]{LineEtal2013apjRetrieval1}
{Line}, M.~R., {Wolf}, A.~S., {Zhang}, X., {et~al.} 2013, \apj, 775, 137,
  \dodoi{10.1088/0004-637X/775/2/137}

\bibitem[{{Madhusudhan}(2018)}]{Madhusudhan2018bookAtmRetrExo}
{Madhusudhan}, N. 2018, {Atmospheric Retrieval of Exoplanets} (Springer
  International Publishing AG), 104, \dodoi{10.1007/978-3-319-55333-7_104}

\bibitem[{{Madhusudhan} \&
  {Seager}(2009)}]{MadhusudhanSeager2009apjAtmRetrMeth}
{Madhusudhan}, N., \& {Seager}, S. 2009, \apj, 707, 24,
  \dodoi{10.1088/0004-637X/707/1/24}

\bibitem[{{Madhusudhan} \&
  {Seager}(2010)}]{MadhusudhanSeager2010apjThermalHotJup}
---. 2010, \apj, 725, 261, \dodoi{10.1088/0004-637X/725/1/261}

\bibitem[{{Madhusudhan} {et~al.}(2011){Madhusudhan}, {\jhauth{Harrington}},
  {\jhstud{Stevenson}}, {\jhstud{Nymeyer}}, {\jhstud{Campo}}, {Wheatley},
  {Deming}, {\jhstud{Blecic}}, {\jhstud{Hardy}}, {\jhstud{Lust}}, {Anderson},
  {Collier Cameron}, {\jhstud{Britt}}, {\jhstud{Bowman}}, {Hebb}, {Hellier},
  {Maxted}, {Pollacco}, \& {West}}]{MadhusudhanEtal2011natWASP12batm}
{Madhusudhan}, N., {\jhauth{Harrington}}, {\jhauth{J}}., {\jhstud{Stevenson}},
  {\jhstud{K.\ B}}., {et~al.} 2011, Nature, 469, 64,
  \dodoi{10.1038/nature09602}

\bibitem[{{Majeau} {et~al.}(2012){Majeau}, {Agol}, \&
  {Cowan}}]{MajeauEtal2012apjl2DMapHD189}
{Majeau}, C., {Agol}, E., \& {Cowan}, N.~B. 2012, \apjl, 747, L20,
  \dodoi{10.1088/2041-8205/747/2/L20}

\bibitem[{{Malik} {et~al.}(2017){Malik}, {Grosheintz}, {Mendon{\c c}a},
  {Grimm}, {Lavie}, {Kitzmann}, {Tsai}, {Burrows}, {Kreidberg}, {Bedell},
  {Bean}, {Stevenson}, \& {Heng}}]{MalikEtal2017ajHELIOS}
{Malik}, M., {Grosheintz}, L., {Mendon{\c c}a}, J.~M., {et~al.} 2017, \aj, 153,
  56, \dodoi{10.3847/1538-3881/153/2/56}

\bibitem[{{Malik} {et~al.}(2018){Malik}, {Grosheintz}, {Mendon{\c c}a},
  {Grimm}, {Lavie}, {Kitzmann}, {Tsai}, {Burrows}, {Kreidberg}, {Bedell},
  {Bean}, {Stevenson}, \& {Heng}}]{MalikEtal2018asclHELIOS}
---. 2018, {HELIOS: Radiative} transfer code for exoplanetary atmospheres,
  Astrophysics Source Code Library.
\newblock \doeprint{1807.009}

\bibitem[{{M{\'a}rquez-Neila} {et~al.}(2018){M{\'a}rquez-Neila}, {Fisher},
  {Sznitman}, \& {Heng}}]{MarquezNeilaEtal2018natHELA}
{M{\'a}rquez-Neila}, P., {Fisher}, C., {Sznitman}, R., \& {Heng}, K. 2018,
  Nature Astronomy, 2, 719, \dodoi{10.1038/s41550-018-0504-2}

\bibitem[{Martins \&
  Coelho(2007)}]{MartinsCoelho2007mnrasStellarLibraryComparisons}
Martins, L.~P., \& Coelho, P. 2007, Monthly Notices of the Royal Astronomical
  Society, 381, 1329, \dodoi{10.1111/j.1365-2966.2007.11954.x}

\bibitem[{Meurer {et~al.}(2017)Meurer, Smith, Paprocki, \v{C}ert\'{i}k,
  Kirpichev, Rocklin, Kumar, Ivanov, Moore, Singh, Rathnayake, Vig, Granger,
  Muller, Bonazzi, Gupta, Vats, Johansson, Pedregosa, Curry, Terrel,
  Rou\v{c}ka, Saboo, Fernando, Kulal, Cimrman, \&
  Scopatz}]{MeurerEtal2017pjcsSYMPY}
Meurer, A., Smith, C.~P., Paprocki, M., {et~al.} 2017, PeerJ Computer Science,
  3, e103, \dodoi{10.7717/peerj-cs.103}

\bibitem[{{Moses} {et~al.}(2013){Moses}, {Madhusudhan}, {Visscher}, \&
  {Freedman}}]{MosesEtl2013ApJC/ORatios}
{Moses}, J.~I., {Madhusudhan}, N., {Visscher}, C., \& {Freedman}, R.~S. 2013,
  \apj, 763, 25, \dodoi{10.1088/0004-637X/763/1/25}

\bibitem[{{National Academies of Sciences, Engineering, and
  Medicine}(2018)}]{NASEM2018NAPNASAOpenSource}
{National Academies of Sciences, Engineering, and Medicine}. 2018, Open Source
  Software Policy Options for {NASA Earth} and {Space Sciences} (Washington,
  DC: The National Academies Press), \dodoi{10.17226/25217}

\bibitem[{Oliphant(2006)}]{Oliphant2006bookNumpy}
Oliphant, T.~E. 2006, A Guide to {NumPy} (Trelgol Publishing USA)

\bibitem[{{Oreshenko} {et~al.}(2017){Oreshenko}, {Lavie}, {Grimm}, {Tsai},
  {Malik}, {Demory}, {Mordasini}, {Alibert}, {Benz}, {Quanz}, {Trotta}, \&
  {Heng}}]{OreshenkoEtal2017apjlWASP12b}
{Oreshenko}, M., {Lavie}, B., {Grimm}, S.~L., {et~al.} 2017, \apjl, 847, L3,
  \dodoi{10.3847/2041-8213/aa8acf}

\bibitem[{{Pierluissi}(1977)}]{PierluissiEtal1977jqsrtVoigt}
{Pierluissi}, J. 1977, \jqsrt, 18, 555, \dodoi{10.1016/0022-4073(77)90056-5}

\bibitem[{Polyansky {et~al.}(2018)Polyansky, Kyuberis, Zobov, Tennyson,
  Yurchenko, \& Lodi}]{PolyanskyEtal2018mnrasExoMolH2O}
Polyansky, O.~L., Kyuberis, A.~A., Zobov, N.~F., {et~al.} 2018, Monthly Notices
  of the Royal Astronomical Society, 480, 2597, \dodoi{10.1093/mnras/sty1877}

\bibitem[{Pr{\v{s}}a {et~al.}(2016)Pr{\v{s}}a, Harmanec, Torres, Mamajek,
  Asplund, Capitaine, Christensen-Dalsgaard, Depagne, Haberreiter, Hekker,
  Hilton, Kopp, Kostov, Kurtz, Laskar, Mason, Milone, Montgomery, Richards,
  Schmutz, Schou, \& Stewart}]{IAU2016apjNominalValues}
Pr{\v{s}}a, A., Harmanec, P., Torres, G., {et~al.} 2016, The Astronomical
  Journal, 152, 41, \dodoi{10.3847/0004-6256/152/2/41}

\bibitem[{{Richard} {et~al.}(2012){Richard}, {Gordon}, {Rothman}, {Abel},
  {Frommhold}, {Gustafsson}, {Hartmann}, {Hermans}, {Lafferty}, {Orton},
  {Smith}, \& {Tran}}]{RichardEtal2012jqsrtHITRANcia}
{Richard}, C., {Gordon}, I.~E., {Rothman}, L.~S., {et~al.} 2012, \jqsrt, 113,
  1276, \dodoi{10.1016/j.jqsrt.2011.11.004}

\bibitem[{{Rojo}(2006)}]{Rojo2006PhD}
{Rojo}, P.~M. 2006, PhD thesis, Cornell University

\bibitem[{Rothman {et~al.}(2009)Rothman, Gordon, Barbe, Benner, Bernath, Birk,
  Boudon, Brown, Campargue, Champion, Chance, Coudert, Dana, Devi, Fally,
  Flaud, Gamache, Goldman, Jacquemart, Kleiner, Lacome, Lafferty, Mandin,
  Massie, Mikhailenko, Miller, Moazzen-Ahmadi, Naumenko, Nikitin, Orphal,
  Perevalov, Perrin, Predoi-Cross, Rinsland, Rotger, Šimečková, Smith, Sung,
  Tashkun, Tennyson, Toth, Vandaele, \& {Vander
  Auwera}}]{RothmanEtal2009jqsrtHITRAN2008}
Rothman, L., Gordon, I., Barbe, A., {et~al.} 2009, Journal of Quantitative
  Spectroscopy and Radiative Transfer, 110, 533,
  \dodoi{https://doi.org/10.1016/j.jqsrt.2009.02.013}

\bibitem[{Rothman {et~al.}(2010)Rothman, Gordon, Barber, Dothe, Gamache,
  Goldman, Perevalov, Tashkun, \& Tennyson}]{RothmanEtal2010jqsrtHITEMP}
Rothman, L.~S., Gordon, I.~E., Barber, R.~J., {et~al.} 2010, \jqsrt, 111, 2139

\bibitem[{Rothman {et~al.}(2013)Rothman, Gordon, Babikov, Barbe, Benner,
  Bernath, Birk, Bizzocchi, Boudon, Brown,
  {et~al.}}]{Rothman2013jqsrtHITRAN2012}
Rothman, L.~S., Gordon, I.~E., Babikov, Y., {et~al.} 2013, \jqsrt, 130, 4

\bibitem[{{Sharp} \& {Burrows}(2007)}]{SharpBurrows2007apjsAMOOpac}
{Sharp}, C.~M., \& {Burrows}, A. 2007, \apjs, 168, 140, \dodoi{10.1086/508708}

\bibitem[{{Southworth}(2010)}]{Southworth2010mnrasPLanetParamsIII}
{Southworth}, J. 2010, \mnras, 408, 1689,
  \dodoi{10.1111/j.1365-2966.2010.17231.x}

\bibitem[{{Stassun} {et~al.}(2018){Stassun}, {Corsaro}, {Pepper}, \&
  {Gaudi}}]{StassunEtal2018AJHD189temp}
{Stassun}, K.~G., {Corsaro}, E., {Pepper}, J.~A., \& {Gaudi}, B.~S. 2018, \aj,
  155, 22, \dodoi{10.3847/1538-3881/aa998a}

\bibitem[{{Stodden}(2009)}]{Stodden2009CSELegalRR}
{Stodden}, V. 2009, Computing in Science and Engineering, 11, 35,
  \dodoi{10.1109/MCSE.2009.19}

\bibitem[{{Swain} {et~al.}(2009){Swain}, {Vasisht}, {Tinetti}, {Bouwman},
  {Chen}, {Yung}, {Deming}, \& {Deroo}}]{SwainEtal2009apjHD189733b}
{Swain}, M.~R., {Vasisht}, G., {Tinetti}, G., {et~al.} 2009, \apjl, 690, L114,
  \dodoi{10.1088/0004-637X/690/2/L114}

\bibitem[{{ter Braak}(2006)}]{Braak2006DifferentialEvolution}
{ter Braak}, C. 2006, Statistics and Computing, 16, 239.
\newblock \url{http://dx.doi.org/10.1007/s11222-006-8769-1}

\bibitem[{{ter Braak} \& {Vrugt}(2008)}]{Braak2008SnookerDEMC}
{ter Braak}, C. J.~F., \& {Vrugt}, J.~A. 2008, Statistics and Computing, 18,
  435, \dodoi{10.1007/s11222-008-9104-9}

\bibitem[{{Todorov} {et~al.}(2014){Todorov}, {Deming}, {Burrows}, \&
  {Grillmair}}]{TodorovEtal2014apjHD189IRS}
{Todorov}, K.~O., {Deming}, D., {Burrows}, A., \& {Grillmair}, C.~J. 2014,
  \apj, 796, 100, \dodoi{10.1088/0004-637X/796/2/100}

\bibitem[{{Torres} {et~al.}(2008){Torres}, {Winn}, \&
  {Holman}}]{TorresEtal2008apjHD189733b}
{Torres}, G., {Winn}, J.~N., \& {Holman}, M.~J. 2008, \apj, 677, 1324,
  \dodoi{10.1086/529429}

\bibitem[{{Triaud} {et~al.}(2009){Triaud}, {Queloz}, {Bouchy}, {Moutou},
  {Collier Cameron}, {Claret}, {Barge}, {Benz}, {Deleuil}, {Guillot},
  {H{\'e}brard}, {Lecavelier Des {\'E}tangs}, {Lovis}, {Mayor}, {Pepe}, \&
  {Udry}}]{TriaudEtal2009A&AHD189params}
{Triaud}, A.~H.~M.~J., {Queloz}, D., {Bouchy}, F., {et~al.} 2009, \aap, 506,
  377, \dodoi{10.1051/0004-6361/200911897}

\bibitem[{{van der Walt} {et~al.}(2011){van der Walt}, {Colbert}, \&
  {Varoquaux}}]{vanderWaltEtal2011numpy}
{van der Walt}, S., {Colbert}, S.~C., \& {Varoquaux}, G. 2011, Computing in
  Science and Engineering, 13, 22, \dodoi{10.1109/MCSE.2011.37}

\bibitem[{{Vats} \& {Knudson}(2018)}]{VatsKnudson2018arxivGelmanRubin}
{Vats}, D., \& {Knudson}, C. 2018, arXiv e-prints, arXiv:1812.09384.
\newblock \doarXiv{1812.09384}

\bibitem[{{Vehtari} {et~al.}(2019){Vehtari}, {Gelman}, {Simpson}, {Carpenter},
  \& {B{\"u}rkner}}]{VehtariEtal2019arxivRhatESS}
{Vehtari}, A., {Gelman}, A., {Simpson}, D., {Carpenter}, B., \& {B{\"u}rkner},
  P.-C. 2019, arXiv e-prints, arXiv:1903.08008.
\newblock \doarXiv{1903.08008}

\bibitem[{{Virtanen} {et~al.}(2020){Virtanen}, {Gommers}, {Oliphant},
  {Haberland}, {Reddy}, {Cournapeau}, {Burovski}, {Peterson}, {Weckesser},
  {Bright}, {van der Walt}, {Brett}, {Wilson}, {Millman}, {Mayorov}, {Nelson},
  {Jones}, {Kern}, {Larson}, {Carey}, {Polat}, {Feng}, {Moore}, {VanderPlas},
  {Laxalde}, {Perktold}, {Cimrman}, {Henriksen}, {Quintero}, {Harris},
  {Archibald}, {Ribeiro}, {Pedregosa}, {van Mulbregt}, \& {SciPy 1.0
  Contributors}}]{VirtanenEtal2020NatMethSciPy1}
{Virtanen}, P., {Gommers}, R., {Oliphant}, T.~E., {et~al.} 2020, Nature
  Methods, 17, 261, \dodoi{https://doi.org/10.1038/s41592-019-0686-2}

\bibitem[{{Waldmann}(2016)}]{Waldmann2016apjDreamingAtmospheres}
{Waldmann}, I.~P. 2016, \apj, 820, 107, \dodoi{10.3847/0004-637X/820/2/107}

\bibitem[{{Waldmann} {et~al.}(2015{\natexlab{a}}){Waldmann}, {Rocchetto},
  {Tinetti}, {Barton}, {Yurchenko}, \& {Tennyson}}]{WaldmannEtal2015apjTauRex2}
{Waldmann}, I.~P., {Rocchetto}, M., {Tinetti}, G., {et~al.} 2015{\natexlab{a}},
  \apj, 813, 13, \dodoi{10.1088/0004-637X/813/1/13}

\bibitem[{{Waldmann} {et~al.}(2015{\natexlab{b}}){Waldmann}, {Tinetti},
  {Rocchetto}, {Barton}, {Yurchenko}, \&
  {Tennyson}}]{WaldmannEtal2015apjTauREx}
{Waldmann}, I.~P., {Tinetti}, G., {Rocchetto}, M., {et~al.} 2015{\natexlab{b}},
  \apj, 802, 107, \dodoi{10.1088/0004-637X/802/2/107}

\bibitem[{Wenger \& Champion(1998)}]{WengerChampion1998jqsrtSTDS}
Wenger, C., \& Champion, J. 1998, Journal of Quantitative Spectroscopy and
  Radiative Transfer, 59, 471,
  \dodoi{https://doi.org/10.1016/S0022-4073(97)00106-4}

\bibitem[{Yurchenko {et~al.}(2020)Yurchenko, Mellor, Freedman, \&
  Tennyson}]{YurchenkoEtal2020mnrasExoMolCO2}
Yurchenko, S.~N., Mellor, T.~M., Freedman, R.~S., \& Tennyson, J. 2020, Monthly
  Notices of the Royal Astronomical Society, 496, 5282,
  \dodoi{10.1093/mnras/staa1874}

\bibitem[{{Yurchenko} \&
  {Tennyson}(2014)}]{YurchenkoTennyson2014mnrasExoMolCH4}
{Yurchenko}, S.~N., \& {Tennyson}, J. 2014, \mnras, 440, 1649,
  \dodoi{10.1093/mnras/stu326}

\bibitem[{{Zhang} {et~al.}(2019){Zhang}, {Chachan}, {Kempton}, \&
  {Knutson}}]{ZhangEtal2019paspPLATON}
{Zhang}, M., {Chachan}, Y., {Kempton}, E. M.~R., \& {Knutson}, H.~A. 2019,
  \pasp, 131, 034501, \dodoi{10.1088/1538-3873/aaf5ad}

\bibitem[{{Zingales} \& {Waldmann}(2018)}]{ZingalesWaldmann2018arxivExoGAN}
{Zingales}, T., \& {Waldmann}, I.~P. 2018, \aj, 156, 268,
  \dodoi{10.3847/1538-3881/aae77c}

\end{thebibliography}

\end{document}